\documentclass[final, preprint, 5p, times, twocolumn]{elsarticle}

\newif\ifreview
\reviewfalse

\usepackage{lineno,hyperref}
\modulolinenumbers[5]

\journal{Acta Astronautica (\url{https://doi.org/10.1016/j.actaastro.2022.03.029})}

\usepackage{float}
\usepackage{mathptmx}
\usepackage{caption}
\usepackage{subcaption}
\usepackage{float}
\usepackage[utf8]{inputenc}
\usepackage{amsmath, amsfonts, amssymb}
\usepackage{etoolbox}
\usepackage{bm}
\usepackage{supertabular}
\usepackage{booktabs}
\usepackage{wasysym}
\usepackage{algorithm}
\usepackage{multirow}
\usepackage{color}
\usepackage{url}
\usepackage{braket}
\usepackage{fancyhdr}

\newcommand{\destiny}{DESTINY$^+$}

\newif\ifminorrev
\minorrevfalse 
\newcommand{\minorblue}[1]{\ifminorrev\textcolor{blue}{#1}\else{#1}\fi}
\newcommand{\minorred}[1]{\ifminorrev\textcolor{red}{#1}\else{#1}\fi}

\begin{document}

\begin{frontmatter}

\title{Mission Design of \destiny: Toward Active Asteroid (3200) Phaethon and Multiple Small Bodies}

\author[affil_ozaki]{Naoya Ozaki\corref{mycorrespondingauthor}}
\cortext[mycorrespondingauthor]{Corresponding author}
\ead{ozaki.naoya@jaxa.jp}

\author[affil_yamamoto]{Takayuki Yamamoto}
\author[affil_sokendai]{Ferran Gonzalez-Franquesa}
\author[affil_sokendai]{Roger Gutierrez-Ramon}
\author[affil_sokendai]{Nishanth Pushparaj}
\author[affil_utokyo]{Takuya Chikazawa}
\author[affil_diogene]{Diogene Alessandro Dei Tos}
\author[affil_onur]{Onur \c{C}elik}
\author[affil_nicola]{Nicola Marmo}
\author[affil_nishiyama]{Yasuhiro Kawakatsu}
\author[affil_arai]{Tomoko Arai}
\author[affil_nishiyama]{Kazutaka Nishiyama}
\author[affil_ozaki]{Takeshi Takashima}

\address[affil_ozaki]{Department of Spacecraft Engineering, Institute of Space and Astronautical Science, Japan Aerospace Exploration Agency, Sagamihara, Kanagawa, 252-5210, Japan.}
\address[affil_yamamoto]{\destiny Project Team, Institute of Space and Astronautical Science, Japan Aerospace Exploration Agency, Sagamihara, Kanagawa, 252-5210, Japan.}
\address[affil_sokendai]{Department of Space and Astronautical Science, The Graduate University for Advanced Studies, SOKENDAI, Sagamihara, Kanagawa, 252-5210, Japan.}
\address[affil_utokyo]{The University of Tokyo, Bunkyo-ku, Tokyo, 113-8656, Japan.}
\address[affil_diogene]{Deimos Space, ESA/ESOC, Darmstadt, 64293, Germany.}
\address[affil_onur]{University of Glasgow, Glasgow G12 8QQ, Scotland, United Kingdom.}
\address[affil_nicola]{Department of Mechanical and Aerospace Engineering, Sapienza University of Rome, Via Eudossiana 18, Rome, Italy.}
\address[affil_arai]{Planetary Exploration Research Center, Chiba Institute of Technology, 275-0016, Japan.}
\address[affil_nishiyama]{Department of Space Flight Systems, Institute of Space and Astronautical Science, Japan Aerospace Exploration Agency, Sagamihara, Kanagawa, 252-5210, Japan.}

\begin{abstract}
\destiny\ is an upcoming JAXA Epsilon medium-class mission to fly by the Geminids meteor shower parent body (3200) Phaethon. It will be the world's first spacecraft to escape from a near-geostationary transfer orbit into deep space using a low-thrust propulsion system. In doing so, \destiny\ will demonstrate a number of technologies that include a highly efficient ion engine system, lightweight solar array panels, and advanced asteroid flyby observation instruments. These demonstrations will pave the way for JAXA's envisioned low-cost, high-frequency space exploration plans. Following the Phaethon flyby observation, \destiny\ will visit additional asteroids as its extended mission. The mission design is divided into three phases: a spiral-shaped apogee-raising phase, a multi-lunar-flyby phase to escape Earth, and an interplanetary and asteroids flyby phase. The main challenges include the optimization of the many-revolution low-thrust spiral phase under operational constraints; the design of a multi-lunar-flyby sequence in a multi-body environment; and the design of multiple asteroid flybys connected via Earth gravity assists. This paper shows a novel, practical approach to tackle these complex problems, and presents feasible solutions found within the mass budget and mission constraints. Among them, the baseline solution is shown and discussed in depth; \destiny\ will spend two years raising its apogee with ion engines, followed by four lunar gravity assists, and a flyby of asteroids (3200) Phaethon and (155140) 2005 UD. Finally, the flight operations plan for the spiral phase and the asteroid flyby phase are presented in detail.
\end{abstract}


\end{frontmatter}


%
%
\section{Introduction}
\label{sec:introduction}
%
%
%

Low-cost and high-frequency missions are revolutionizing deep space exploration. To advance the deep space exploration technologies that enable such missions, JAXA is developing the \destiny\ (Demonstration and Experiment of Space Technology and INterplanetary voYage, Phaethon fLyby and dUst Science) mission\cite{Toyota2017}. \destiny\ will be launched by the low-cost Epsilon~S launch vehicle in the mid-2020s. Epsilon~S has the capability to insert the spacecraft into a near-geostationary transfer orbit. The spacecraft will raise its orbit by means of solar electric propulsion in a spiral-shaped trajectory that will take it to deep space. \destiny\ is the world's first mission to escape from a near-geostationary transfer orbit into deep space using a low-thrust propulsion system. This will be accomplished by highly-efficient solar electric propulsion employing an upgraded version of the $\mu 10$~ion thruster mounted on the Hayabusa2 spacecraft\cite{Nishiyama2016, Nishiyama2020, TSUDA2013356}, as well as light-weight, high-efficiency solar array panels. A simplified prototype of the latter was used briefly on the technology demonstration satellite RAPIS-1 in 2019. The demonstration of these solar electric propulsion integrated with solar array systems is among the technological objectives of \destiny.
For its nominal science mission, the spacecraft will perform a high-speed flyby observation of the active asteroid (3200) Phaethon, the parent object of the Geminids meteor shower\cite{Jewitt2006}. As an extended mission, multiple other asteroid flybys are currently under consideration.


%
%
%

The \destiny\ mission design involves the technical challenges of low-thrust many-revolution trajectory optimization under a multi-body dynamical system. To tackle the complex trajectory design, we divide the entire mission into three phases: the Spiral Orbit-Raising (SOR) phase, the Moon Flyby (MFB) phase, and the Interplanetary Transfer (IPT) phase. The SOR trajectory design needs to optimize a low-thrust many-revolutions trajectory considering the system constraints\cite{Watanabe2017, DeiTos2020SciTech, Aziz2018}. ESA's SMART-1 mission also relied on spiral orbit raising to insert the spacecraft into orbit around the Moon\cite{DICARA2005250}. However, \destiny's case has additional difficulty in that it must perform a series of lunar flybys at the right time to achieve the desired escape conditions. The MFB exploits multi-body gravitational effects of the Sun, Earth, and Moon to increase the escape energy from the Earth efficiently\cite{Belbruno2000, UESUGI1991347}. The multi-lunar-flyby trajectories have been employed for the interplanetary transfer in past missions such as JAXA's Nozomi\cite{NozomiMission} and NASA's STEREO\cite{Dunham2009}. However, the systematic design methods have not been established yet\cite{Yarnoz2016, Suda2017AAS}. Also, \destiny\ involves a new challenge to connect the IPT trajectory and the SOR trajectory by the moon flybys. In the IPT, the spacecraft will fly by multiple asteroids utilizing low-thrust maneuvers and Earth gravity assist maneuvers\cite{Sarli2015, Celik2021, Ozaki2021}. NASA's CONTOUR mission\cite{Veverka1995} \minorblue{would have} employed multiple Earth gravity assists to visit multiple small bodies. \minorblue{NASA's Lucy mission, launched in 2021, also plans to use several Earth gravity assists to visit multiple Jupiter trojans\cite{Englander2019, Olkin_2021}.} \destiny\ equips a low-thrust propulsion system with larger maneuverability, making the trajectory design more complex. Although previous studies tackled some parts of the trajectory design\cite{Watanabe2017, DeiTos2020SciTech, Yarnoz2016, Suda2017AAS, Sarli2015, Celik2021}, the overall mission design and its methodology that patches all segments together have not been reported yet.

This paper presents the mission design and flight operation overview of the 
\ifreview
\\
\fi
\destiny\ mission, thoroughly explaining the trajectory design approach for each phase. More in detail, in the SOR we implement a multi-objective evolutionary algorithm that minimizes flight time, fuel consumption, and the duration of the longest eclipse subject to system and operation constraints. In the MFB, we generate a Moon flyby trajectory database under a high-fidelity dynamical system, where these are obtained via backward propagation from the escape conditions until crossing the SOI of the Moon. In the IPT, we introduce the trajectory design method of asteroid flyby cyclers\cite{Ozaki2021} to generate initial guess trajectories and then perform low-thrust optimization to translate them to full ephemeris. After discussing of each phase of the mission, the results of the optimized end-to-end low-thrust trajectory under a multi-body dynamical system are shown. This paper also shows a preliminary planning of the flight operation, particularly for the SOR phase and the Phaethon flyby segment of the IPT. Detailed operation analyses, including navigation analyses, planetary protection analyses, and missed thrust analyses\cite{Rayman2006, Frank2015}, are not presented in this paper.

The paper is therefore structured as follows: in Section \ref{sec:destiny}, we show an overview of the mission and the spacecraft, along with requirements and specifications on the operations and the spacecraft systems; in Section \ref{sec:mission_analysis}, we describe the results of the current baseline mission design; in Section \ref{sec:trajectory_design}, we present our trajectory design approach for each phase and a patched trajectory; finally, in Section \ref{sec:flight_operation}, we discuss the flight operation plan and summarize the practical issues involved.

%
%
\section{\destiny}
\label{sec:destiny}

%
%
\subsection{Mission Objectives}


\destiny\ is an engineering and science mission with distinct but complementary objectives in both disciplines. In line with JAXA's vision to realize low-cost and high-frequency deep space exploration utilizing small launch vehicles and high-performance deep space probe platforms\cite{Kawakatsu2013}, \destiny\ will demonstrate several necessary space technologies.

The primary engineering objectives include 1) to develop spaceflight technologies using electric propulsion and expand the range of its utilization, and 2) to expand the opportunities for small body exploration by acquiring advanced asteroid flyby exploration technologies. \destiny\ will perform spiral apogee-raising and multiple lunar gravity assists in order to escape the Earth efficiently, and will then employ low-thrust and Earth gravity-assist maneuvers to fly by multiple asteroids. The spacecraft also demonstrates various advanced hardware technologies, namely an upgraded Ion Engine System (IES), thin-film light-weight Solar Array Panels (SAPs), advanced thermal control devices\cite{Akizuki2020}, and high-performance miniaturized components. The demonstration of these astrodynamics methods and spacecraft systems will allow for their use in future deep missions utilizing low-cost launch vehicles.

The primary scientific objectives are 1) to obtain better estimates of the physical properties (velocity, arrival direction, and mass distribution) and chemical composition of meteoroidal and interplanetary dust brought to Earth, and 2) to understand the geology and the dust ejection mechanism of the active asteroid (3200) Phaethon\cite{Arai2018, Arai2021}. Phaethon is the parent body of Geminid meteor shower\cite{Whipple1983, Williams1993} and recurrently ejects dust near the Sun\cite{Jewitt2010, Jewitt2013}. We plan to fly by Phaethon and other small bodies, including the asteroid (155140) 2005 UD, which is a possible break-up body from Phaethon\cite{Jewitt2006, Ohtsuka2006}, as a part of the extended mission.

%
%
\subsection{Spacecraft System Overview}

\begin{figure}[htb]
\centering
\ifreview
\includegraphics[width=\textwidth]{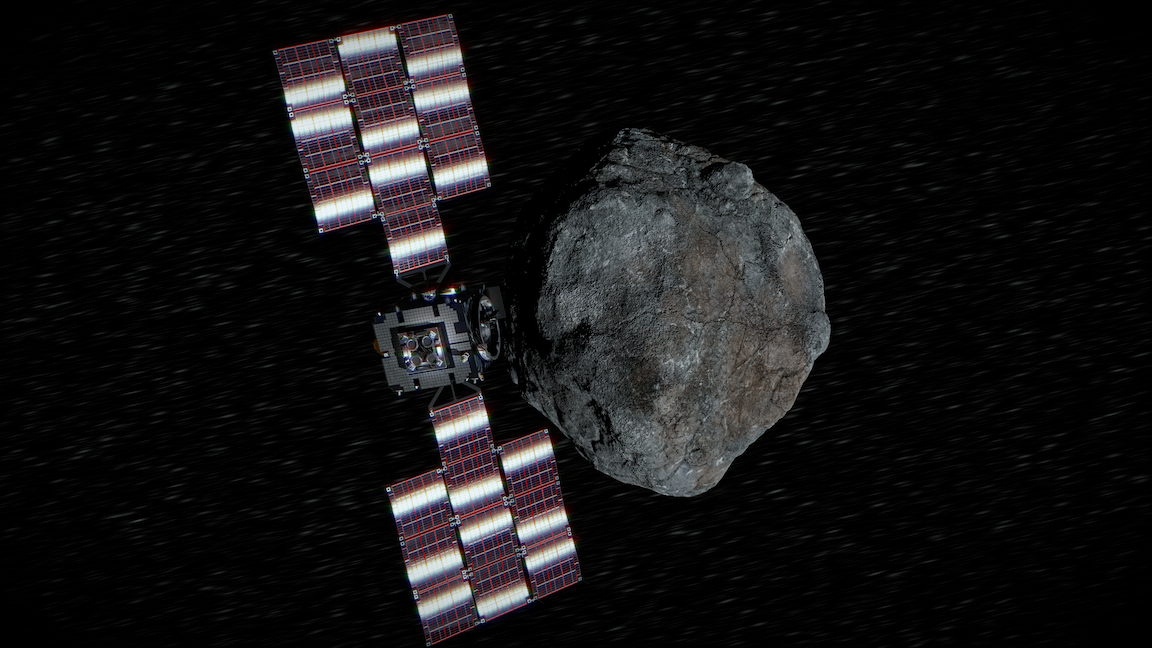}
\else
\includegraphics[width=0.45\textwidth]{img/3_destiny_spacecraft.png}
\fi
\caption{\destiny spacecraft overview.}
\label{fig:destiny_spacecraft}
\end{figure}

\destiny\ is a 480~kg spacecraft with 60~kg of Xe propellant, capable of providing 4~km/s $\Delta V$ by IES. The dry mass and the propellant mass are defined by multi-objective optimization, taking into account the launch capability of the Epsilon~S 
\ifreview
\\
\fi
launch vehicle\cite{Zuiani2013, Yam2014, Yamamoto2015}. The current baseline configuration of the spacecraft is shown in Fig.~\ref{fig:destiny_spacecraft}, and the baseline specifications of the spacecraft systems are presented in Table~\ref{tab:destiny_spec}.

Nominal scientific observations are carried out during the Phaethon flyby at a maximum relative velocity of 36~km/s and a nominal closest approach distance of 500~km from the surface. The spacecraft is equipped with three scientific instruments to observe the surface properties of Phaethon and other potential targets. They will also observe the physical and chemical properties of the dust they produce. Two onboard cameras, a Telescopic Camera for Phaethon (TCAP) and a Multi-band Camera for Phaethon (MCAP)\cite{Ishibashi2021}, will perform optical observations during flybys. The former uses a single-axis motor to drive the telescope mirror to follow the target, whereas the latter has no moving parts and has a wide field of view. TCAP is also used as an optical navigation sensor to improve the accuracy of orbit determination during the flyby. The spacecraft will study the asteroid's dust ejection processes\cite{Jewitt2006}, as well as the properties of interplanetary dust throughout its journey in space. The \destiny\ Dust Analyzer (DDA), developed by the University of Stuttgart, counts the dust collected along with its impact state\cite{KRUGER201922}. 

\begin{table}[h!]
	\caption{\destiny\ specifications}\label{tab:destiny_spec}
	\centering 
    \ifreview
    \scalebox{0.9}{
    \fi
	\begin{tabular}{@{}l  l @{}} 
		\toprule
		\multicolumn{2}{c}{\textbf{Spacecraft} } \\ 
		\midrule 
		Initial mass (wet)      & 480~kg  \\
		Power generated & 2.6~kW~@~EOL \\
		& thin-film lightweight SAP\\
		& single-axis gimbal  \\
		Battery capability & Eclipse duration $<$ 90 min.\\ 
		Attitude control & 3-axis stabilization \\
		Lifetime & $>$ 6.2 years\\
		\midrule
		\multicolumn{2}{c}{\textbf{Ion Engine System} } \\ 
		\midrule
		Thrust & 40~mN (4~units in operation) \\
		 & 36~mN (3~units in operation) \\
		Specific impulse $I_{sp}$ & 3,000~s   \\
		Propellant mass & approx. 60~kg    \\
		\midrule
		\multicolumn{2}{c}{\textbf{Scientific Instruments} } \\ 
		\midrule
		TCAP & 1-axis rotatable telescope \\
		 \ \ \ Bandwidth & $\lambda=400$ to 800~nm \\
		 \ \ \ FOV & 0.80~deg$\times$0.80 deg\\
		 \ \ \ Image sensor pixel & 2048$\times$2028\\
		MCAP & Multiband camera \\
		 \ \ \ Bandwidth & $\lambda=425/550/700/850$~nm \\
		 \ \ \ FOV & 6.5~deg$\times$6.5~deg\\
	     \ \ \ Image sensor pixel & 2048$\times$2028\\
		DDA & Dust analyzer \\
		 \ \ \ Measurement range & $10^{-16}$ to $10^{-6}$~g\\
		\midrule
		\multicolumn{2}{c}{\textbf{Launch Conditions} } \\ 
		\midrule
		Launcher & Epsilon~S with kick stage   \\
		Perigee/Apogee altitude & 230~km/37,000~km\\
		Inclination & 31~deg\\
		Nominal launch window & 2024 Jul-Sep \\
		\bottomrule
	\end{tabular}
    \ifreview
    }
    \fi
\end{table}

%
%
\section{Baseline Mission Design}
\label{sec:mission_analysis}

This paper presents the comprehensive mission design through all mission phases, including the extended mission. We first clarify the definition of mission phases, present the mission design options, and show the one that is selected as the baseline trajectory. 

In this study, we use JPL DE430 ephemerides to obtain the state vectors of the planets and Moon as well as the SPICE generic kernels pck00010.tpc, gm\_de431.tpc, and naif0012.tls for the other astrodynamics parameters. \minorblue{The orbital elements of the asteroids at their corresponding flyby epoch are obtained from JPL Horizon's online query system (as of November 30, 2021).}

%
%
\subsection{Mission Phases}

\begin{figure}[htb]
\centering
\ifreview
\includegraphics[width=\textwidth]{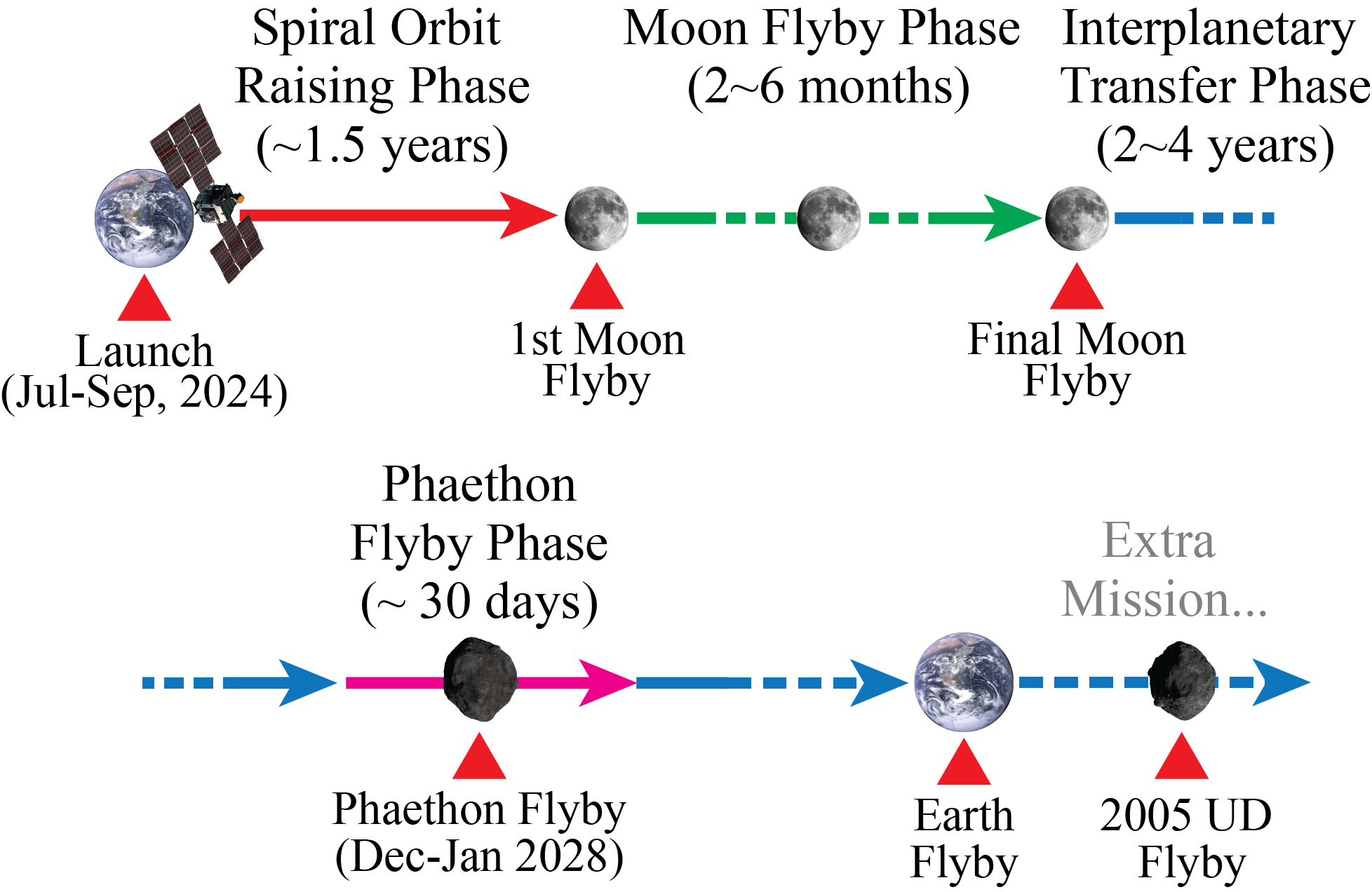}
\else
\includegraphics[width=0.4\textwidth]{img/mission_scenario.png}
\fi
\caption{Mission scenario of \destiny.}
\label{fig:mission_scenario}
\end{figure}

We divide the entire mission into several phases for the convenience of the operation studies. Figure~\ref{fig:mission_scenario} shows the definition and approximate duration of these phases. For mission design purposes, the Phaethon flyby is included in the interplanetary transfer phase. 

The three phases of the mission are the following:

\begin{description}
    \item 1. Spiral Orbit-Raising phase (SOR): From launch to the first Moon flyby.
    \item 2. Moon Flyby phase (MFB): From the first Moon flyby to the exit of the Earth's SOI.
    \item 3. Interplanetary Transfer phase (IPT):  From the exit of the Earth's SOI to the end of the mission.
\end{description}

%
%
\subsection{Mission Options}

This subsection discusses the mission design options considering different launch dates and different flyby sequences. The detailed trajectory design approach is instead explained in the next section.

\begin{figure}[h!]
\centering
\ifreview
\includegraphics[width=\textwidth]{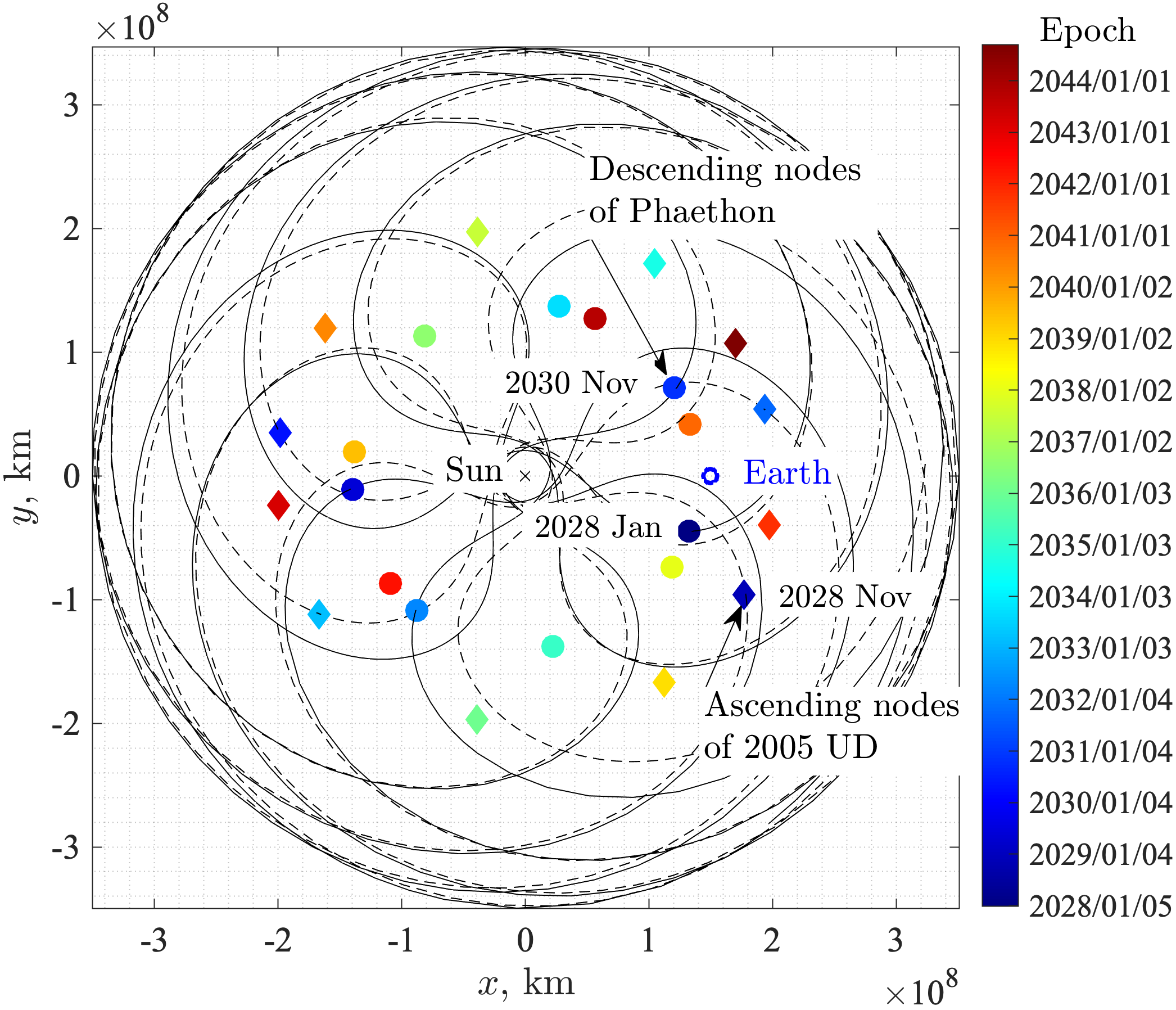}
\else
\includegraphics[width=0.45\textwidth]{img/phaethon_flyby_opportunities.png}
\fi
\caption{Flyby opportunities to Phaethon \minorblue{(solid line) and} 2005 UD \minorblue{(dashed line)} in the Sun-centered, Sun-Earth line fixed rotational frame.}
\label{fig:flyby_opportunities}
\end{figure}

Our target asteroids, Phaethon and 2005 UD, have highly inclined elliptical orbits. The preliminary analyses\cite{Celik2021} found that \destiny\ can perform flybys in the vicinity of either the descending or ascending point of the asteroid's orbits. Note that one of the nodes is located inside Mercury's orbit, which is difficult to access with the spacecraft's propulsive capabilities, while the other node is in the proximity of Earth orbit, making it the perfect candidate node for a flyby. Figure \ref{fig:flyby_opportunities} plots the accessible descending and ascending nodes of Phaethon and 2005 UD nodes for the next few decades in the Sun-centered, Sun-Earth line fixed rotational frame. Due to communication requirements that constrain the Earth distance, flybys of Phaethon are possible either in January 2028 or November 2030.

\begin{figure*}[h!]
\centering
\ifreview
\includegraphics[width=\textwidth]{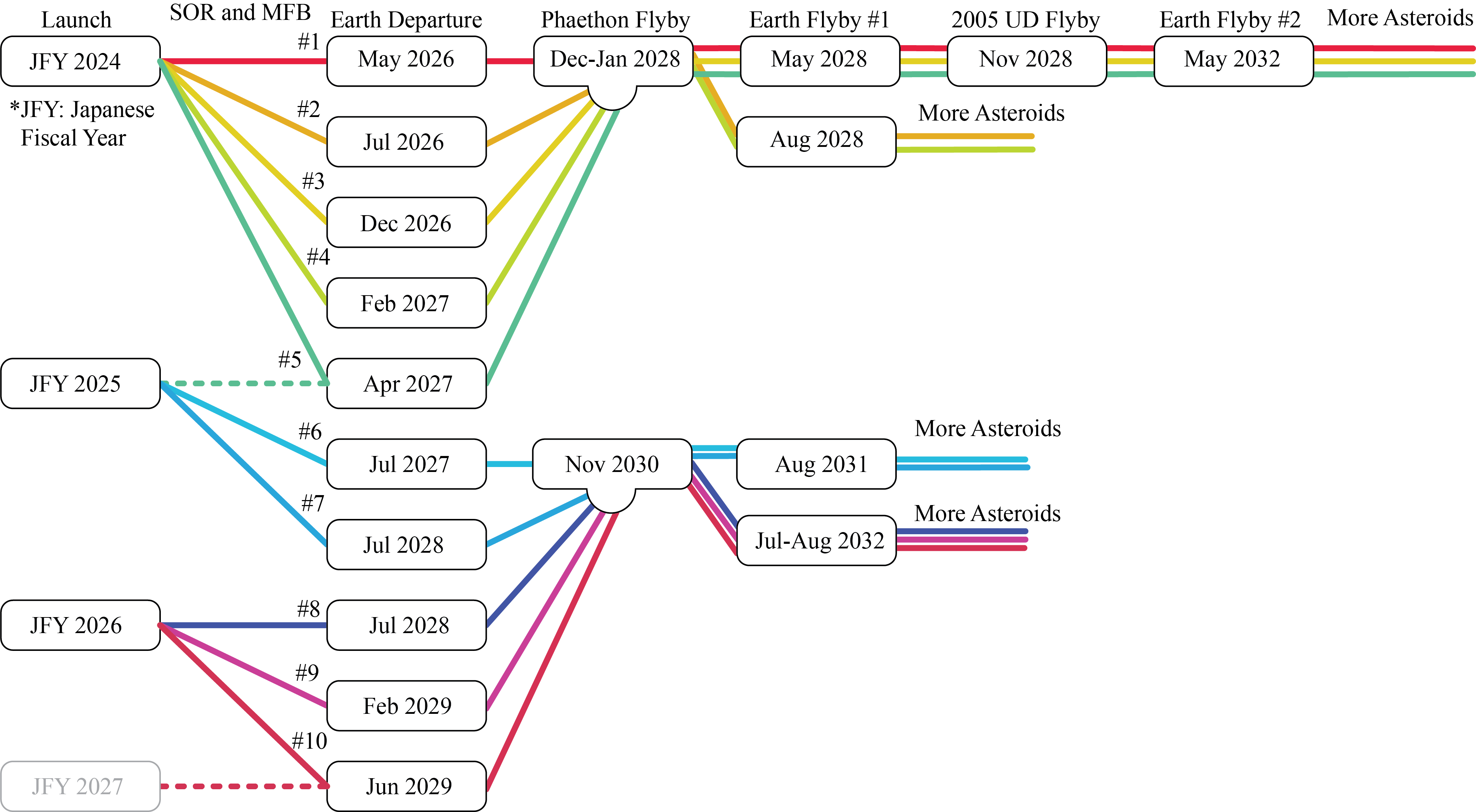}
\else
\includegraphics[width=0.9\textwidth]{img/metromap.png}
\fi
\caption{Mission scenario tree of \destiny.}
\label{fig:metro_map}
\end{figure*}

Figure~\ref{fig:metro_map} depicts the mission scenario tree for representative launch dates and flyby sequences. The case IDs correspond to the those of the IPT trajectories in defined in Section \ref{sec:ipt}. Due to the constraints of the spacecraft development schedule and the availability of the launch site, the nominal launch window is between July and September 2024, the earliest possible launch date. If the development schedule is delayed, additional launch windows are available in 2025, 2026, and 2027 (Japanese fiscal years, \minorblue{running from April through March of the next natural year}).

Cases \#1, \#3, and \#5 are the only solutions that \minorred{allow the spacecraft to visit both Phaethon and 2005 UD, perform the Earth flyby after 2005 UD, and then fly by various asteroids later. For the additional launch window cases, including Case \#2 and \#4, we must choose either stopping by 2005 UD or visiting more asteroids.} Among Cases \#1, \#3, and \#5, Case \#1 is not a realistic solution considering the SOR operation, and Case \#3 is the earliest option among the realistic solutions. Therefore, Case \#3 is selected as the baseline scenario.

%
%
\subsection{Baseline Trajectory}

\begin{figure}[h!]
\centering
\ifreview
\includegraphics[width=\textwidth]{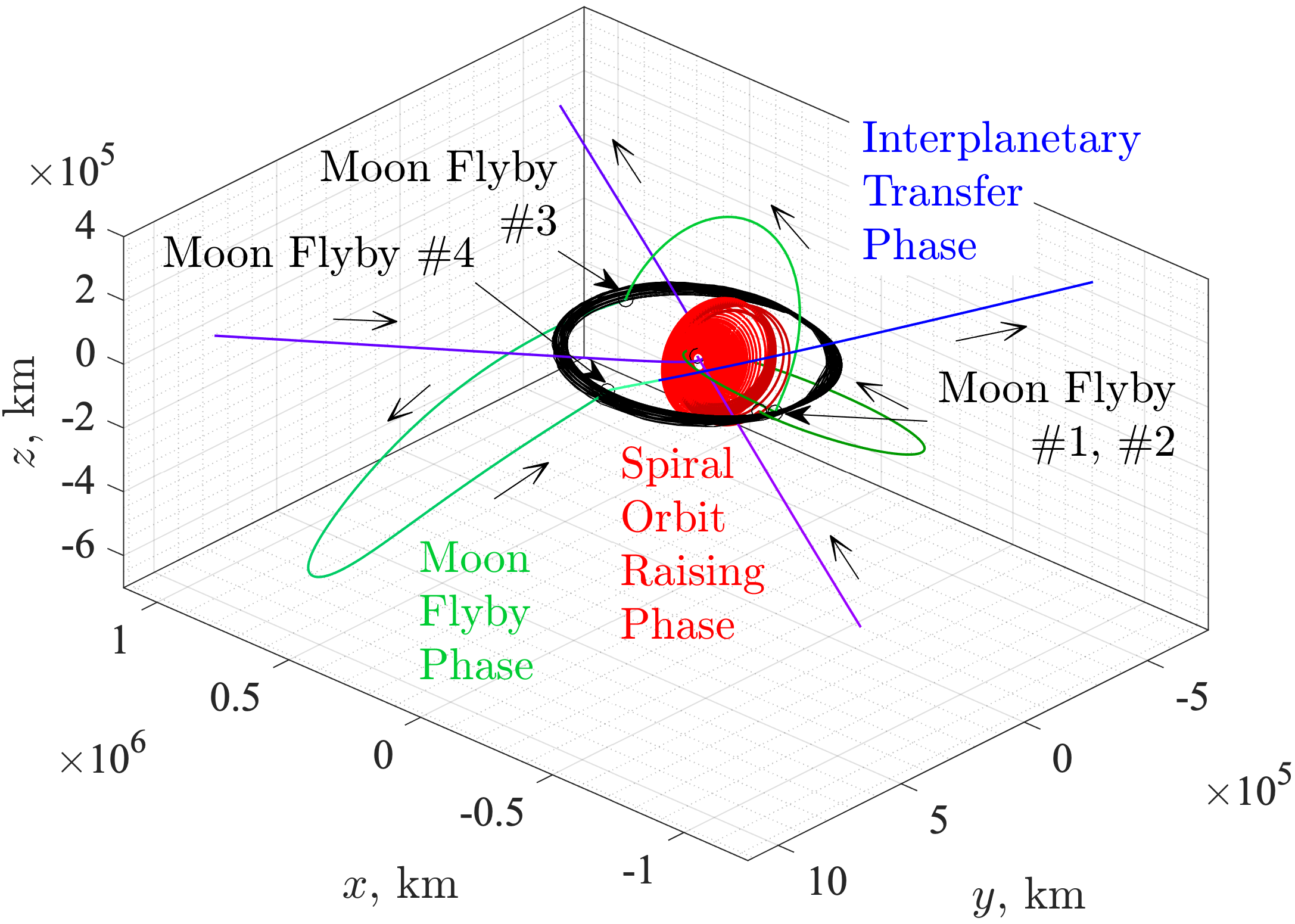}
\else
\includegraphics[width=0.45\textwidth]{img/destiny_trj_iner_ec.png}
\fi
\caption{Baseline near Earth trajectory of \destiny\ in the Earth-centered ECLIPJ2000 inertial frame.}
\label{fig:baseline_earthescape_trajectory}
\end{figure}

\begin{figure}[h!]
\centering
\ifreview
\includegraphics[width=\textwidth]{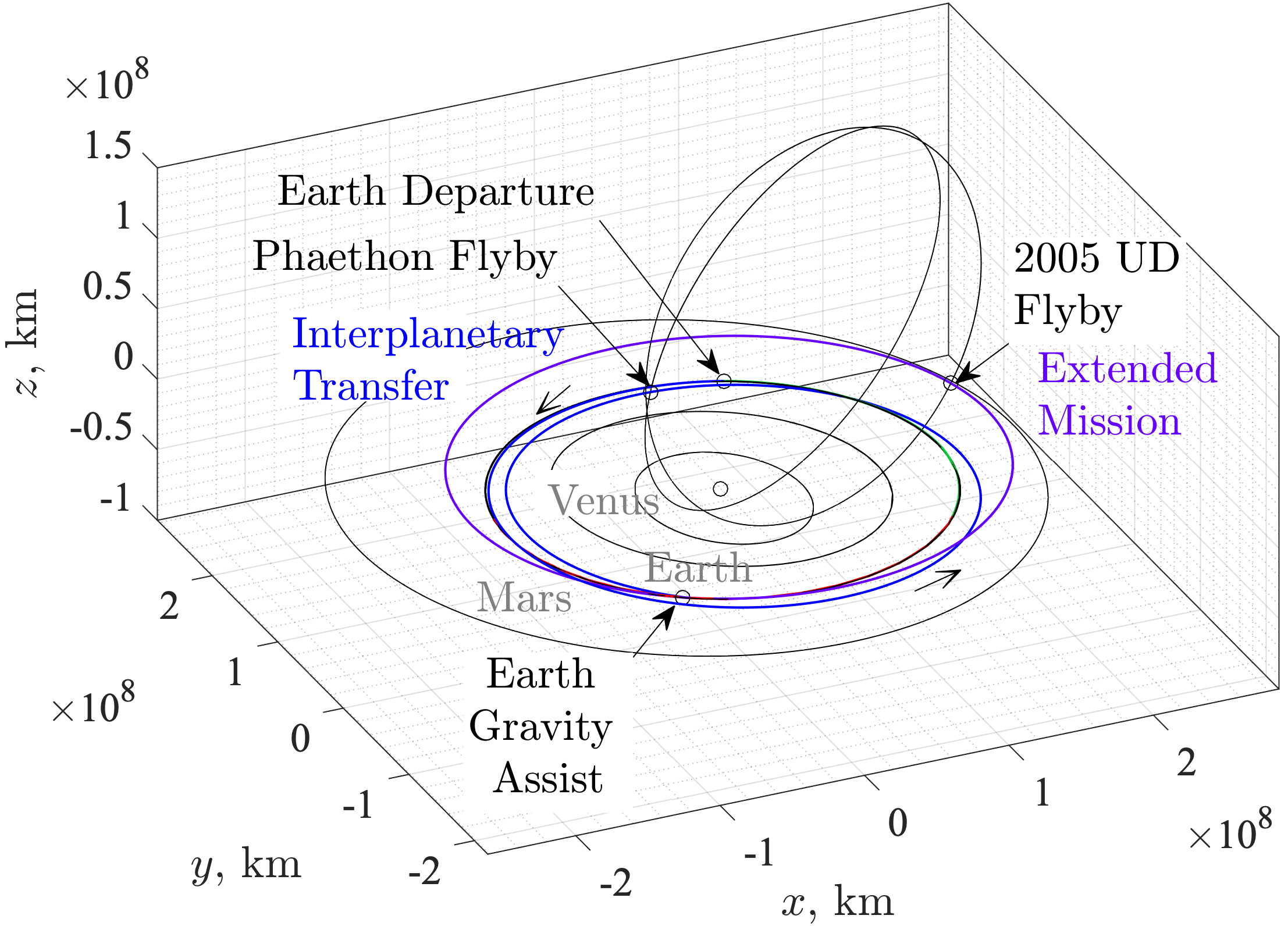}
\else
\includegraphics[width=0.45\textwidth]{img/destiny_trj_iner_sc.png}
\fi
\caption{Baseline interplanetary trajectory of \destiny\ in the Sun-centered ECLIPJ2000 inertial frame.}
\label{fig:baseline_interplanetary_trajectory}
\end{figure}

Figures~\ref{fig:baseline_earthescape_trajectory}~and~\ref{fig:baseline_interplanetary_trajectory} illustrate the baseline trajectory (Case \#3), and Table~\ref{tab:baseline_trajectory} describes the sequence of events along this baseline trajectory. In this baseline trajectory, 
\ifreview
\\
\fi
\destiny\ spends 2~years and consumes about 80\% of the propellant in the SOR, performs 4 Moon flybys achieving the Earth escape $V_{\infty}$ of 1.624 km/s, flies by Phaethon on January 4 in 2028 in the nominal mission, and visits 2005 UD and more asteroids in the extended mission. Some deviation from the baseline trajectory is expected in the presence of uncertainties that originate from navigation errors and missed thrust, particularly in the SOR. As shown in the spacecraft masses and the event date and time in Table~\ref{tab:baseline_trajectory}, we currently allocate large fuel consumption (46.8~kg) with long time of flight (25~months) in the SOR including margins for uncertainties. \minorblue{Because we assign the forced coast period, we set an IES duty cycle of 100\% in the nominal trajectory design of the SOR; in the nominal trajectory of the IPT, we impose an IES duty cycle of 80\% without a forced coast period.}


\begin{table*}[htb]
\caption{\label{tab:baseline_trajectory} Sequence of events in the baseline trajectory. Velocity values are with respect to their corresponding gravity assist or flyby bodies.}
\ifreview
\scalebox{0.75}{
\fi
\centering
\begin{tabular}{lllll}
\toprule
\textbf{Date and time (UTC)} & \textbf{Event} & \textbf{Spacecraft mass $\mathbf{m}$ (kg)} & $\mathbf{V}_{\infty}$ \textbf{(km/s)} & $\mathbf{V}_{\textrm{rel}}$ \textbf{(km/s)} \\\midrule
2024 JUL 01 17:00:00 & Launch & 480.0 & - \\
2025 FEB 23 13:47:51 & Escape from radiation belt & 455.9 & - & - \\
2026 AUG 14 05:56:53 & Moon flyby \#1 & 433.2 & 0.9761 & -\\
2026 SEP 11 03:37:26 & Moon flyby \#2 & 433.2 & 1.0551 & -\\
2026 SEP 25 07:30:42 & Moon flyby \#3 & 433.2 & 1.0457 & -\\
2026 NOV 26 09:40:12 & Moon flyby \#4 & 433.2 & 1.3843 & -\\
2026 NOV 27 04:41:19 & Earth closest approach & 433.2 & 1.6240 & - \\
2028 JAN 04 10:34:03 & Phaethon flyby & 422.9 & - & 33.7923 \\
2028 MAY 18 00:47:28 & Earth flyby & 421.7 & 2.5260 & -\\
2028 NOV 11 23:57:11 & 2005 UD flyby & 421.7 & -& 23.7479\\
2032 MAY 17 07:43:53 & Earth flyby & 421.7 & 2.5454 & -\\
\bottomrule
\end{tabular}
\ifreview
}
\fi
\end{table*}
%
%
\section{Trajectory Design Approach}
\label{sec:trajectory_design}

\begin{figure}[h!]
\centering
\ifreview
\includegraphics[width=\textwidth]{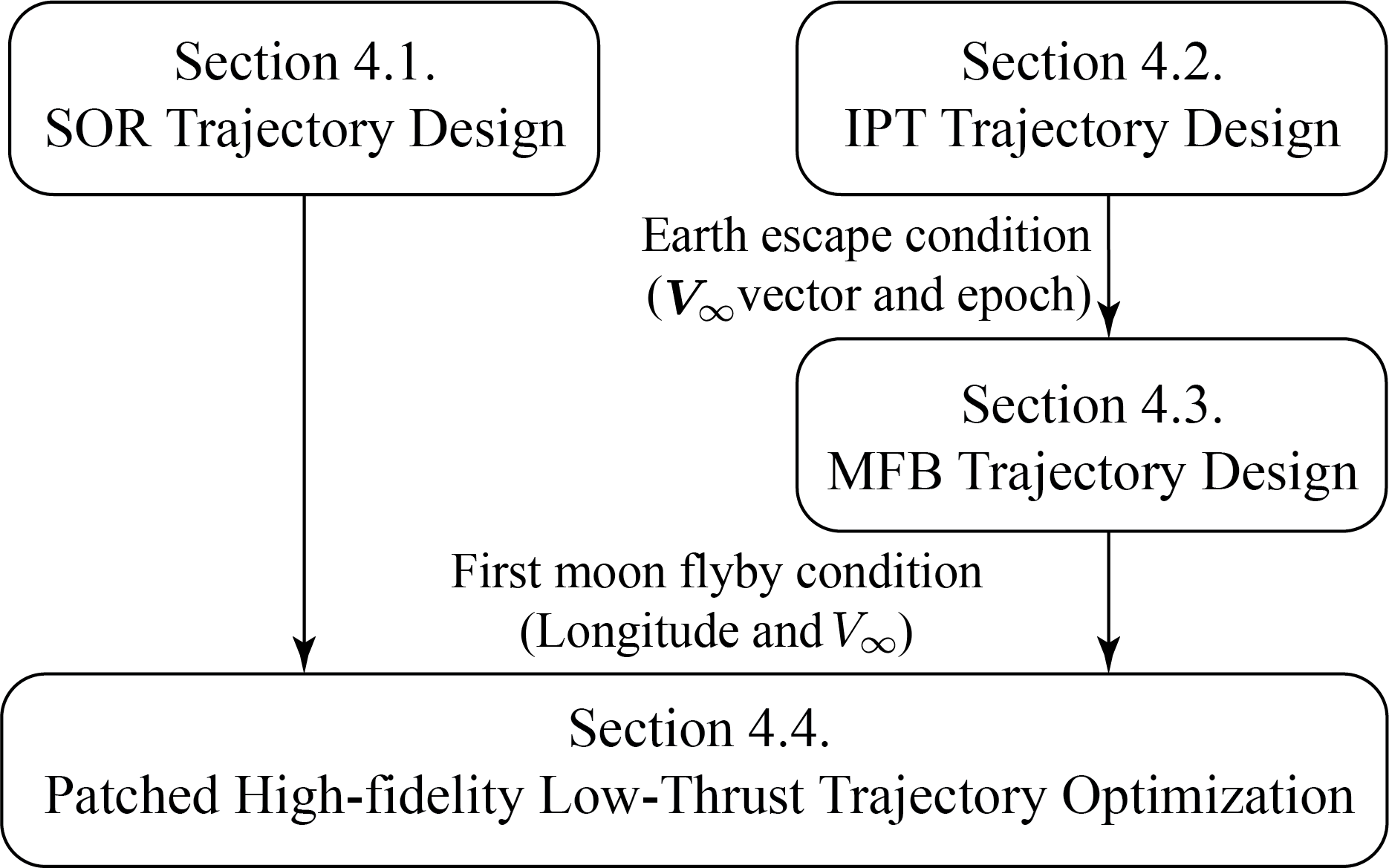}
\else
\includegraphics[width=0.43\textwidth]{img/trjdesign_approach.png}
\fi
\caption{Trajectory design approach overview.}
\label{fig:trjdesign_approach}
\end{figure}

This section presents the trajectory design approach for each phase, followed by the optimization of the complete low-thrust trajectory in \minorblue{the full ephemeris model. This model considers the gravities of the Earth, Moon, Sun, Venus, Mars, and Jupiter, obtained from the JPL DE430 model. Solar radiation pressure and higher order effects of Earth's and Moon's gravity are not taken into account}. Figure~\ref{fig:trjdesign_approach} summarizes the design procedure, which is structured as follows: first, the SOR and IPT trajectories are generated separately; after designing the IPT trajectory, the MFB trajectory is obtained by propagating backward the initial conditions of the ITP segment up to the possible encounter with the Moon; finally, the first Moon flyby conditions of the SOR and MFB trajectories are compared to find a suitable pair that should be optimized. 

The three phases are presented in the order they are designed. That is, the SOR and IPT are presented first, and the MFB is discussed afterwards. Although the MFB phase takes place chronologically in between the SOR and IPT, it is designed last because the multi-flyby trajectory is chosen following the constraints at the interfaces with the SOR and IPT. The following sections describe the details of each process.

%
%
\subsection{Spiral Orbit-Raising Phase} 

\minorred{The SOR trajectory design significantly impacts the spacecraft system design, including radiation tolerance, power budget during eclipses, lifetime, and IES $\Delta V$ budget. In the preliminary analyses, we apply a Multi-Objective Evolutionary Algorithm (MOEA)\cite{Watanabe2017} to the entire SOR trajectory considering the radiation environment, eclipse duration, time of flight, and fuel consumption. The result showed that the trajectory could be divided into three sub-phases for an enhanced optimization. The first sub-phase, SOR-1, is coasting for the initial checkout operation and is expected to last 30 days. In the second sub-phase, SOR-2, we continuously operate the ion engine \minorblue{except during the eclipse period} so that \destiny can escape from the radiation belt (altitude~$<$~20,000~km) as quickly as possible. In the third sub-phase, SOR-3, we initiate coasting arcs in order to reduce the fuel consumption and design low-thrust many-revolution trajectories. To solve the SOR-3 problem, we employ an averaging method to propagate the low-thrust many-revolutions trajectory\cite{Zuiani2015} with the MOEA under three objective functions:  time of flight, IES fuel consumption, and duration of the longest eclipse. The spacecraft will experience a number of eclipses throughout this phase; the duration of the longest eclipse is minimized.}

\begin{figure}[htb]
\centering
\ifreview
\includegraphics[width=0.6\textwidth]{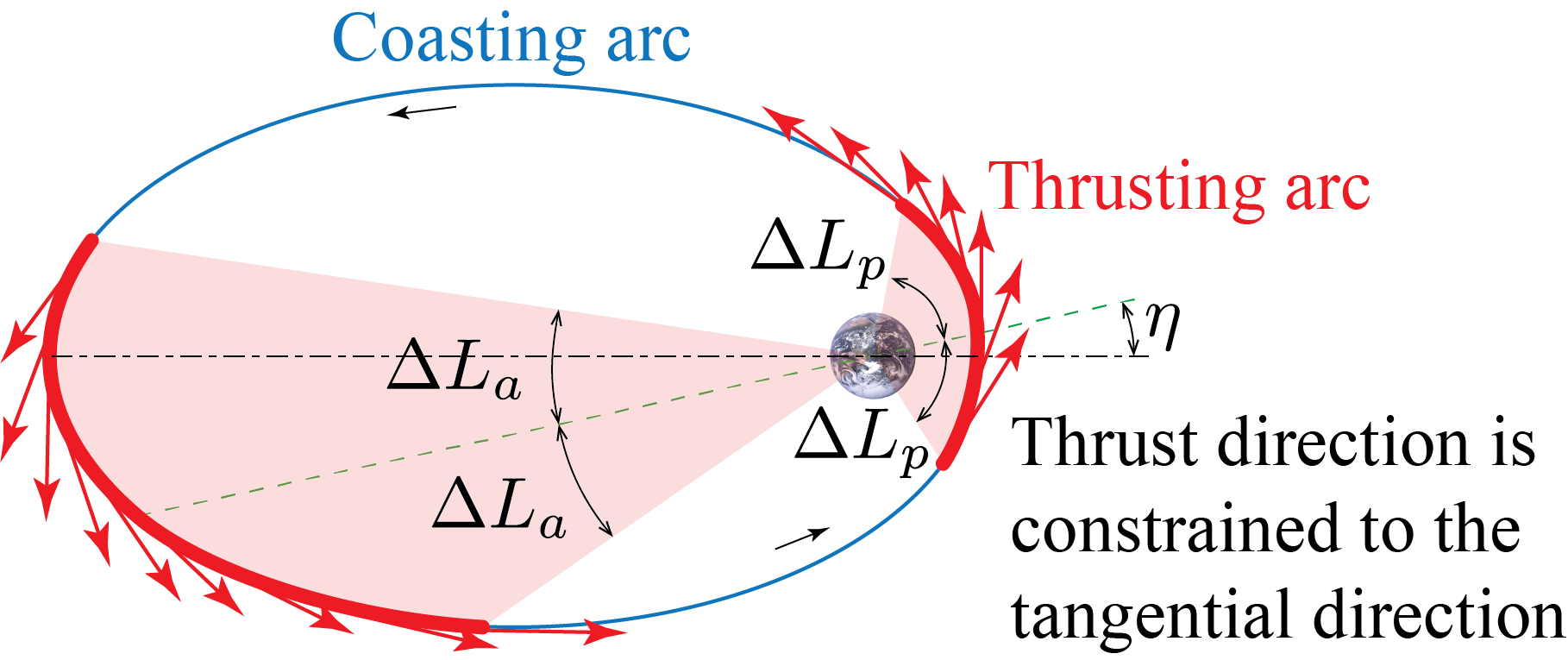}
\else
\includegraphics[width=0.35\textwidth]{img/spiral_parameters.png}
\fi
\caption{Design parameters of spiral trajectory.}
\label{fig:spiral_parameters}
\end{figure}

In the SOR-3 trajectory design, we split the trajectory into ten segments with the same \minorblue{duration} and apply the control policy defined by the design parameters 
\ifreview
\\
\fi
$(\Delta L_p, \Delta L_a, \eta)$ \minorblue{defined by true anomaly, as} shown in Fig. \ref{fig:spiral_parameters} for each segment. \minorblue{That is, the spacecraft accelerates when its true anomaly is in the range of $[-\Delta L_p+\eta, \Delta L_p+\eta]$ and $[\pi-\Delta L_a+\eta, \pi+\Delta L_a+\eta]$.} The perigee thrusting arc changes the apogee altitude, the apogee thrusting arc changes the perigee altitude, and $\eta$ will change the major axis direction which helps avoiding eclipses. In the thrusting arcs, the thrust direction is constrained to the tangential direction\cite{Zuiani2015}. The IES acceleration stops during the eclipse.

\begin{figure}[htb]
\centering
\ifreview
\includegraphics[width=\textwidth]{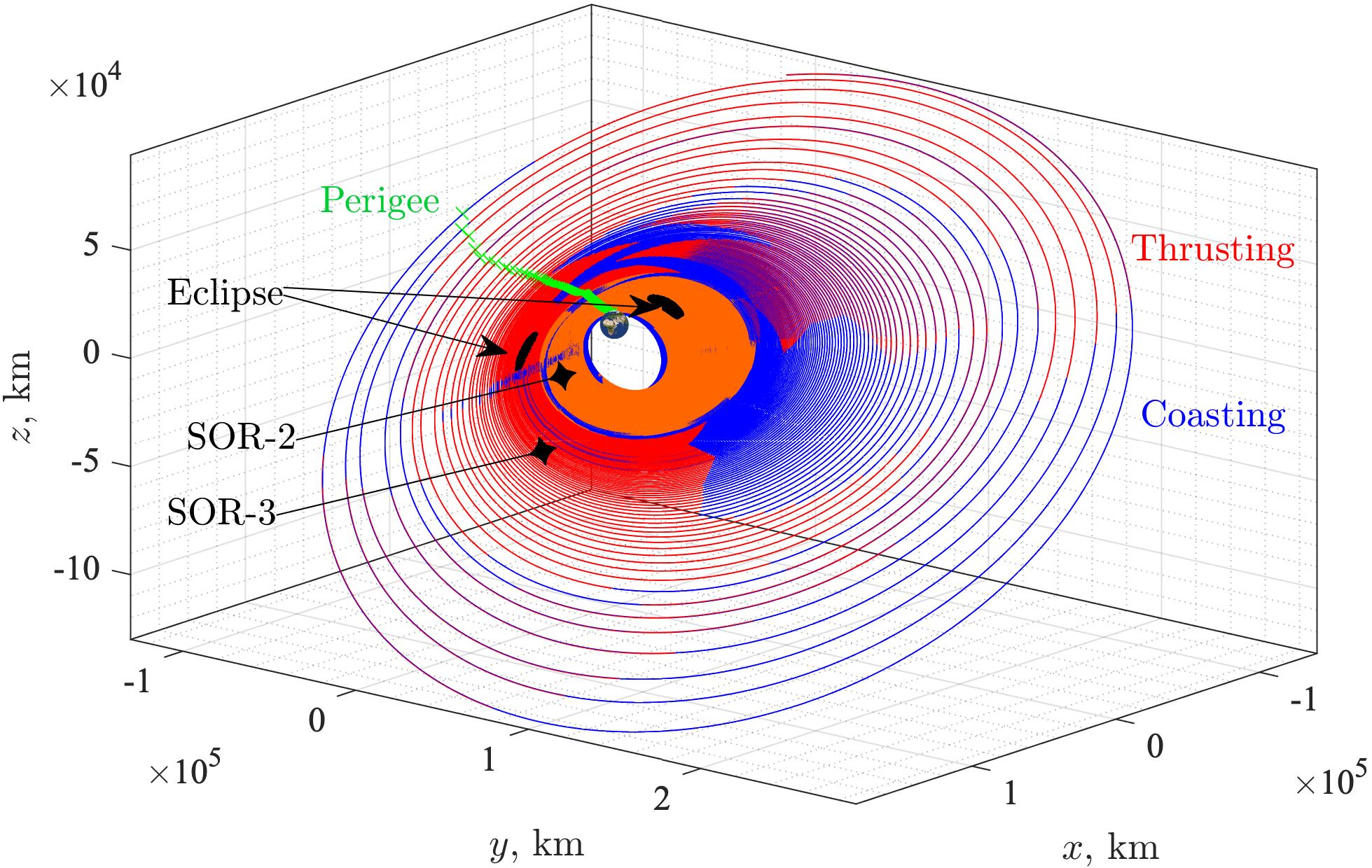}
\else
\includegraphics[width=0.46\textwidth]{img/9875_Traj.png}
\fi
\caption{Example of the SOR trajectory (IES thrust magnitude$ = 40$~mN).}
\label{fig:spiral_example}
\end{figure}

Table~\ref{tab:spiral_parameters} summarizes the objective function, design parameters, terminal boundary condition, and propagation condition of MOEA. Figure \ref{fig:spiral_example} shows one example trajectory where the time of flight is 534~days and fuel consumption is 45.239~kg. 

\begin{table}[htb]
\caption{\label{tab:spiral_parameters} SOR-3 multi-objective evolutionary algorithm settings.}
\centering
\begin{tabular}{l}
\toprule
\textbf{Objective function}\\
 \ \ \ 1. Minimize the time of flight\\
 \ \ \ 2. Minimize the fuel consumption\\
 \ \ \ 3. Minimize the longest eclipse duration\\
\midrule
\textbf{Design parameters}\\
 \ \ \ 1. Perigee thrust arc $\Delta L_{p,i}, i\in\{1,2,...,10\}$\\
 \ \ \ 2. Apogee thrust arc $\Delta L_{a,i}, i\in\{1,2,...,10\}$\\
 \ \ \ 3. Asymmetric angle $\eta_i, i\in\{1,2,...,10\}$\\
 \ \ \ 4. Launch date $d_{0}$ and launch time $t_{0}$\\
\midrule
\textbf{Terminal boundary condition}\\
 \ \ \ The ascending or descending node radius w.r.t.\\
 \ \ \ Moon orbital plane is more than 385,000km altitude.\\
\midrule
\textbf{Propagation condition}\\
 \ \ \ 1. Thrust direction is along velocity direction\\
 \ \ \ 2. Ion engine is suspended during eclipse\\
\bottomrule
\end{tabular}
\end{table}

Figure~\ref{fig:spiral_tof_vs_dv} illustrates the trade-off between the time of flight and propellant consumption for different IES thrust magnitudes. In these figures, the propellant consumption is between 42~kg and 48~kg, while the onboard propellant mass is 60~kg. Comparing Figs.~\ref{fig:spiral_tof_vs_dv_36mN}~and~\ref{fig:spiral_tof_vs_dv_40mN} shows that the change IES thrust magnitude is less sensitive to propellant consumption, but more sensitive to time of flight. Although the spacecraft is capable of 40~mN IES operation, to increase the robustness against uncertainties we also design the trajectory with various thrust magnitudes, including 36~mN, and ensure sufficient timing margin \minorblue{(more than several months)} until the first Moon flyby.

\ifreview
\begin{figure}
     \centering
     \begin{subfigure}[htb]{\textwidth}
         \centering
         \includegraphics[width=\textwidth]{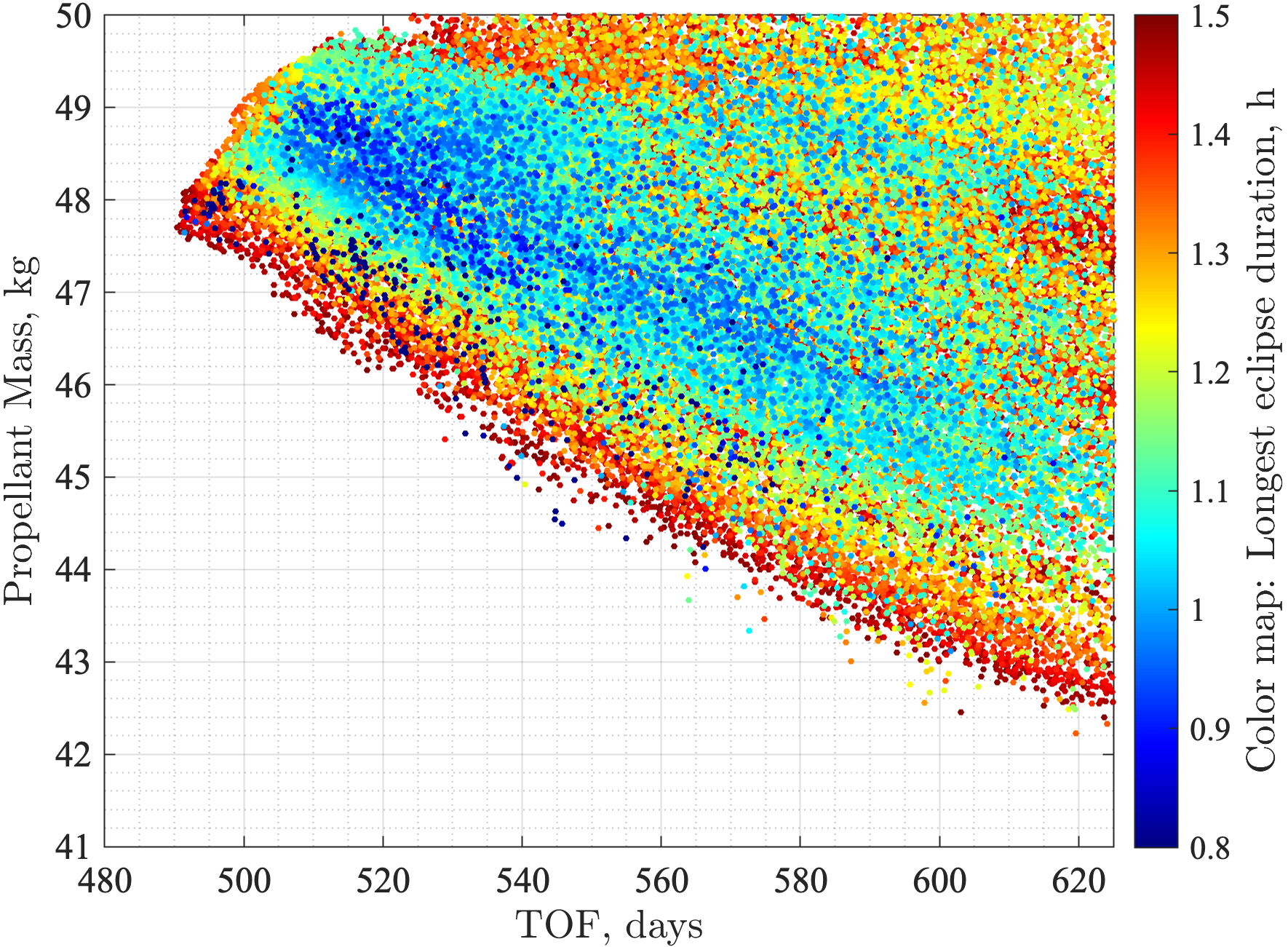}
         \caption{IES thrust magnitude $ = 36$mN}
         \label{fig:spiral_tof_vs_dv_36mN}
     \end{subfigure}
     \begin{subfigure}[htb]{\textwidth}
         \centering
         \includegraphics[width=\textwidth]{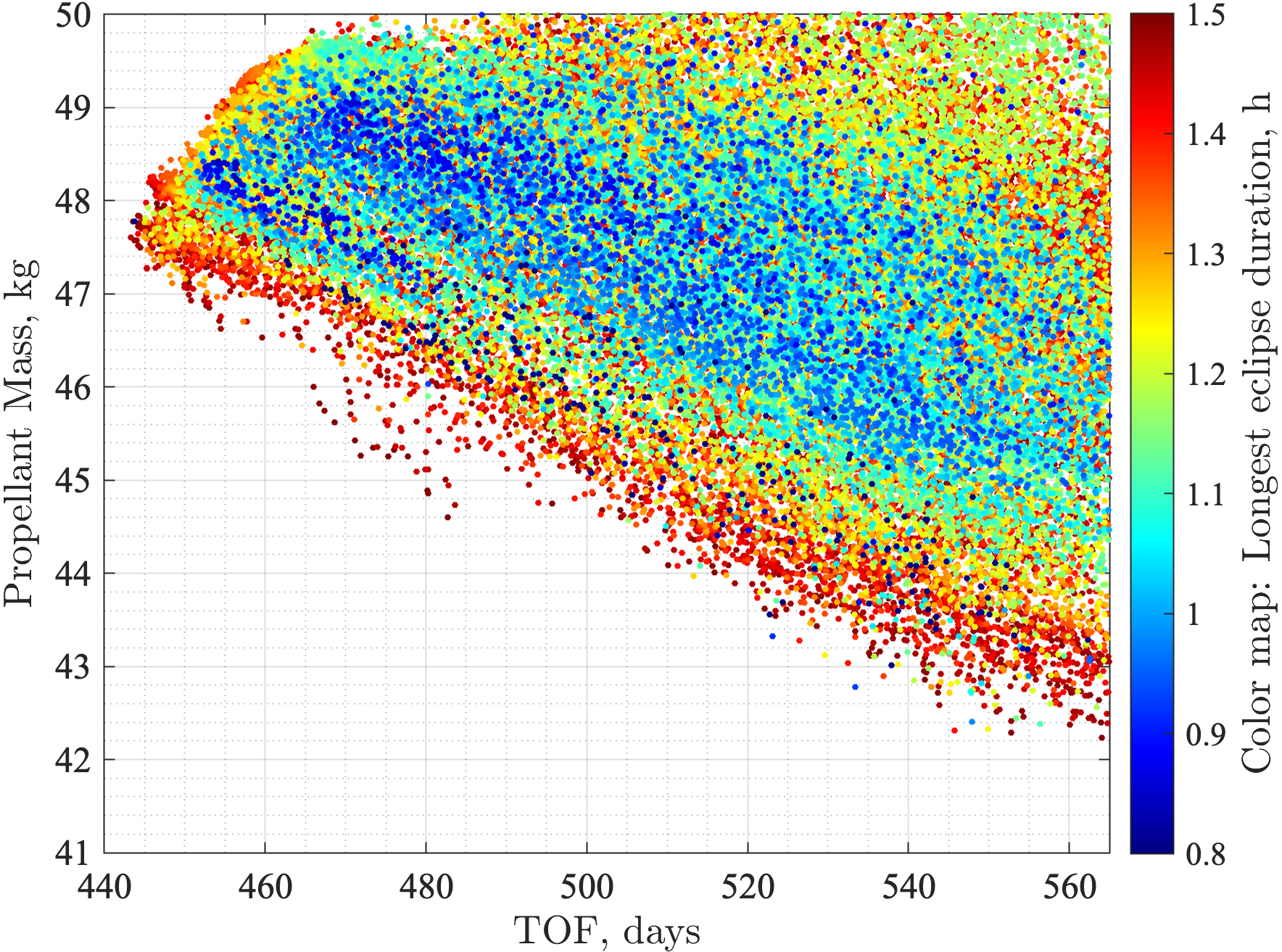}
         \caption{IES thrust magnitude $ = 40$mN}
         \label{fig:spiral_tof_vs_dv_40mN}
     \end{subfigure}
     \caption{Spiral trajectories.}
     \label{fig:spiral_tof_vs_dv}
\end{figure}
\else
\begin{figure}
     \centering
     \begin{subfigure}[htb]{0.46\textwidth}
         \centering
         \includegraphics[width=\textwidth]{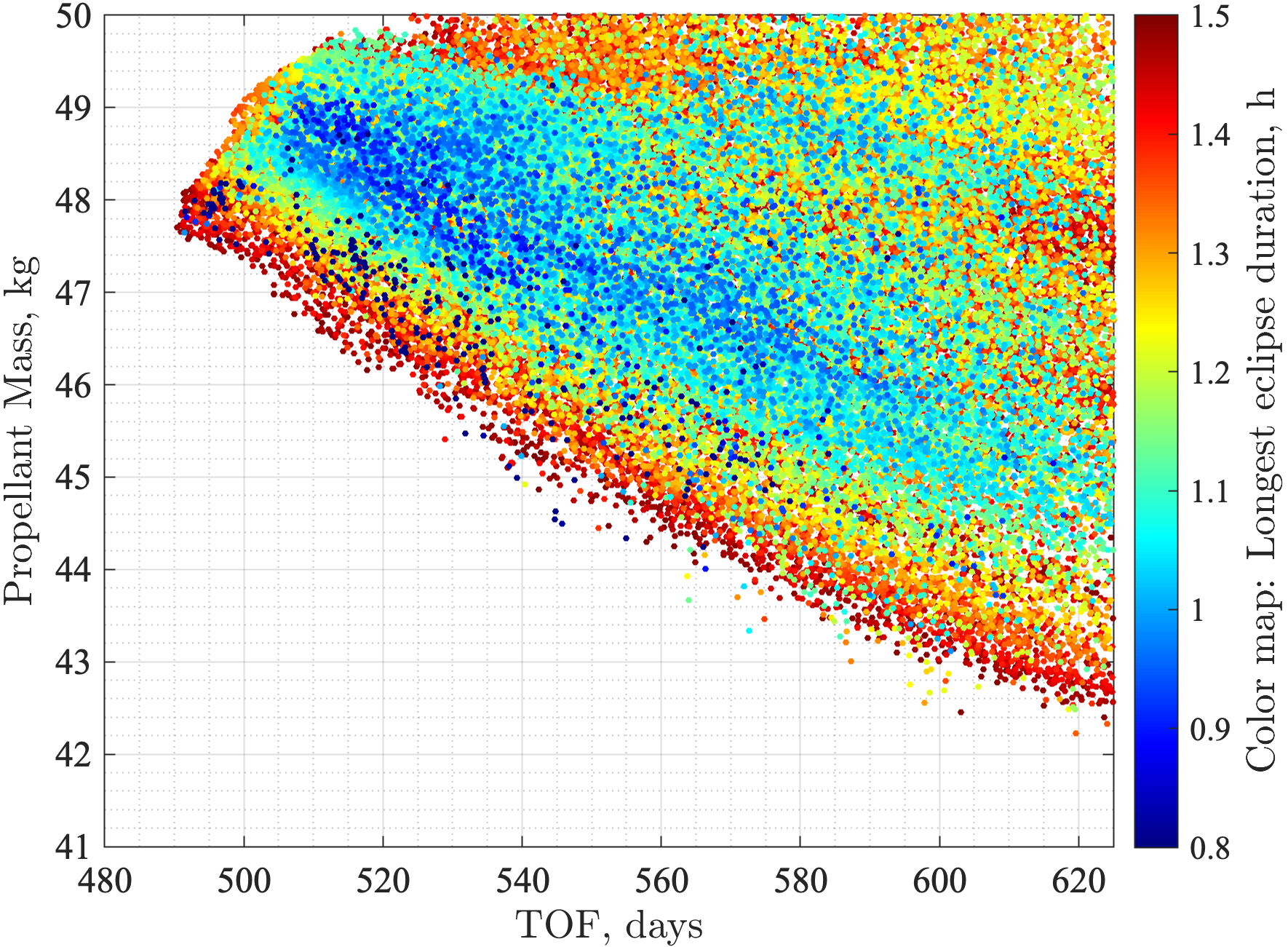}
         \caption{IES thrust magnitude $ = 36$mN}
         \label{fig:spiral_tof_vs_dv_36mN}
     \end{subfigure}
     \begin{subfigure}[htb]{0.46\textwidth}
         \centering
         \includegraphics[width=\textwidth]{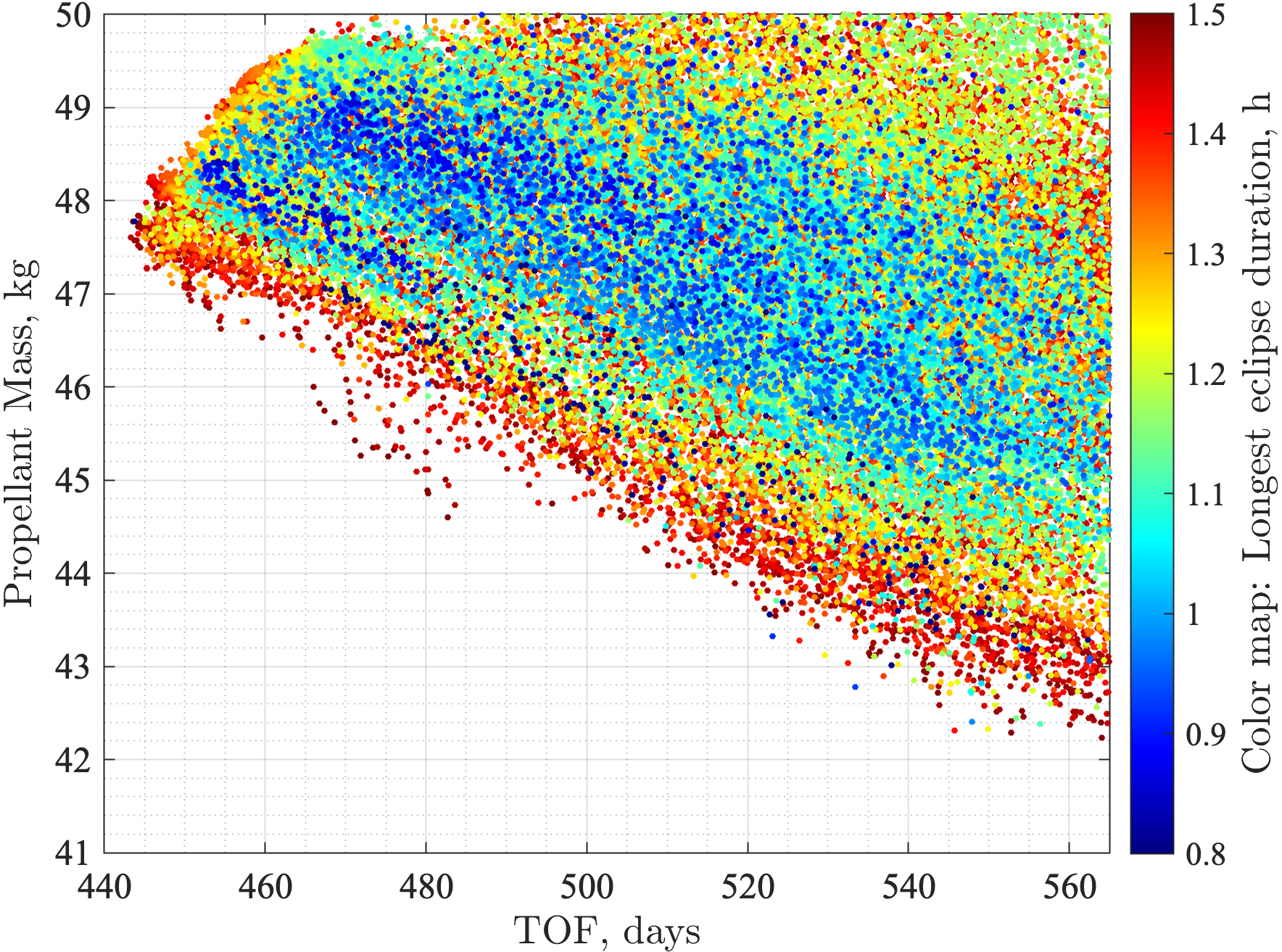}
         \caption{IES thrust magnitude $ = 40$mN}
         \label{fig:spiral_tof_vs_dv_40mN}
     \end{subfigure}
     \caption{Spiral trajectories.}
     \label{fig:spiral_tof_vs_dv}
\end{figure}
\fi

Figure \ref{fig:spiral_launchwindow_all} plots the feasible launch windows for different IES thrust magnitudes, with contours of the maximum eclipse duration. These figures indicate that 
\ifreview
\\
\fi
\destiny can be launched on any day as long as the maximum eclipse duration allowed is 1.5 hours, which is within the design specifications of the battery capacity.

\ifreview
\begin{figure}
     \centering
     \begin{subfigure}[htb]{\textwidth}
         \centering
         \includegraphics[width=\textwidth]{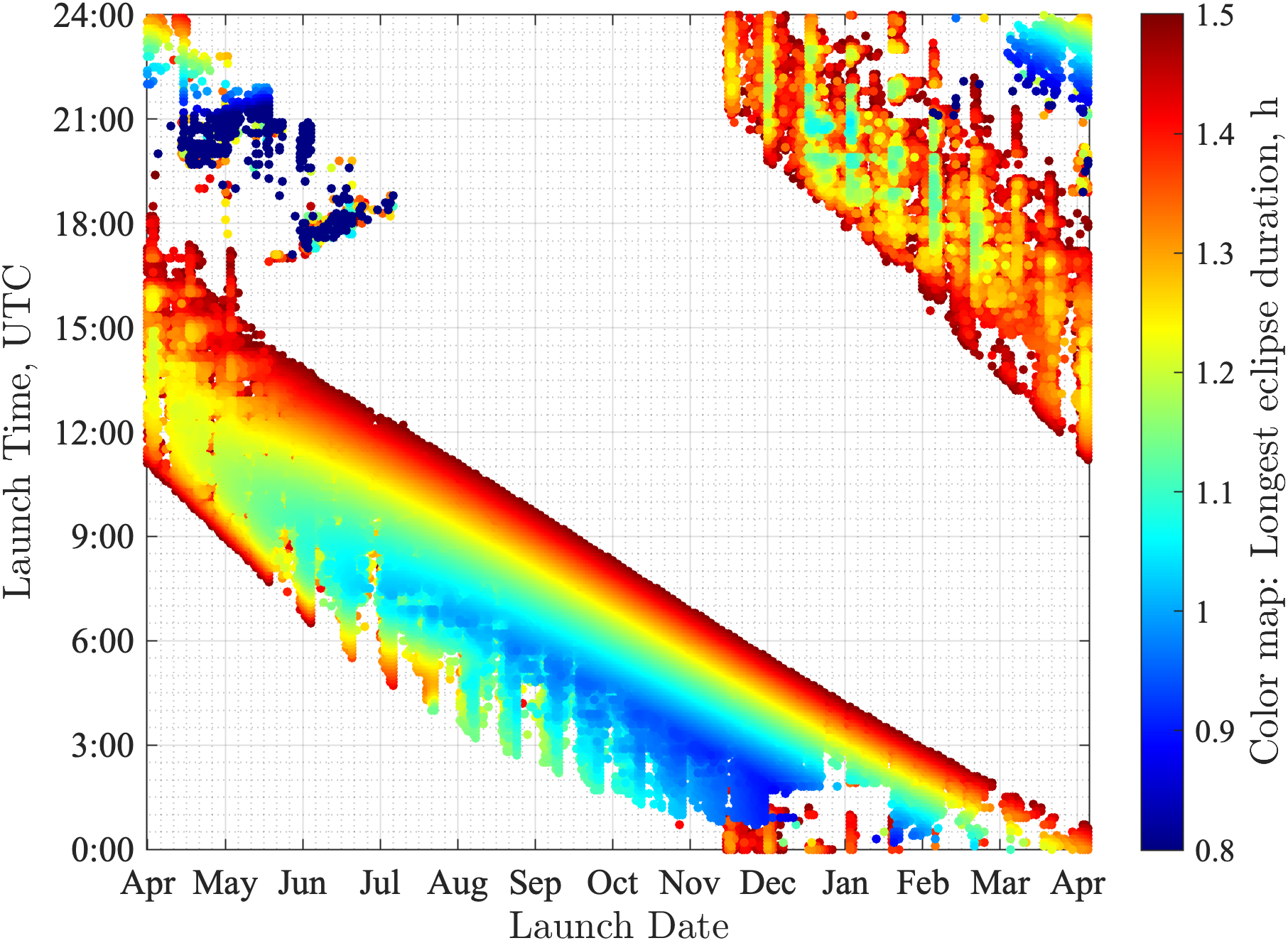}
         \caption{IES thrust magnitude $ = 36$mN}
         \label{fig:spiral_launchwindow_36mN}
     \end{subfigure}
     \begin{subfigure}[htb]{\textwidth}
         \centering
         \includegraphics[width=\textwidth]{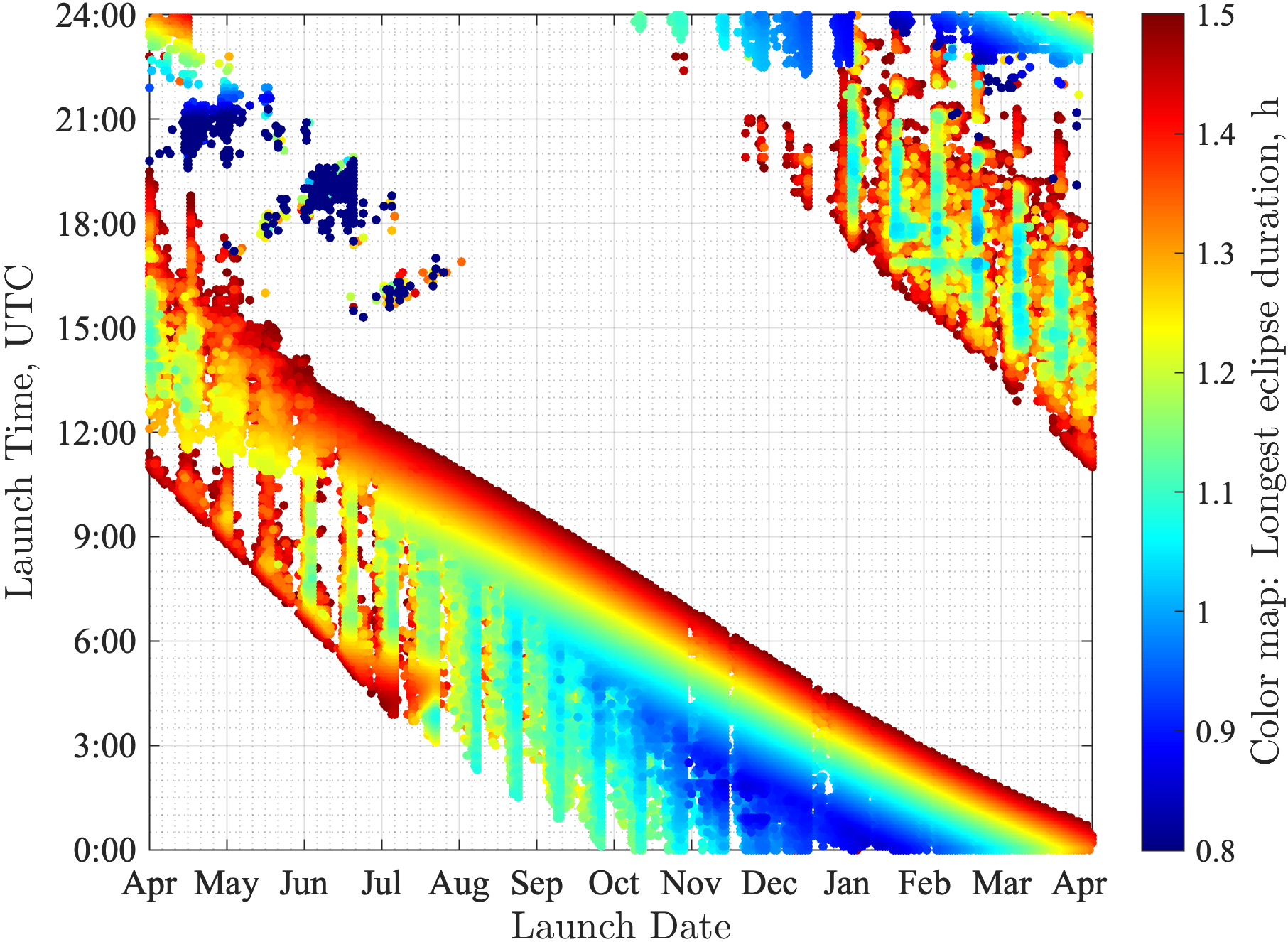}
         \caption{IES thrust magnitude $ = 40$mN}
         \label{fig:spiral_launchwindow_40mN}
     \end{subfigure}
     \caption{Launch window in SOR.}
     \label{fig:spiral_launchwindow_all}
\end{figure}
\else
\begin{figure}
     \centering
     \begin{subfigure}[htb]{0.46\textwidth}
         \centering
         \includegraphics[width=\textwidth]{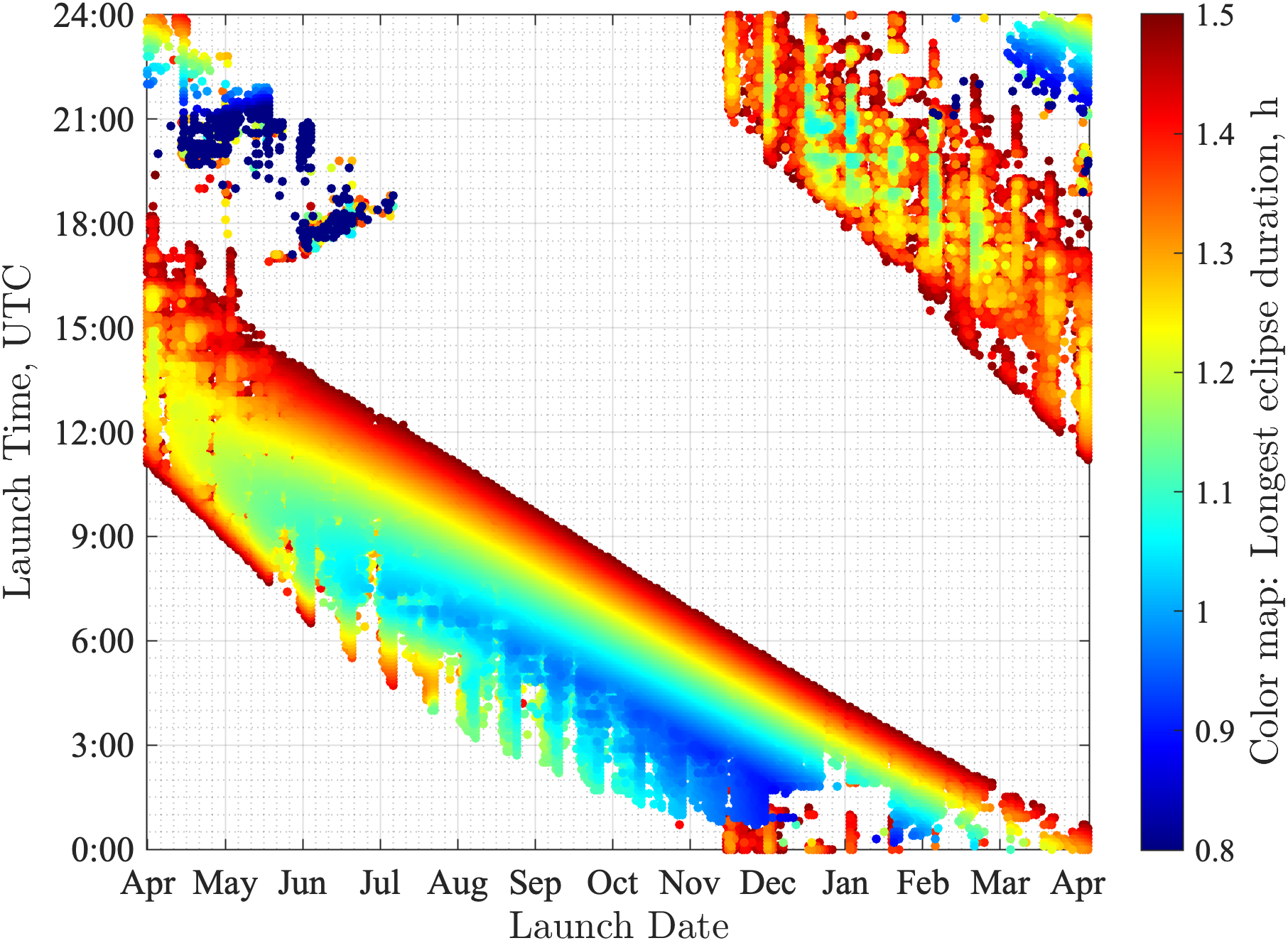}
         \caption{IES thrust magnitude $ = 36$mN}
         \label{fig:spiral_launchwindow_36mN}
     \end{subfigure}
     \begin{subfigure}[htb]{0.46\textwidth}
         \centering
         \includegraphics[width=\textwidth]{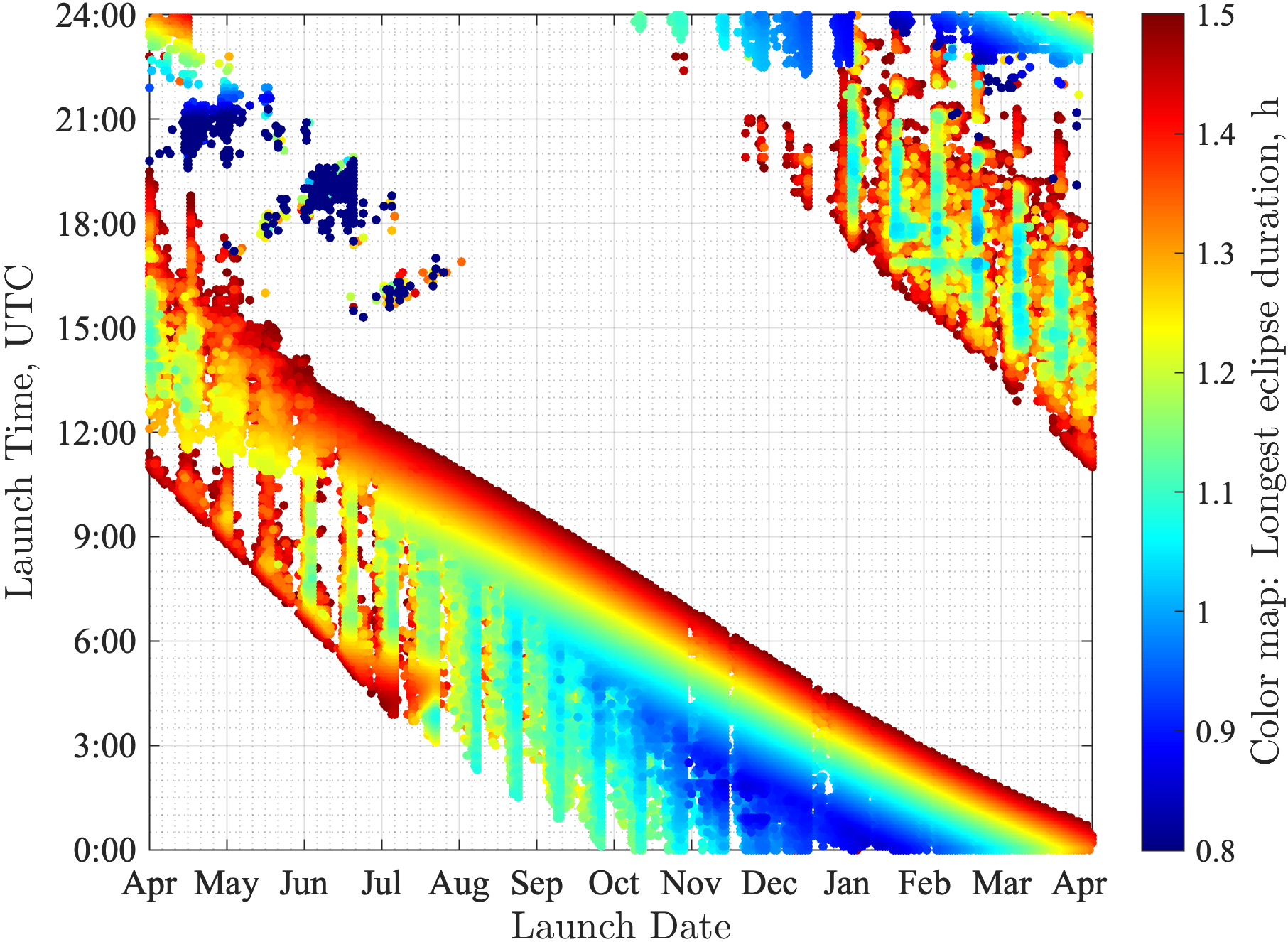}
         \caption{IES thrust magnitude $ = 40$mN}
         \label{fig:spiral_launchwindow_40mN}
     \end{subfigure}
     \caption{Launch window in SOR.}
     \label{fig:spiral_launchwindow_all}
\end{figure}
\fi

%
%
\subsection{Interplanetary Transfer Phase}\label{sec:ipt}

In the interplanetary transfer phase (IPT), we first design the Earth-Phaethon-Earth trajectory, which is necessary to accomplish the nominal mission, and then extend the trajectory for flybys of multiple small bodies. We assume a zero-radius SOI patched-conics model and employ asteroid flyby cyclers\cite{Englander2019, Ozaki2021}, a special case of free-return cyclers\cite{Russell2005}, to solve the IPT trajectory design problem. Free-return cyclers are periodic trajectories that shuttle a spacecraft between two or more celestial bodies, in our case, the Earth and asteroids. We design asteroid flyby cyclers using Earth free-return trajectories and patch them together with Earth gravity-assist maneuvers. 

\begin{figure}[h!]
\centering
\ifreview
\includegraphics[width=\textwidth]{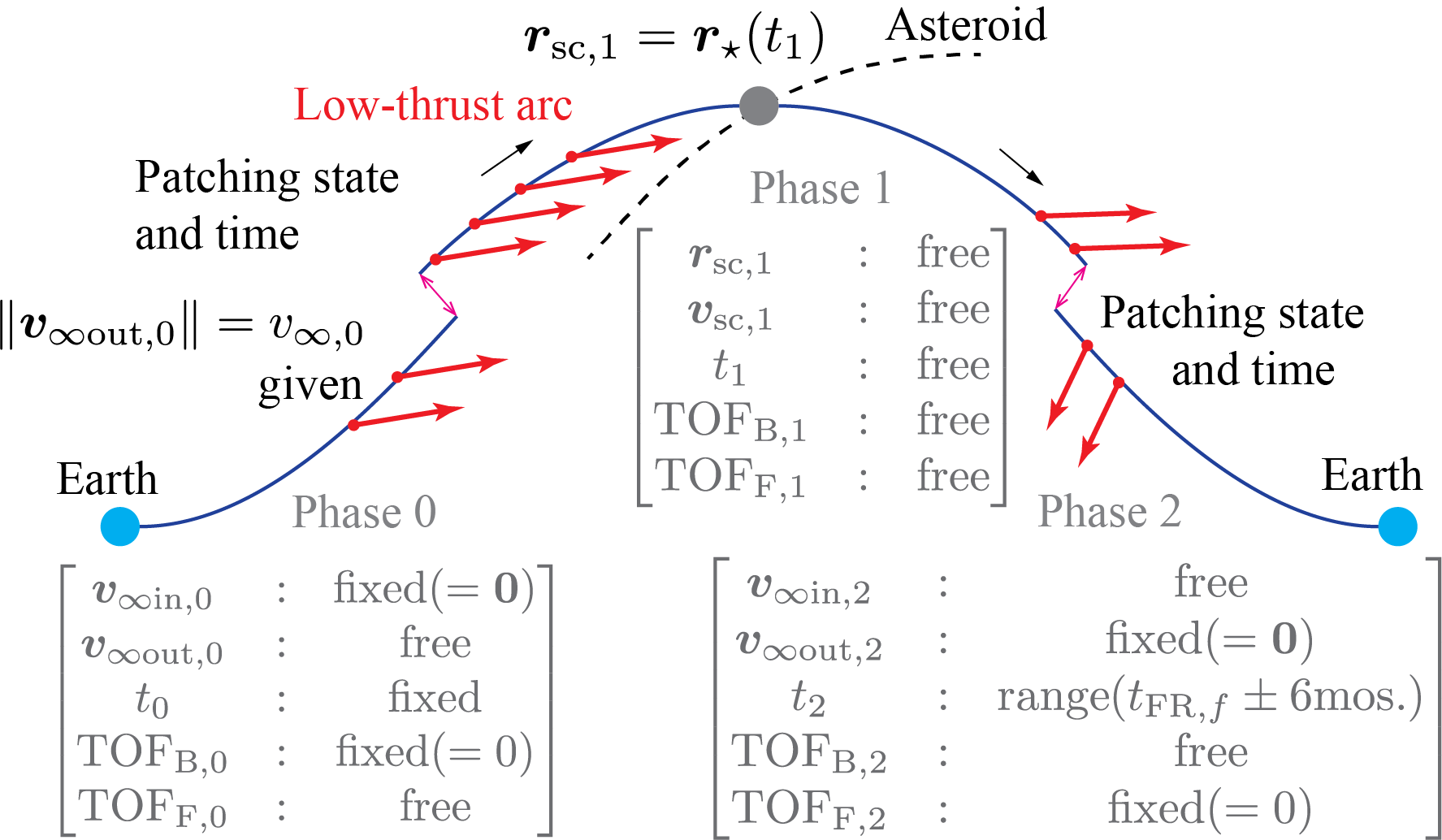}
\else
\includegraphics[width=0.46\textwidth]{img/multipleshooting_formulation.png}
\fi
\caption{Formulation of low-thrust multiple shooting trajectory design}
\label{fig:multipleshooting_formulation}
\end{figure}

To design the Earth-asteroid-Earth trajectory, we first generate the Earth free-return trajectories\cite{Russell2005, Russell2009} for a given Earth departure epoch and $V_{\infty}$ magnitude. We then solve two consecutive Lambert's problems\cite{Izzo2015}, one from the Earth to the asteroid and the other from the asteroid to the Earth, to create the initial guess trajectories for the trajectory optimization. We perform a grid search by repeatedly solving Lambert's problems by changing the asteroid flyby epoch and fixing the initial and final epochs at the Earth flybys to those of the free-return trajectory. In the grid search, we evaluate the sum of two $\Delta V$s; the first $\Delta V$ is the difference between the allowable $V_{\infty} (\leq 1.5$ km/s$)$ and the initial velocity of the first Lambert's problem, and the second $\Delta V$ is the difference between the initial velocity of the second Lambert's problem and the final velocity of the first Lambert's problem. We store the results as the initial guess if the total $\Delta V$ is less than a threshold, which is set to 10km/s in our study. Using the initial guess, we solve the three-phase trajectory optimization problem under the MGA-1DSM model\cite{Vinko2008, Ozaki2021} and then under the low-thrust model\cite{Campagnola2015} \minorblue{in the case of $\Delta V \leq 2$ km/s in the MGA-1DSM model}. The objective functions are the total $\Delta V$ and the propellant consumption, respectively. Figure \ref{fig:multipleshooting_formulation} illustrates the details of the low-thrust trajectory optimization via a direct multiple shooting method.

In the IPT trajectory design, we assume that the maximum Earth departure $V_\infty$ is 1.5~km/s, the remaining mass is 430~kg, the IES thrust magnitude is 24~mN (two units in operation), with the additional 80\% duty cycle to improve the robustness against missed-thrust, and Phaethon flyby occurs on either January 2028 or November 2030. Figures \ref{fig:interplanetary_family2028} and \ref{fig:interplanetary_family2030} show possible Earth-Phaethon-Earth transfer solutions. The result shows that, for the Jan 2028 Phaethon flyby case, the solution with $\Delta V$ of less than 1 km/s \minorblue{(in the low-thrust model)} exists at any time until the departure from Earth in June 2027. We select 10 representative solutions with short flight time and small $\Delta V$ ($<$ 1 km/s) and illustrate the trajectories in Fig.\ref{fig:interp_all} where the case IDs correspond to those in Fig. \ref{fig:metro_map}.

\ifreview
\begin{figure}
     \centering
     \begin{subfigure}[htb]{\textwidth}
         \centering
         \includegraphics[width=\textwidth]{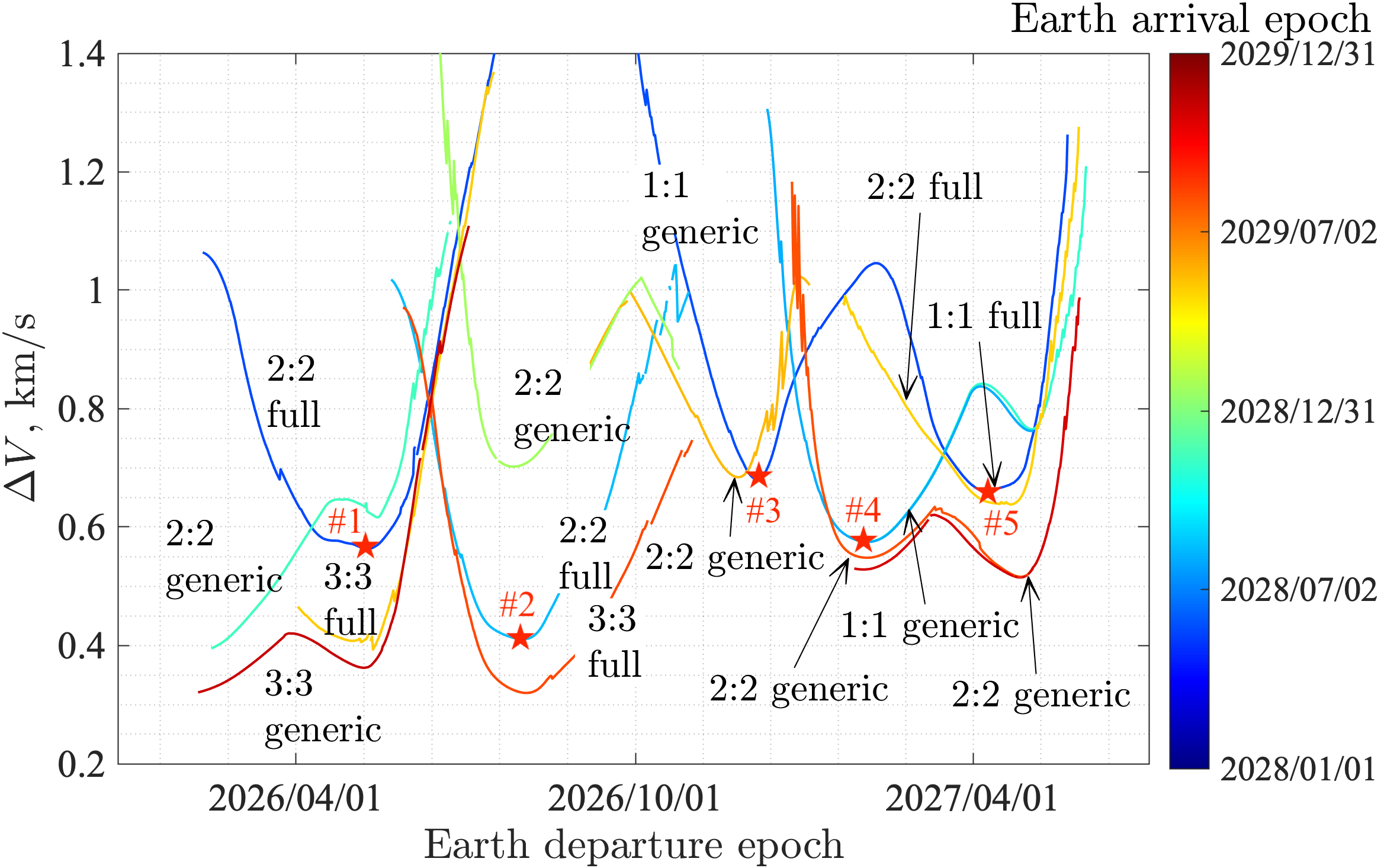}
         \caption{Phaethon flyby in Jan 2028}
         \label{fig:interplanetary_family2028}
     \end{subfigure}
     \begin{subfigure}[htb]{\textwidth}
         \centering
         \includegraphics[width=\textwidth]{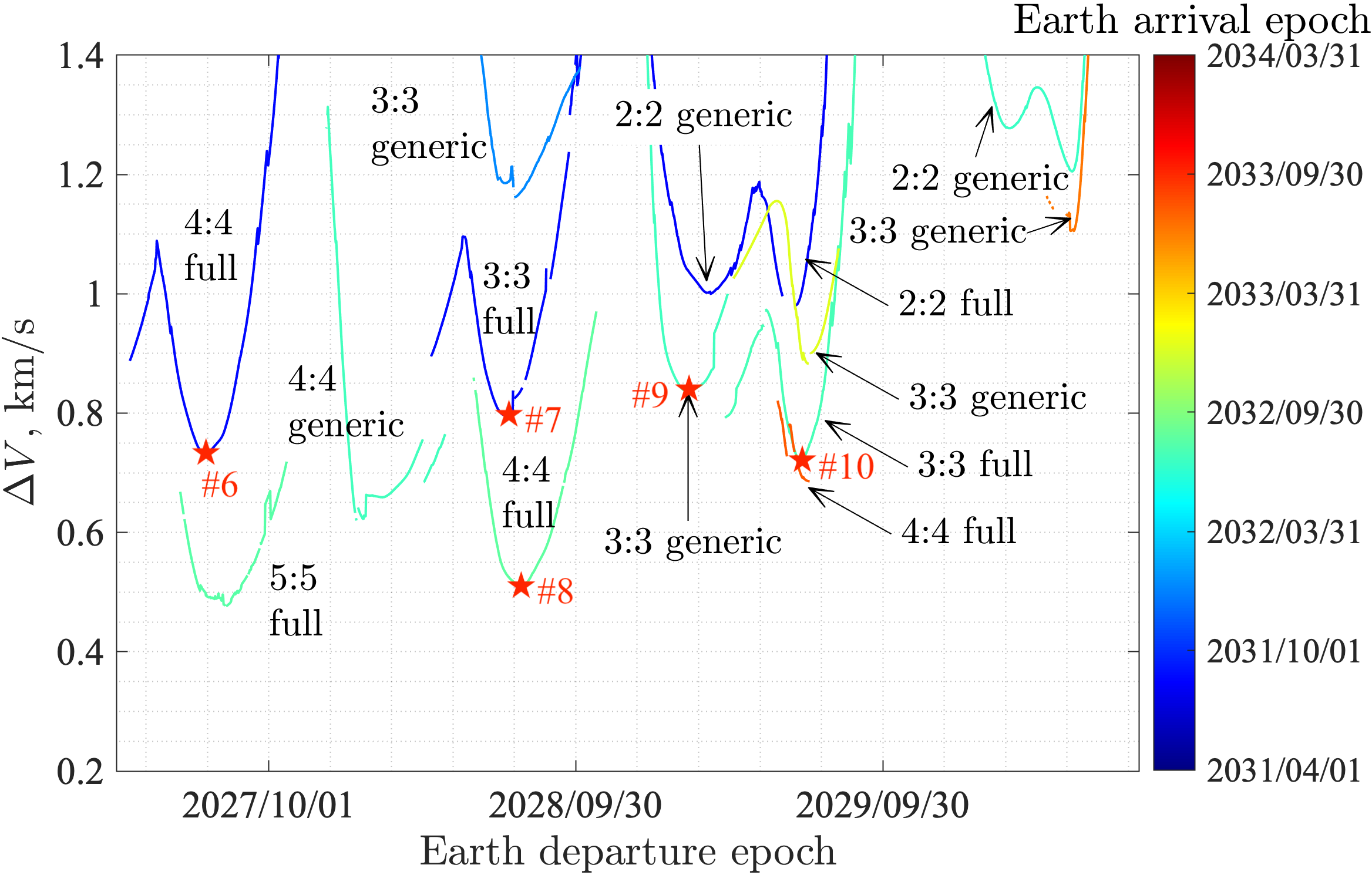}
         \caption{Phaethon flyby in Nov 2030}
         \label{fig:interplanetary_family2030}
     \end{subfigure}
     \caption{Earth-Phaethon-Earth trajectory design solutions.}
     \label{fig:interplanetary_family_all}
\end{figure}
\else
\begin{figure}
     \centering
     \begin{subfigure}[htb]{0.46\textwidth}
         \centering
         \includegraphics[width=\textwidth]{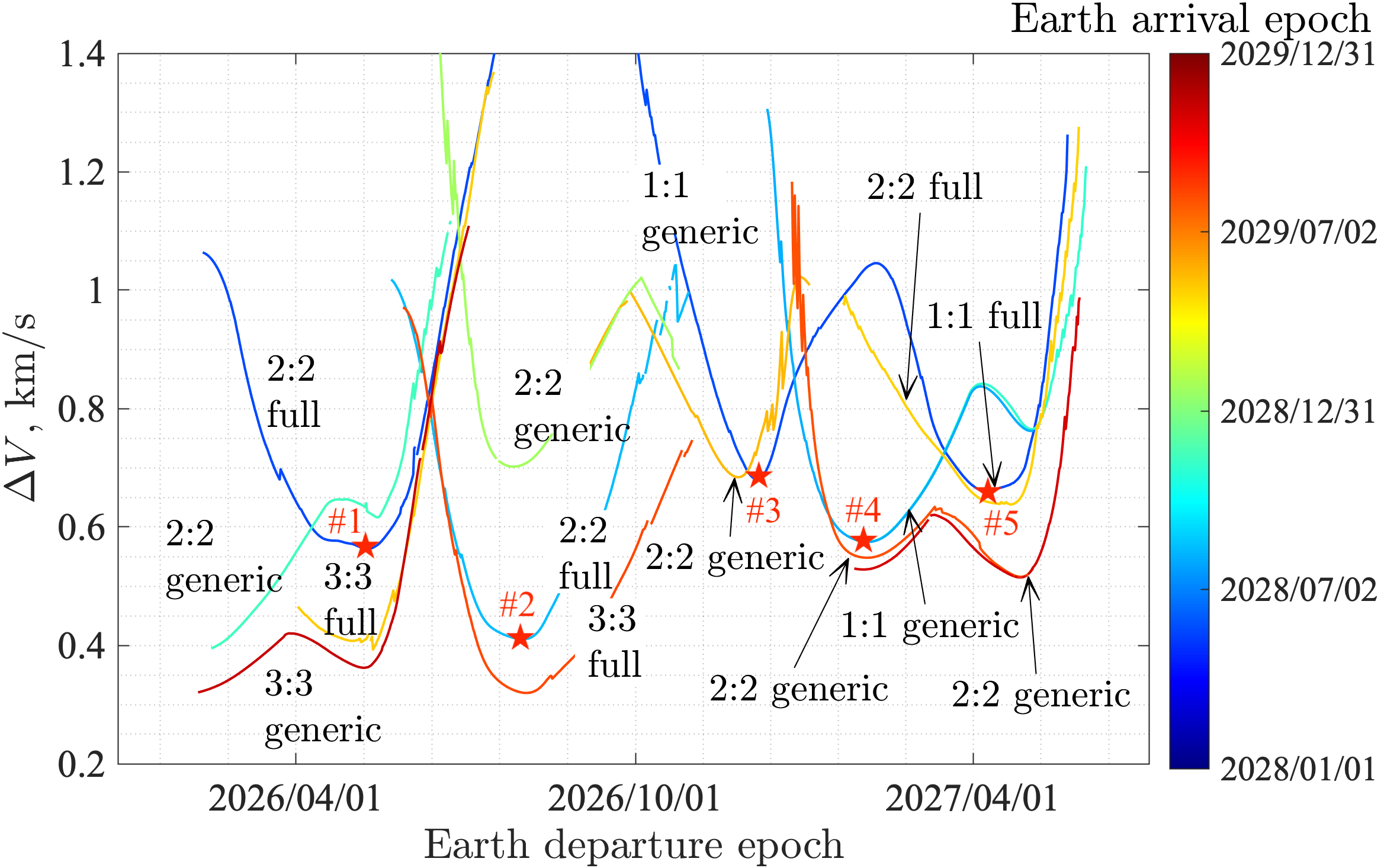}
         \caption{Phaethon flyby in Jan 2028}
         \label{fig:interplanetary_family2028}
     \end{subfigure}
     \begin{subfigure}[htb]{0.46\textwidth}
         \centering
         \includegraphics[width=\textwidth]{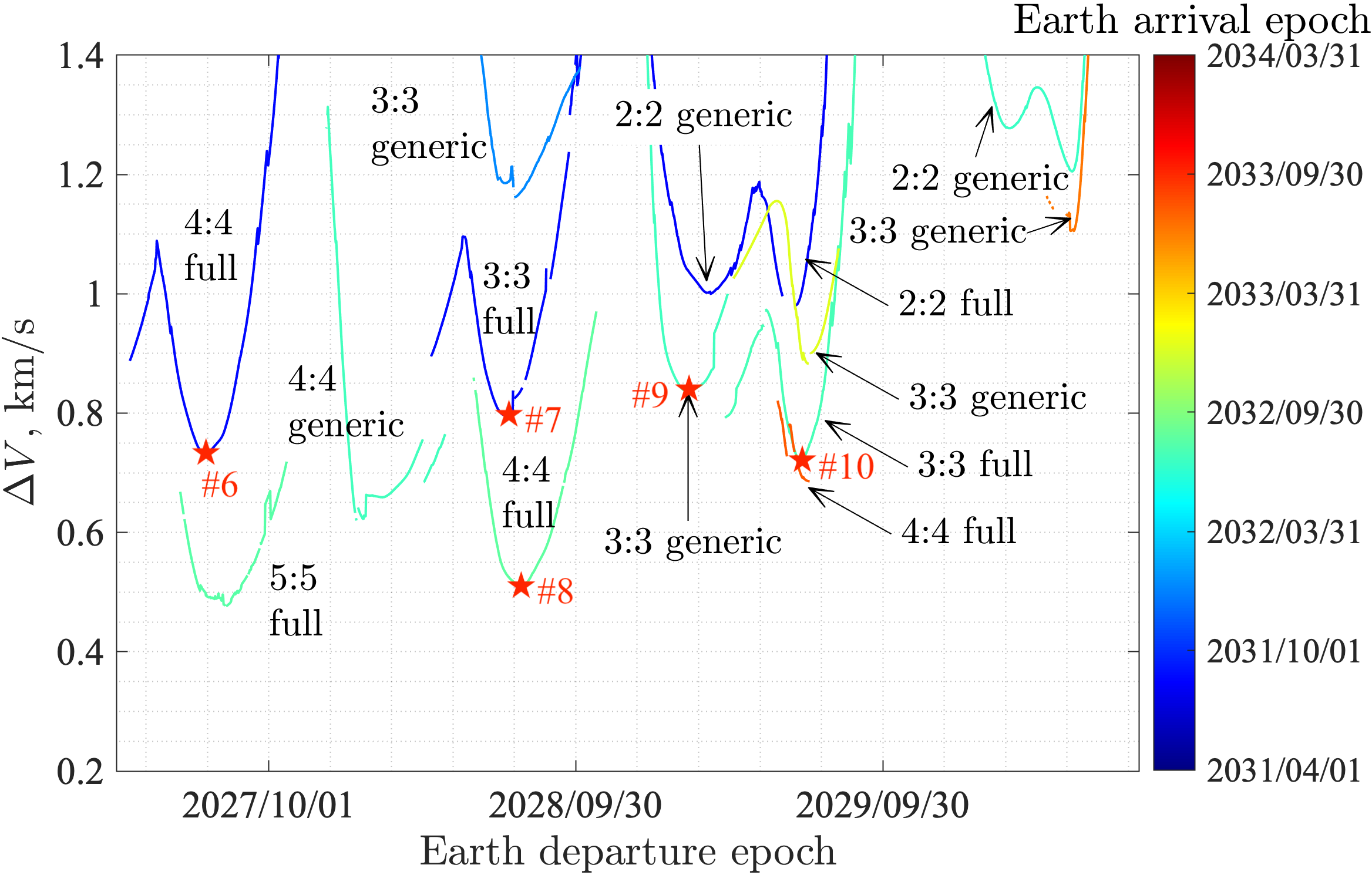}
         \caption{Phaethon flyby in Nov 2030}
         \label{fig:interplanetary_family2030}
     \end{subfigure}
     \caption{Earth-Phaethon-Earth trajectory design solutions.}
     \label{fig:interplanetary_family_all}
\end{figure}
\fi

\ifreview
\begin{figure}
     \centering
     \begin{subfigure}[b]{0.32\textwidth}
         \centering
         \includegraphics[width=\textwidth]{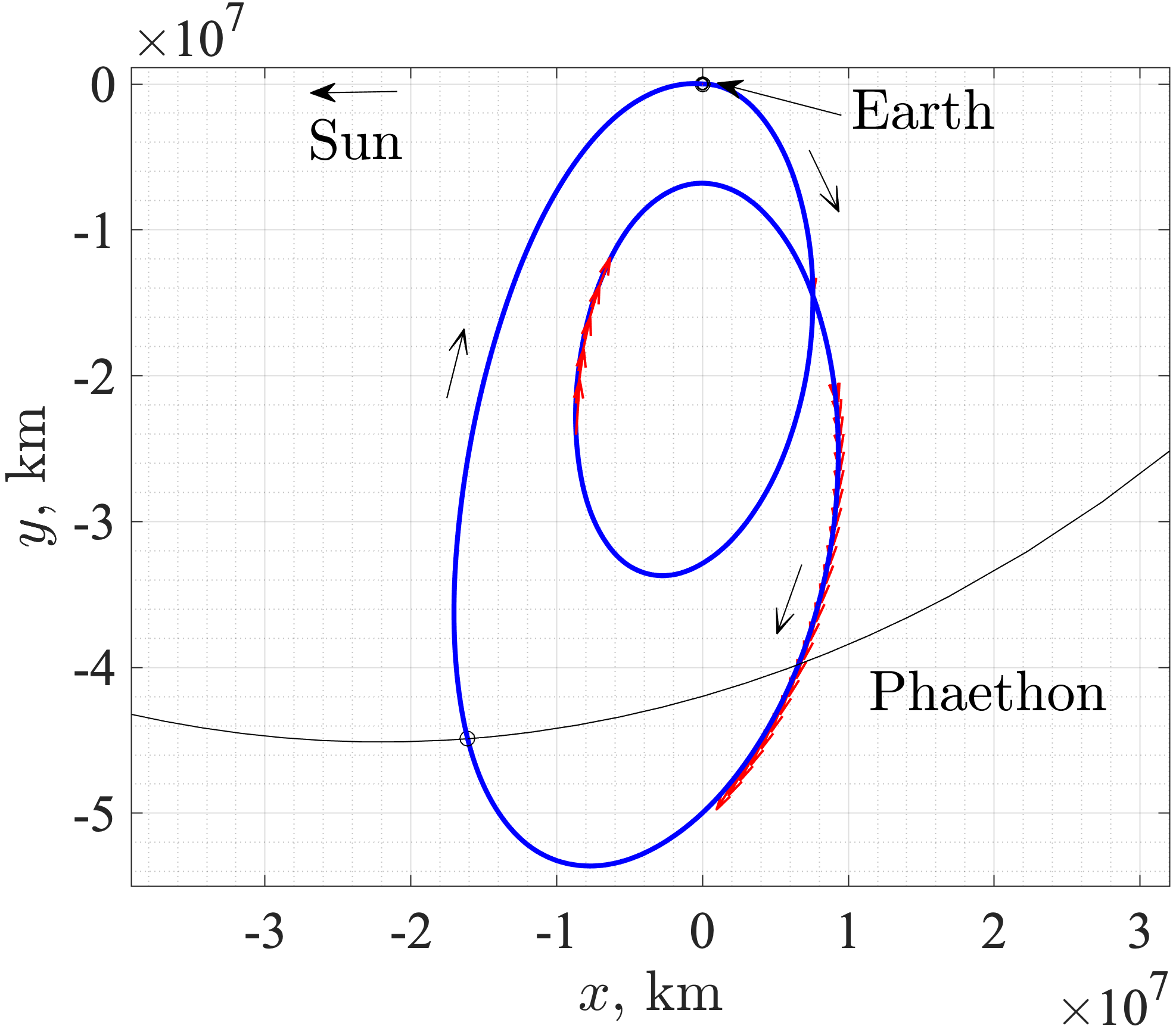}
         \caption{Case \#1, 2:2 full}
         \label{fig:interp_2028_case1}
     \end{subfigure}
     \begin{subfigure}[b]{0.32\textwidth}
         \centering
         \includegraphics[width=\textwidth]{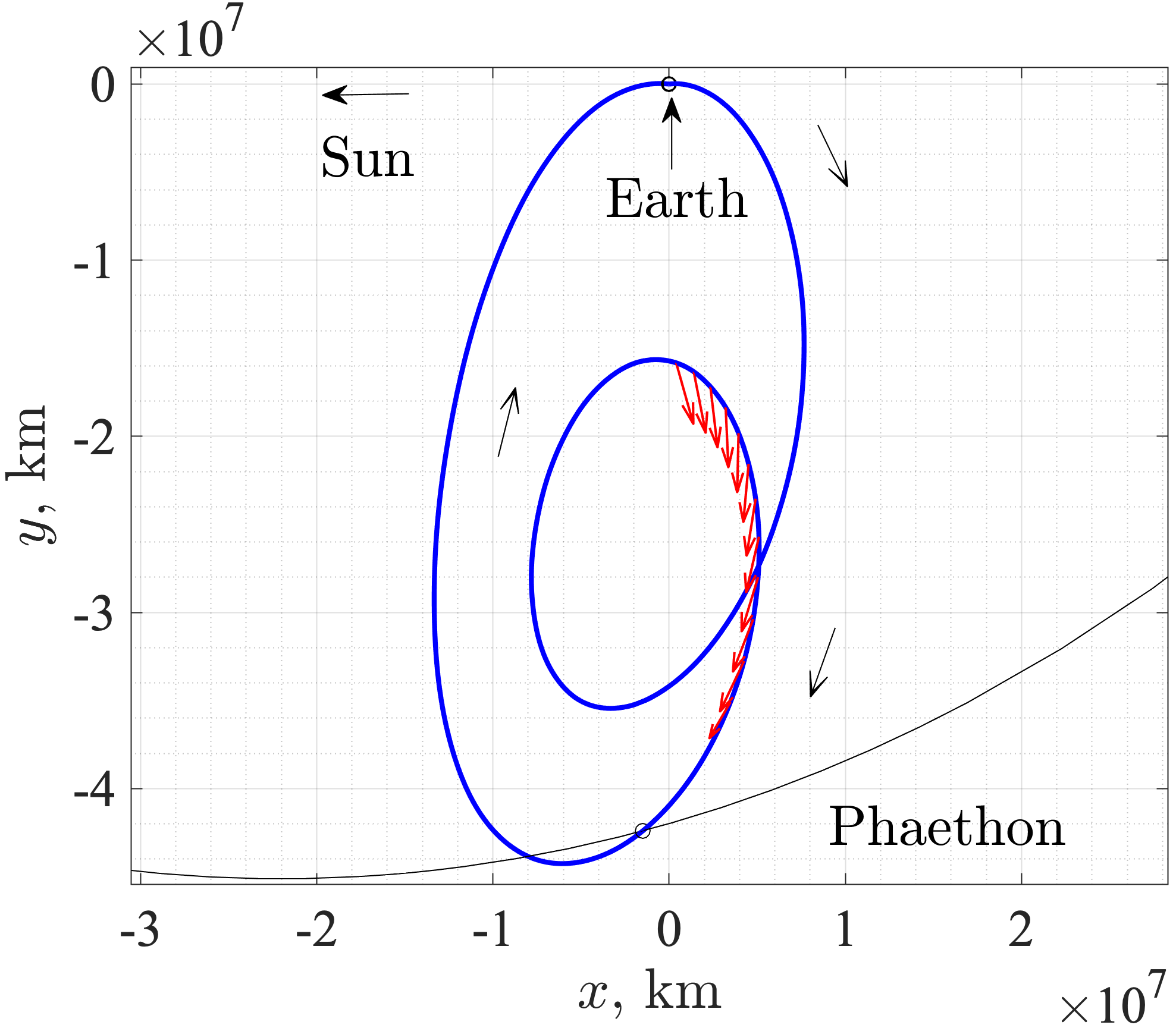}
         \caption{Case \#2, 2:2 full}
         \label{fig:interp_2028_case2}
     \end{subfigure}
     \begin{subfigure}[b]{0.32\textwidth}
         \centering
         \includegraphics[width=\textwidth]{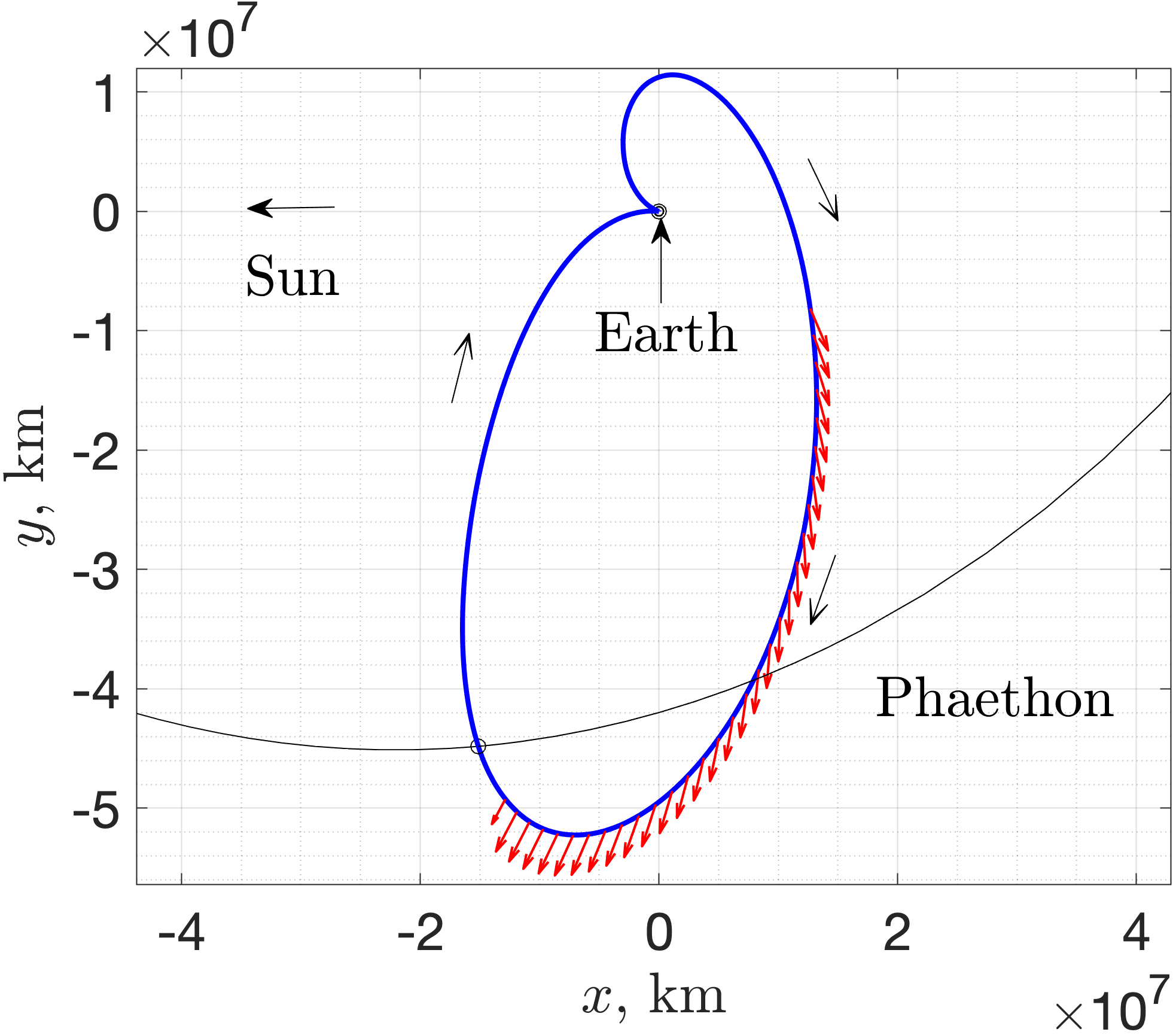}
         \caption{Case \#3, 1:1 generic}
         \label{fig:interp_2028_case3}
     \end{subfigure}
     \begin{subfigure}[b]{0.32\textwidth}
         \centering
         \includegraphics[width=\textwidth]{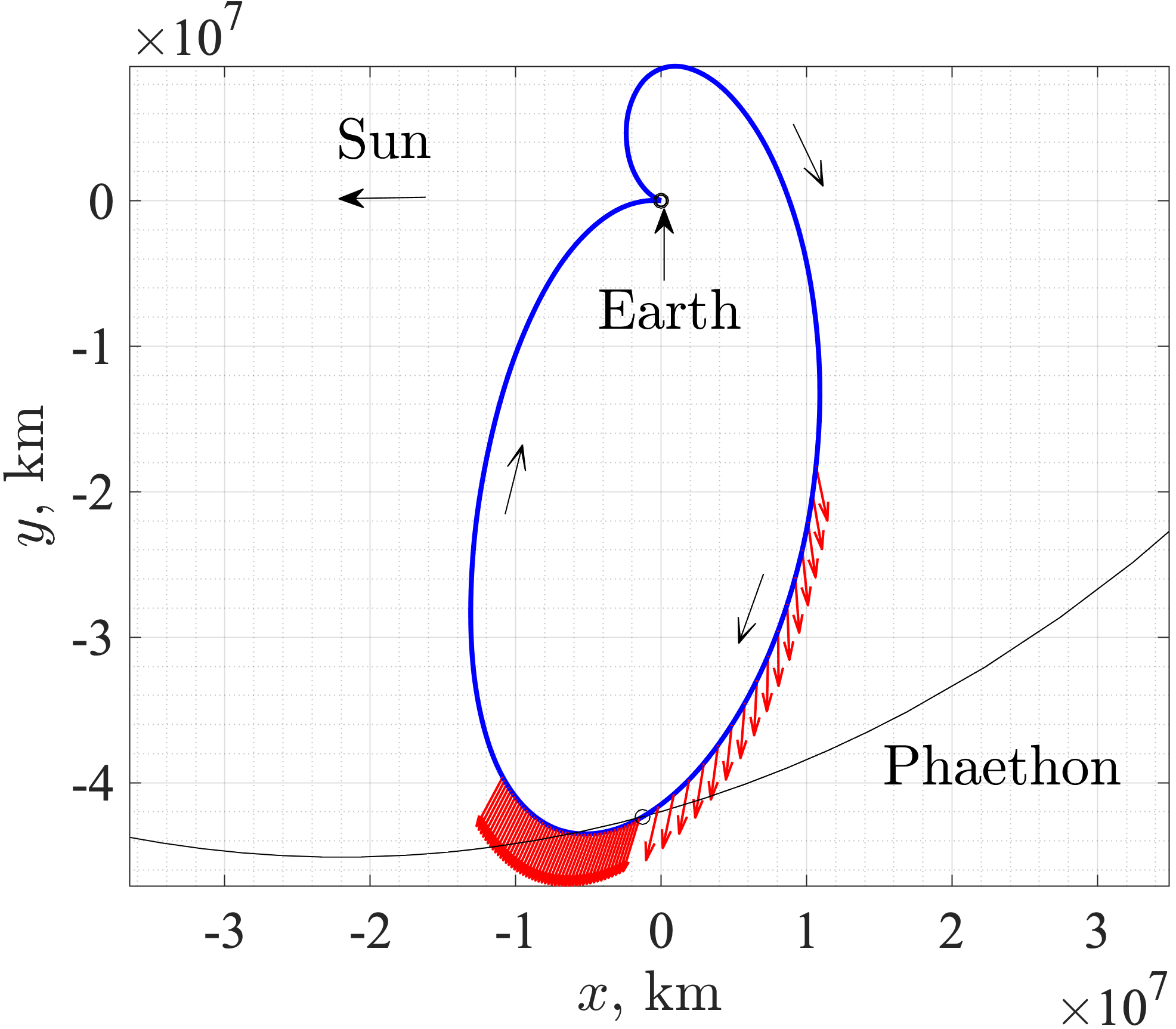}
         \caption{Case \#4, 1:1 generic}
         \label{fig:interp_2028_case4}
     \end{subfigure}
     \begin{subfigure}[b]{0.32\textwidth}
         \centering
         \includegraphics[width=\textwidth]{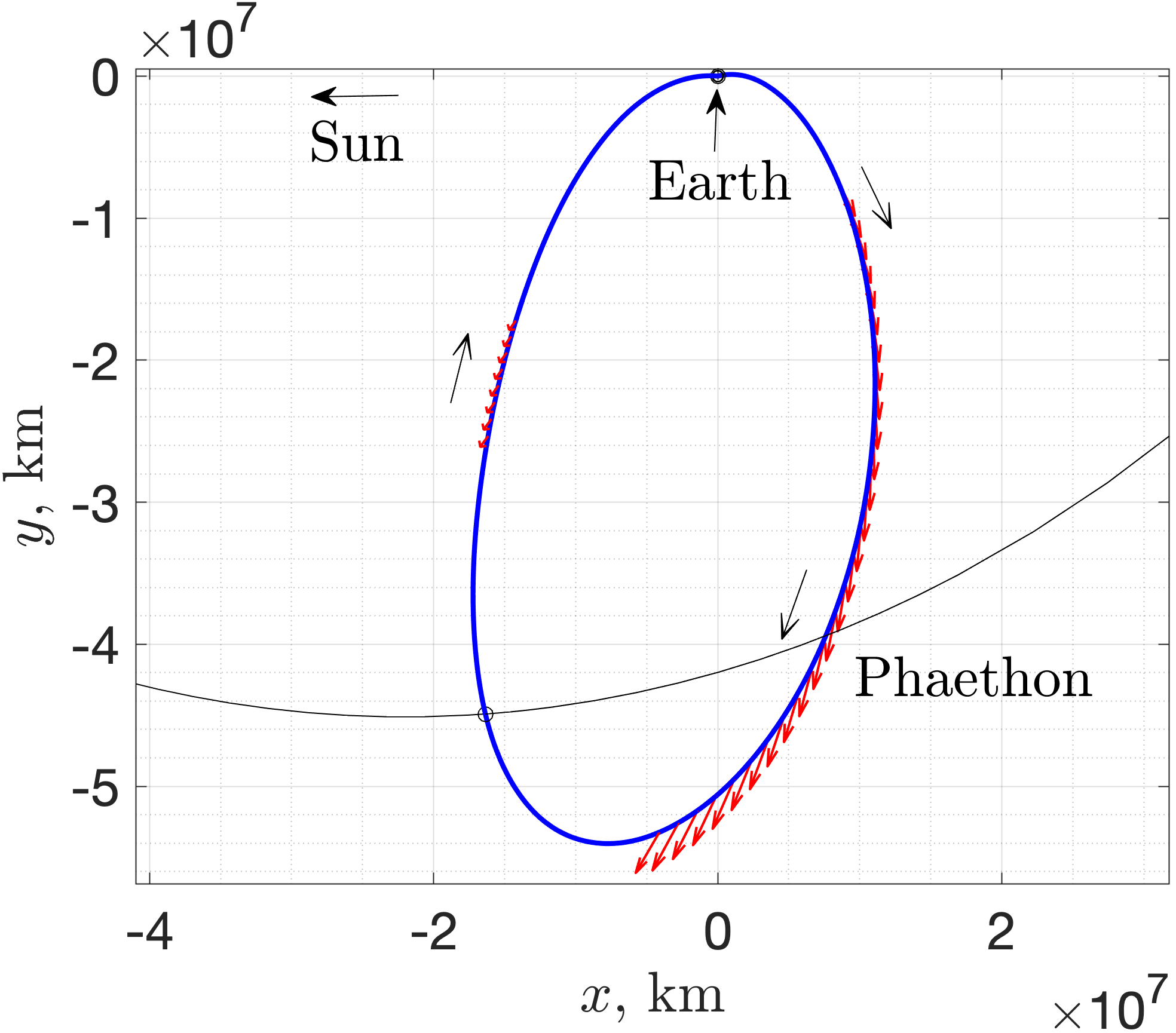}
         \caption{Case \#5, 1:1 full}
         \label{fig:interp_2028_case5}
     \end{subfigure}
     \begin{subfigure}[b]{0.32\textwidth}
         \centering
         \includegraphics[width=\textwidth]{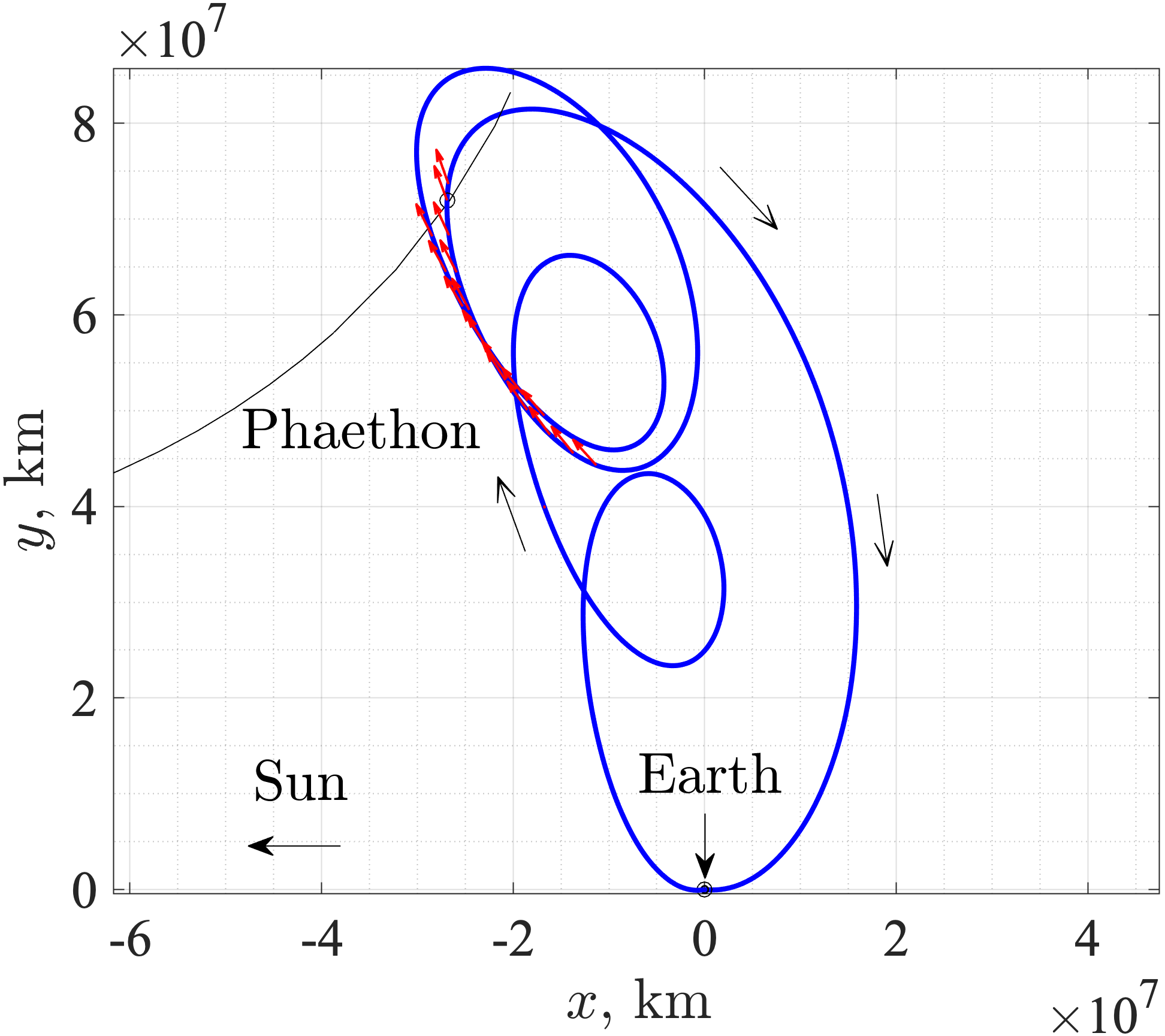}
         \caption{Case \#6, 4:4 full}
         \label{fig:interp_2030_case6}
     \end{subfigure}
     \begin{subfigure}[b]{0.32\textwidth}
         \centering
         \includegraphics[width=\textwidth]{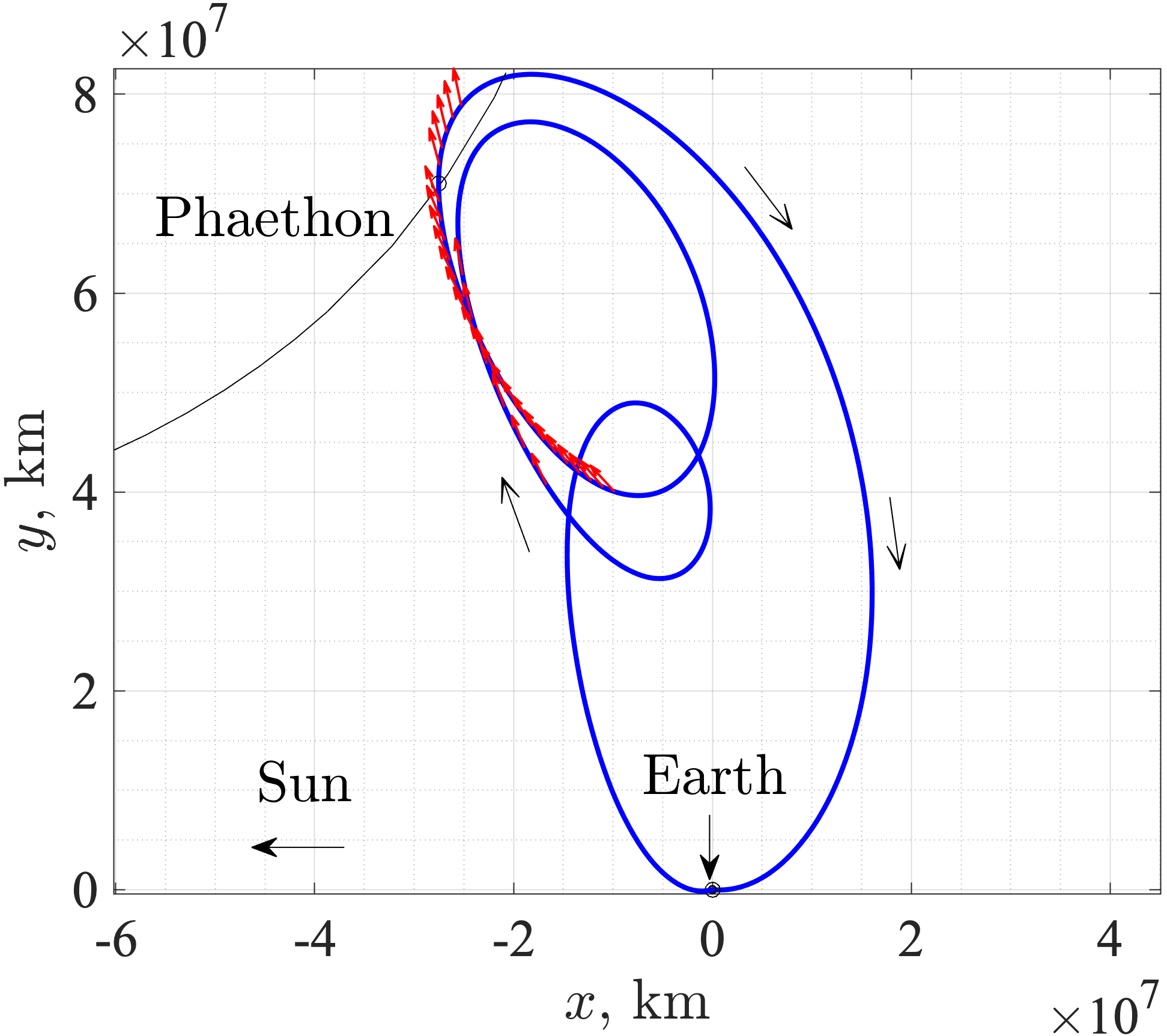}
         \caption{Case \#7, 3:3 full}
         \label{fig:interp_2030_case7}
     \end{subfigure}
     \begin{subfigure}[b]{0.32\textwidth}
         \centering
         \includegraphics[width=\textwidth]{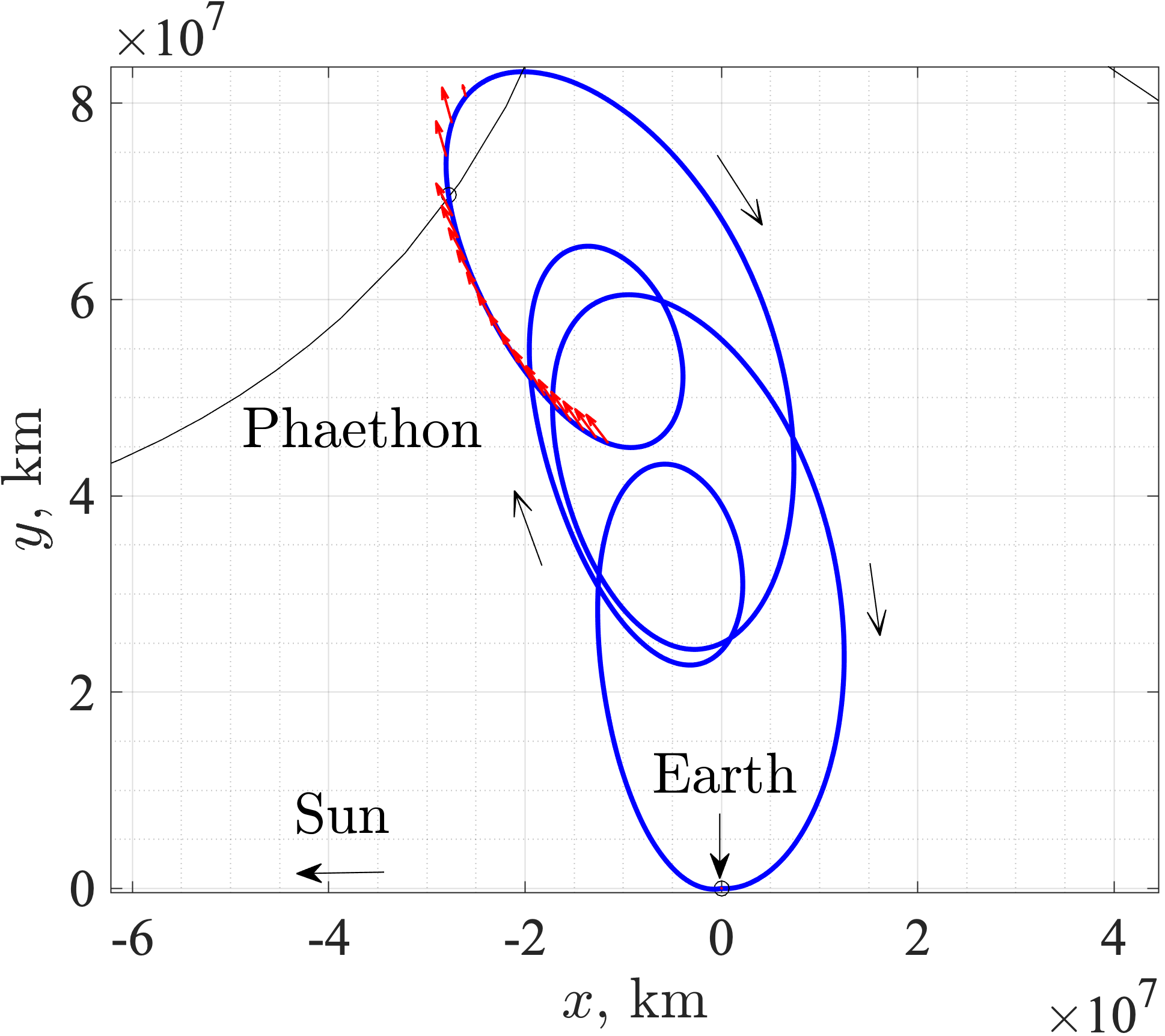}
         \caption{Case \#8, 4:4 full}
         \label{fig:interp_2030_case8}
     \end{subfigure}
     \begin{subfigure}[b]{0.32\textwidth}
         \centering
         \includegraphics[width=\textwidth]{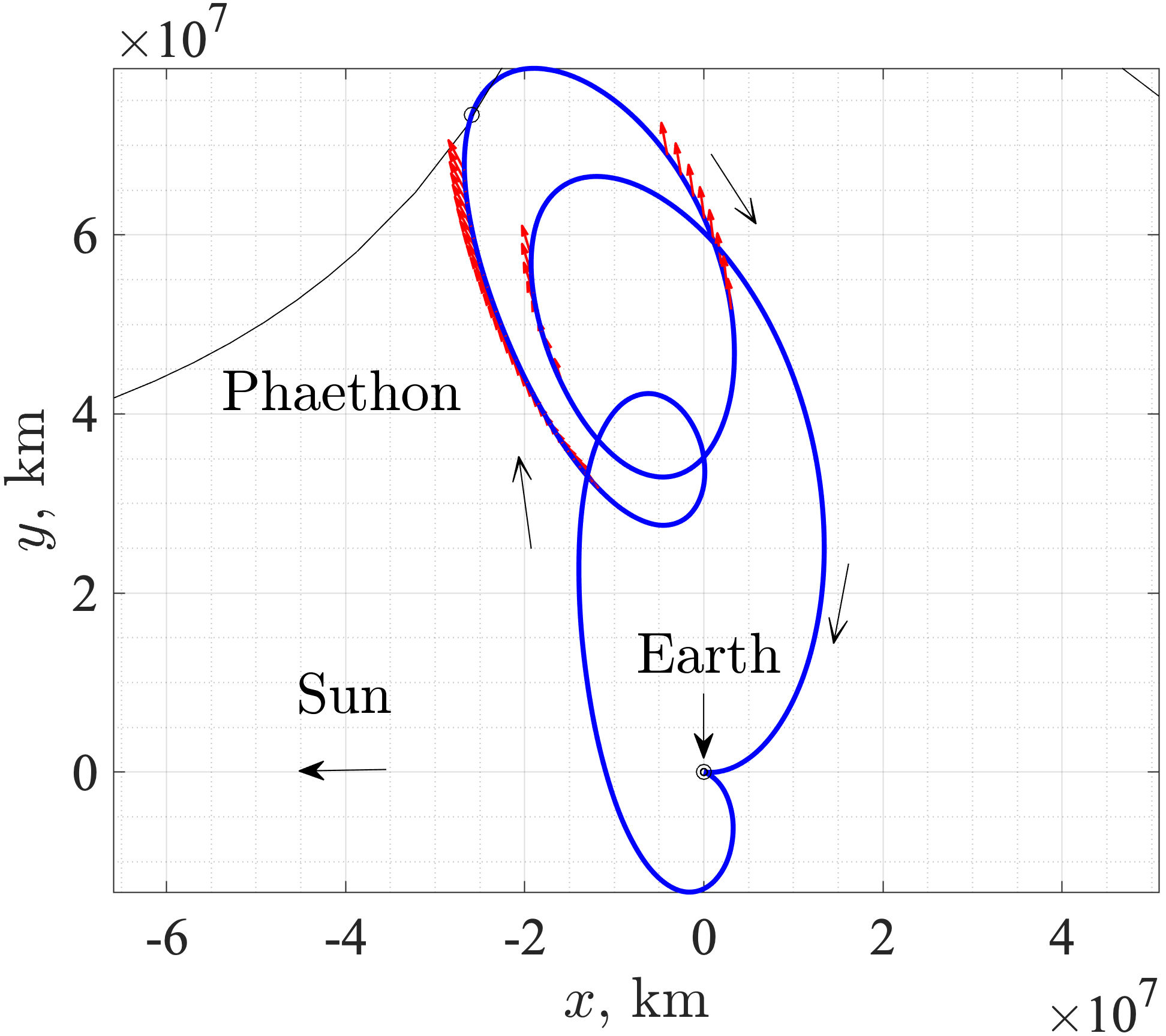}
         \caption{Case \#9, 3:3 generic}
         \label{fig:interp_2030_case9}
     \end{subfigure}
     \begin{subfigure}[b]{0.32\textwidth}
         \centering
         \includegraphics[width=\textwidth]{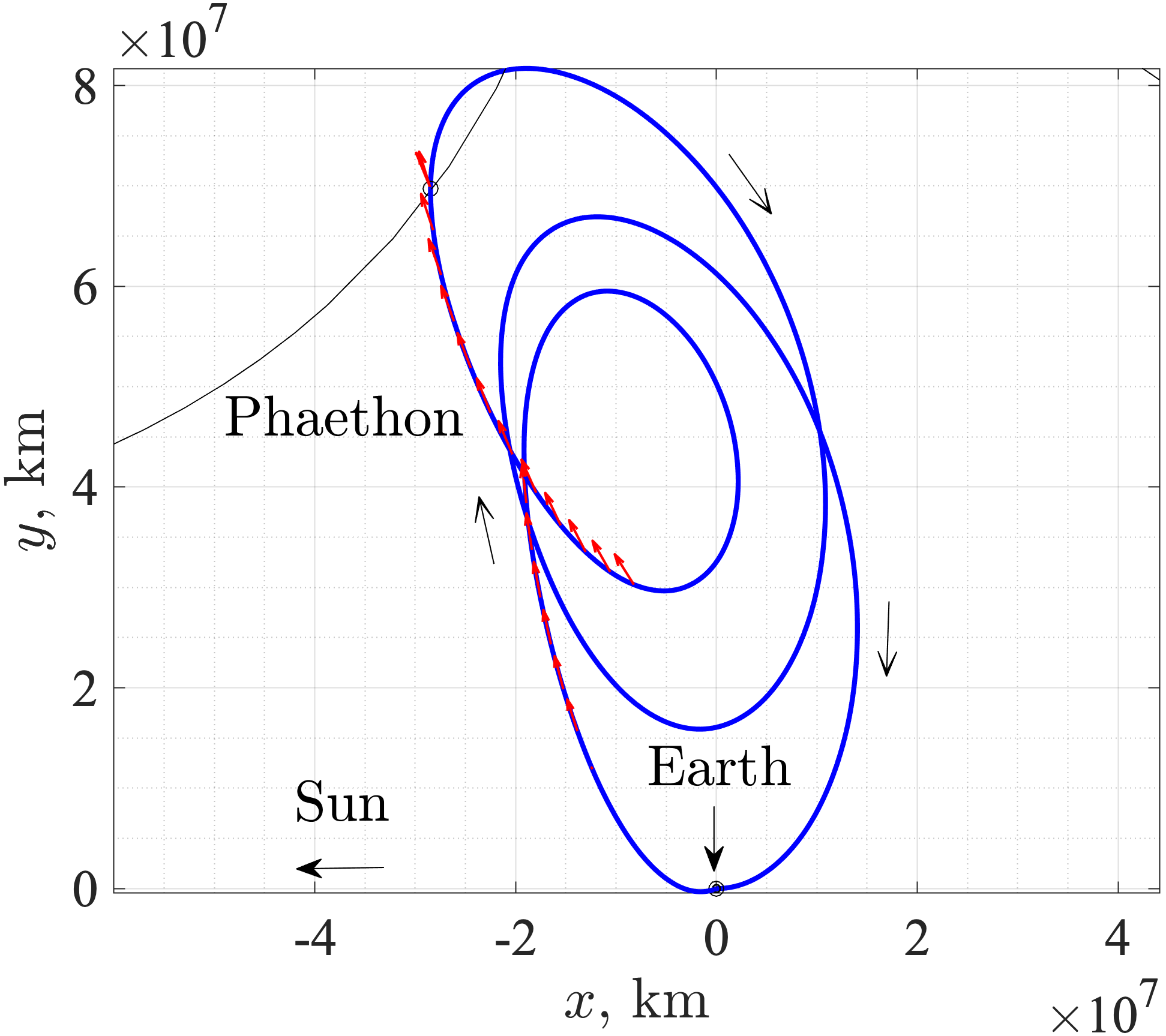}
         \caption{Case \#10, 3:3 full}
         \label{fig:interp_2030_case10}
     \end{subfigure}
     \caption{Representative IPT trajectories (Earth-centered, Sun-Earth line fixed rotational frame. Black solid line denotes Phaethon trajectory. Blue solid line describes \destiny trajectory. Red arrow indicates thrust vectors.)}
     \label{fig:interp_all}
\end{figure}

\else
\begin{figure}
     \centering
     \begin{subfigure}[b]{0.23\textwidth}
         \centering
         \includegraphics[width=\textwidth]{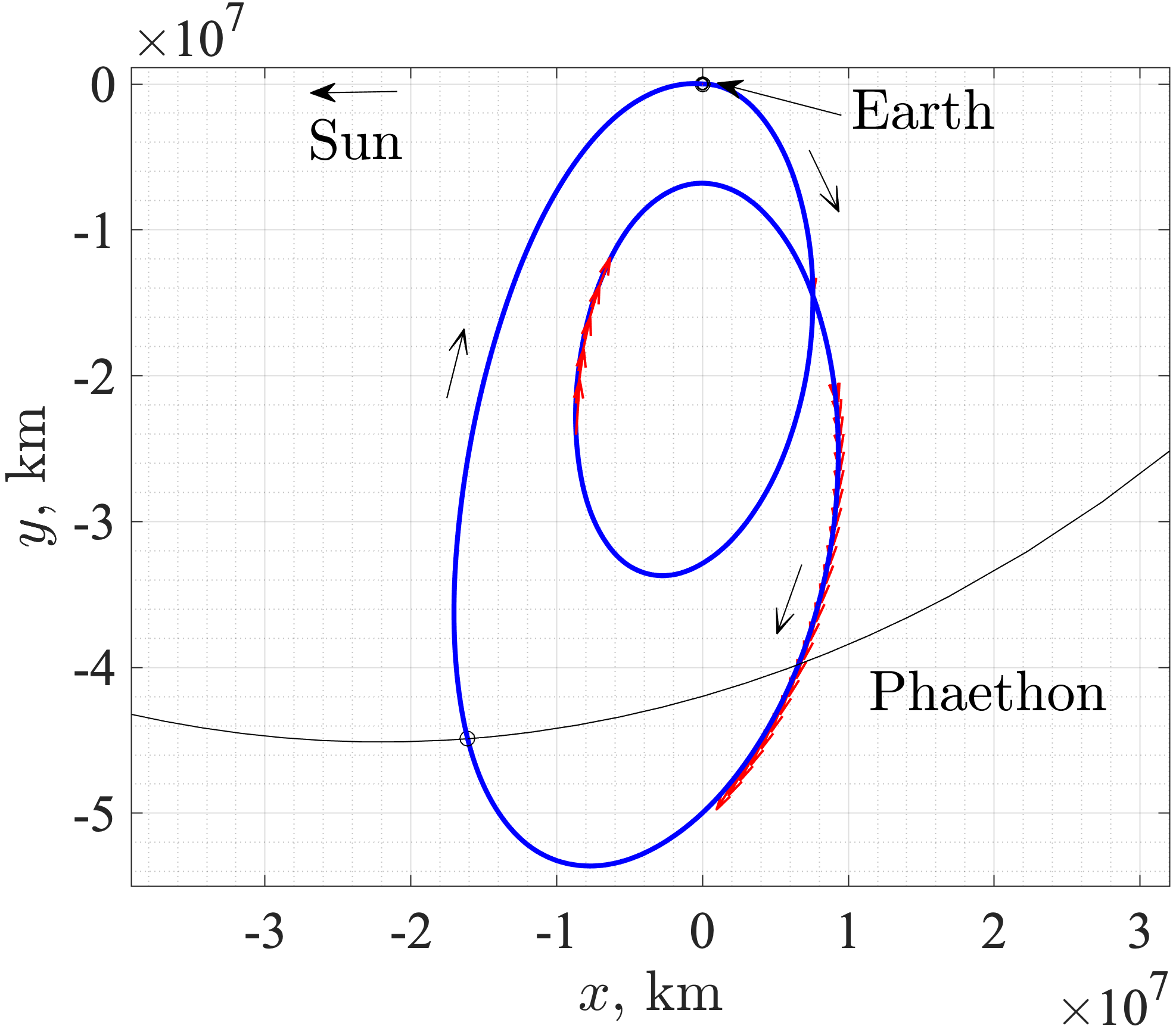}
         \caption{Case \#1, 2:2 full}
         \label{fig:interp_2028_case1}
     \end{subfigure}
     \begin{subfigure}[b]{0.23\textwidth}
         \centering
         \includegraphics[width=\textwidth]{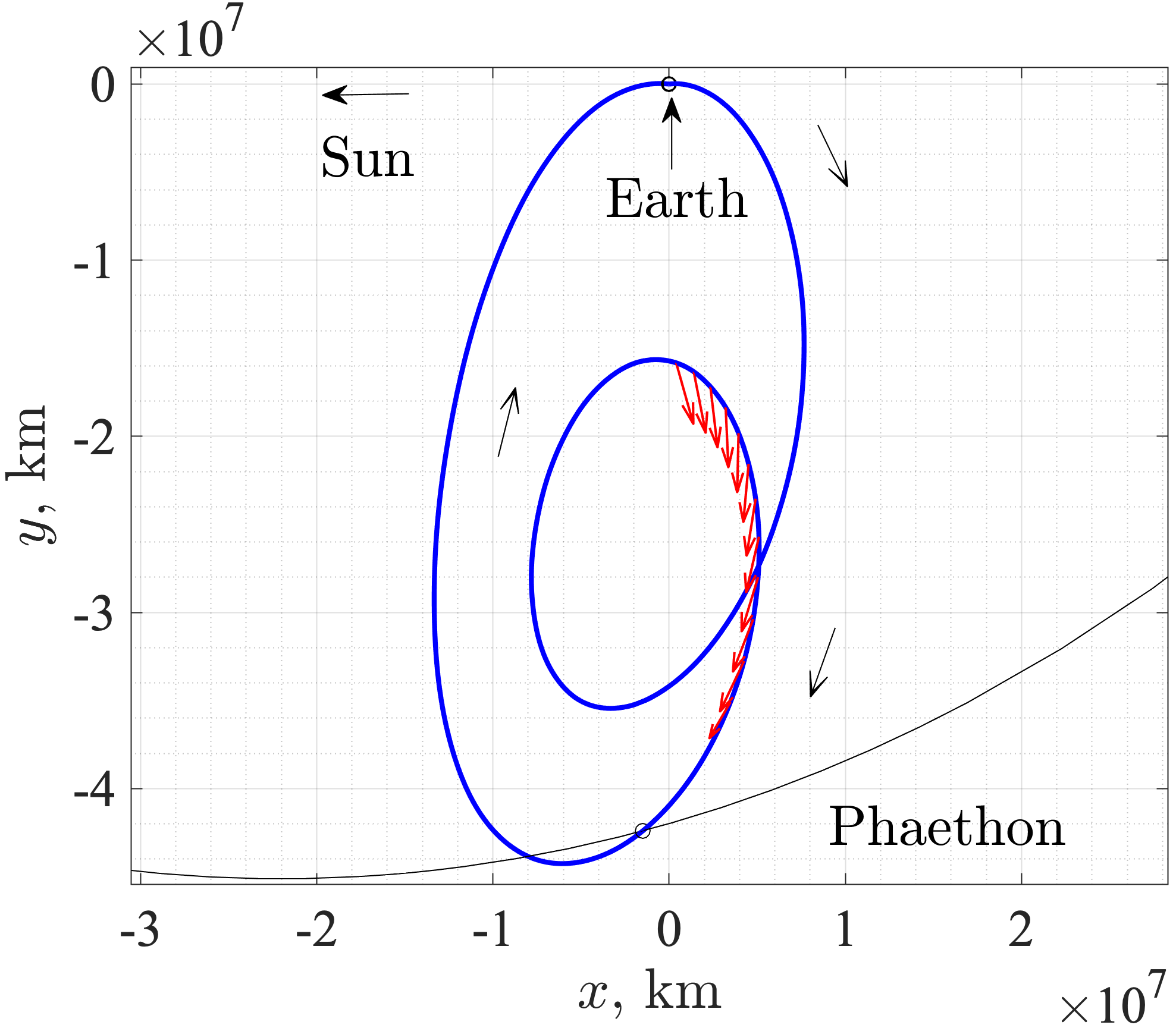}
         \caption{Case \#2, 2:2 full}
         \label{fig:interp_2028_case2}
     \end{subfigure}
     \begin{subfigure}[b]{0.23\textwidth}
         \centering
         \includegraphics[width=\textwidth]{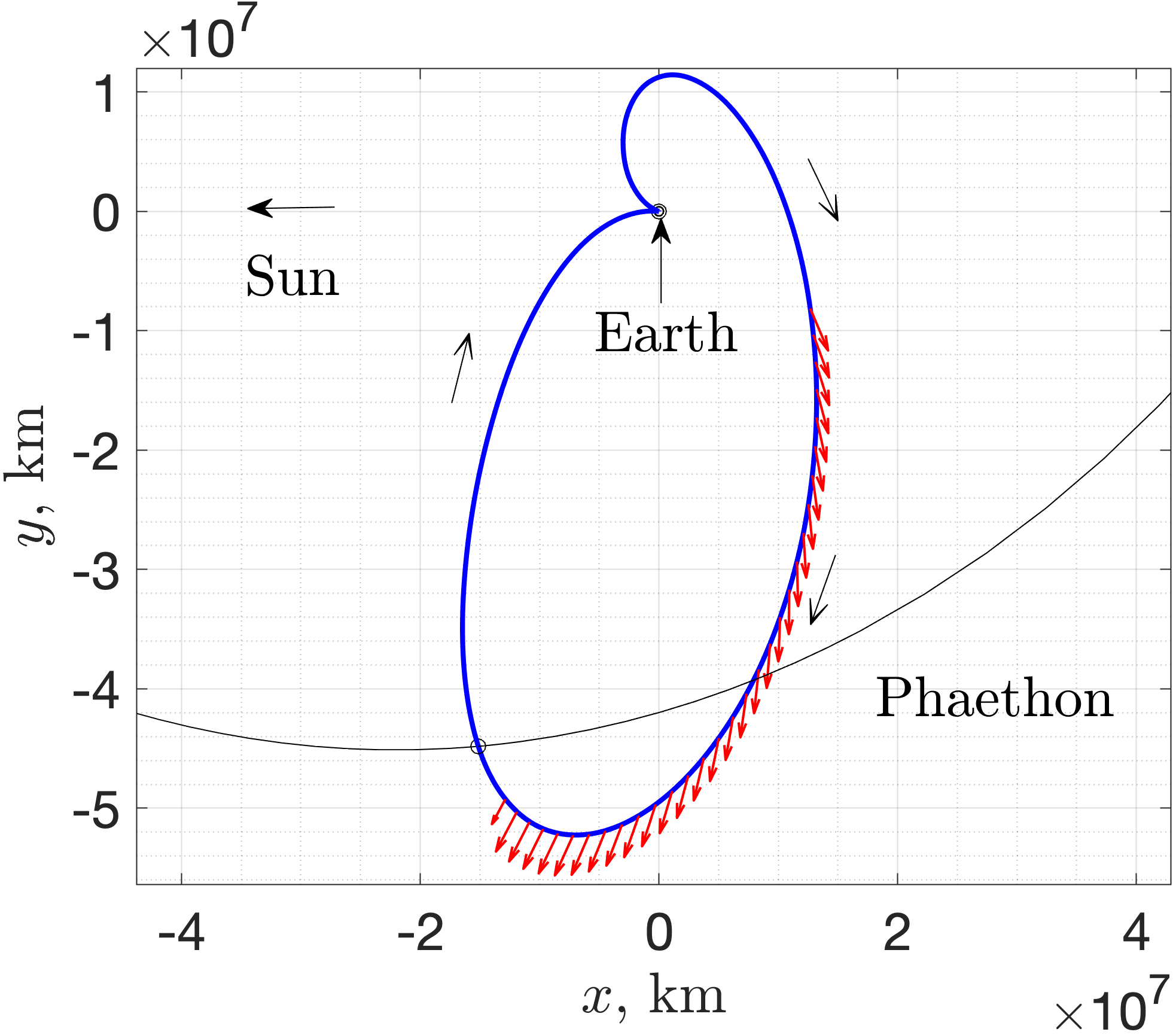}
         \caption{Case \#3, 1:1 generic}
         \label{fig:interp_2028_case3}
     \end{subfigure}
     \begin{subfigure}[b]{0.23\textwidth}
         \centering
         \includegraphics[width=\textwidth]{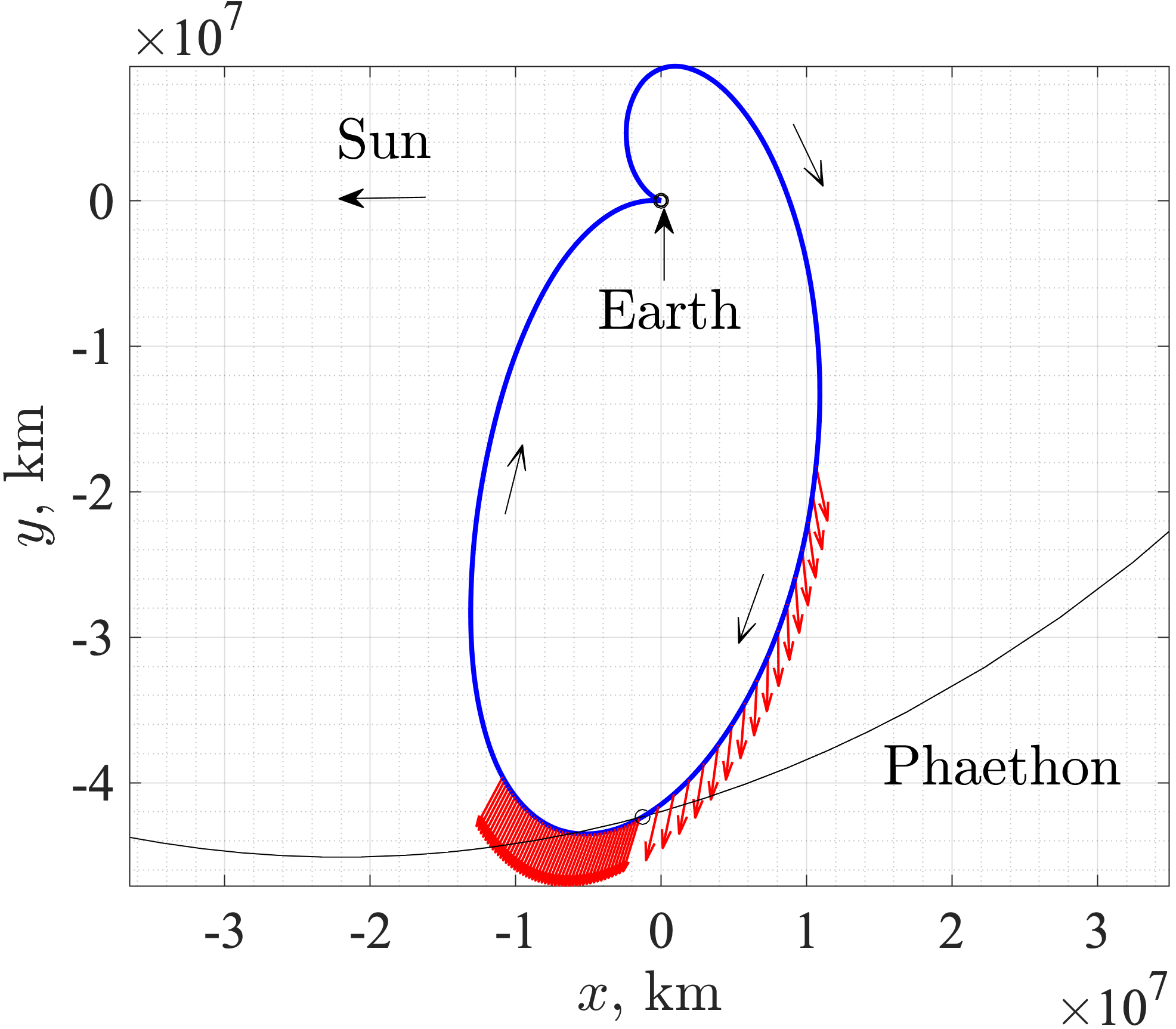}
         \caption{Case \#4, 1:1 generic}
         \label{fig:interp_2028_case4}
     \end{subfigure}
     \begin{subfigure}[b]{0.23\textwidth}
         \centering
         \includegraphics[width=\textwidth]{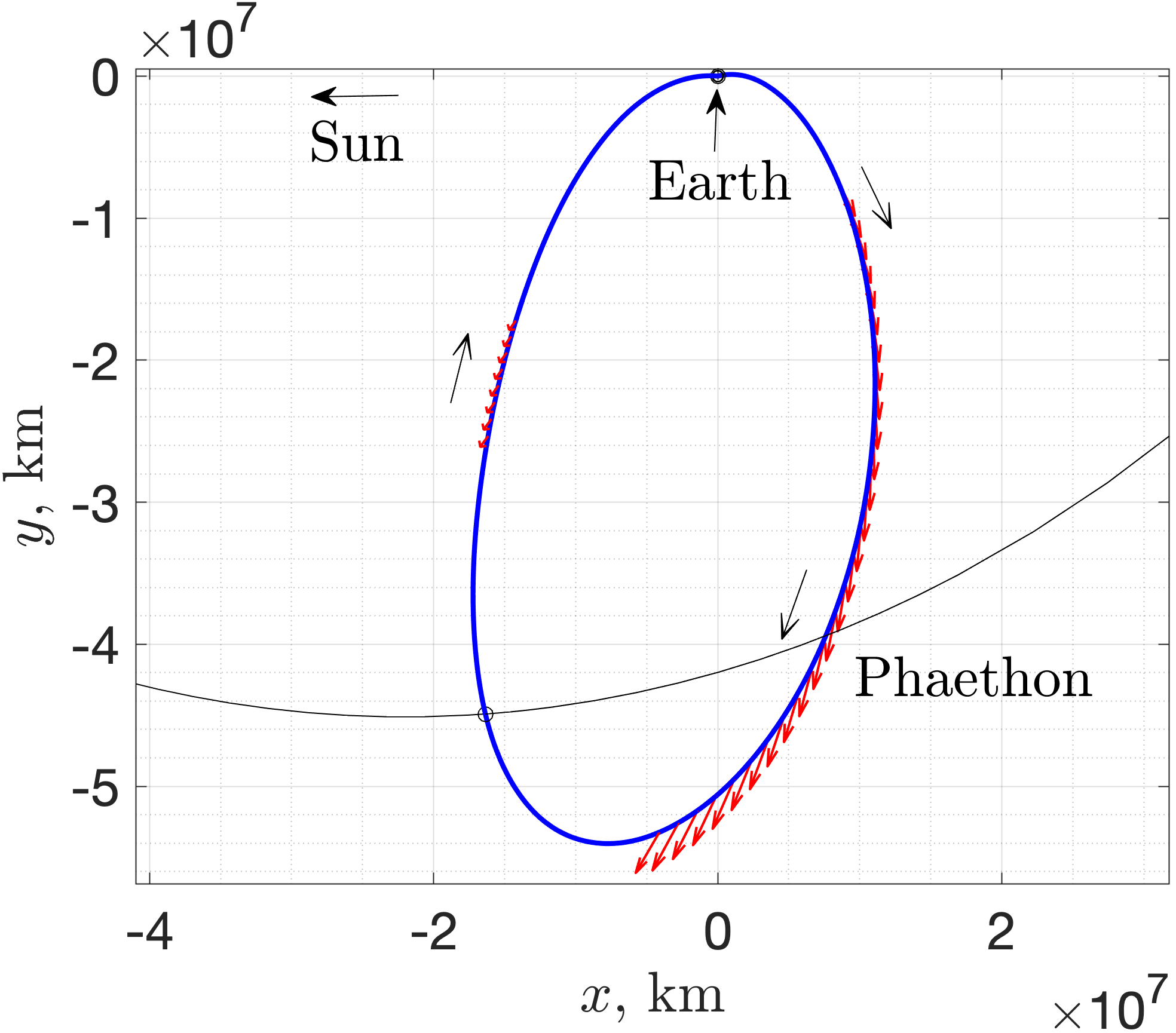}
         \caption{Case \#5, 1:1 full}
         \label{fig:interp_2028_case5}
     \end{subfigure}
     \begin{subfigure}[b]{0.23\textwidth}
         \centering
         \includegraphics[width=\textwidth]{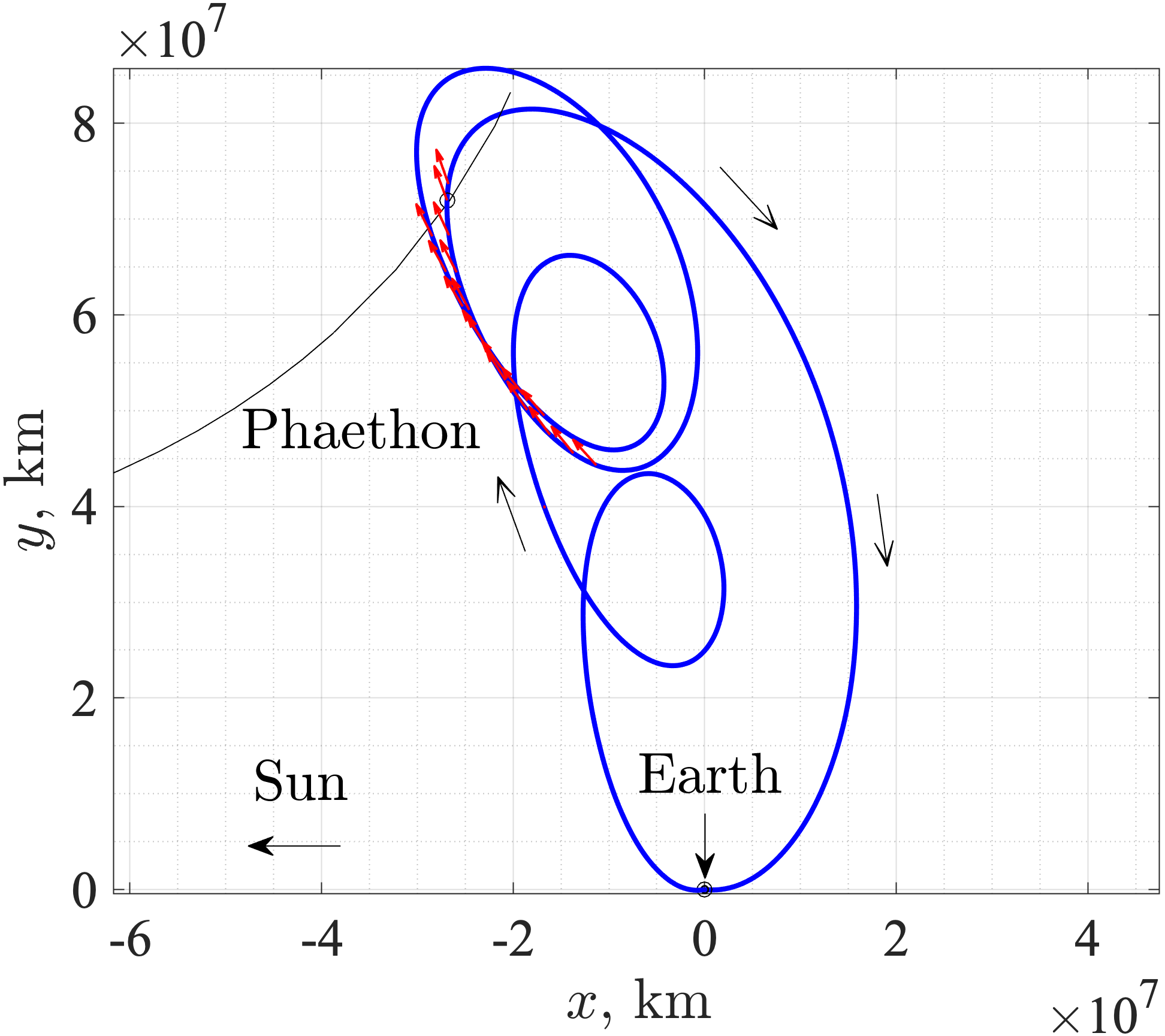}
         \caption{Case \#6, 4:4 full}
         \label{fig:interp_2030_case6}
     \end{subfigure}
     \begin{subfigure}[b]{0.23\textwidth}
         \centering
         \includegraphics[width=\textwidth]{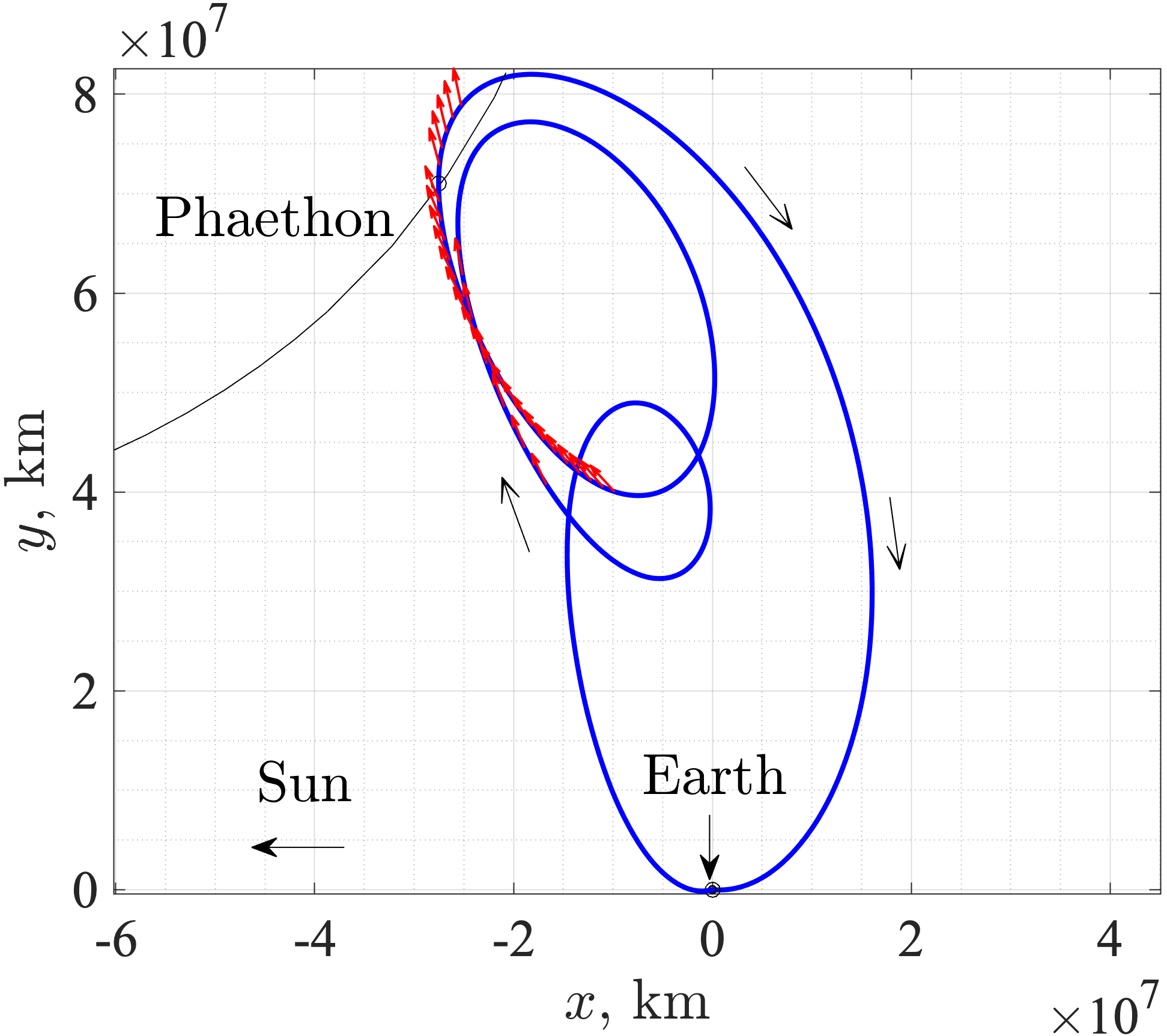}
         \caption{Case \#7, 3:3 full}
         \label{fig:interp_2030_case7}
     \end{subfigure}
     \begin{subfigure}[b]{0.23\textwidth}
         \centering
         \includegraphics[width=\textwidth]{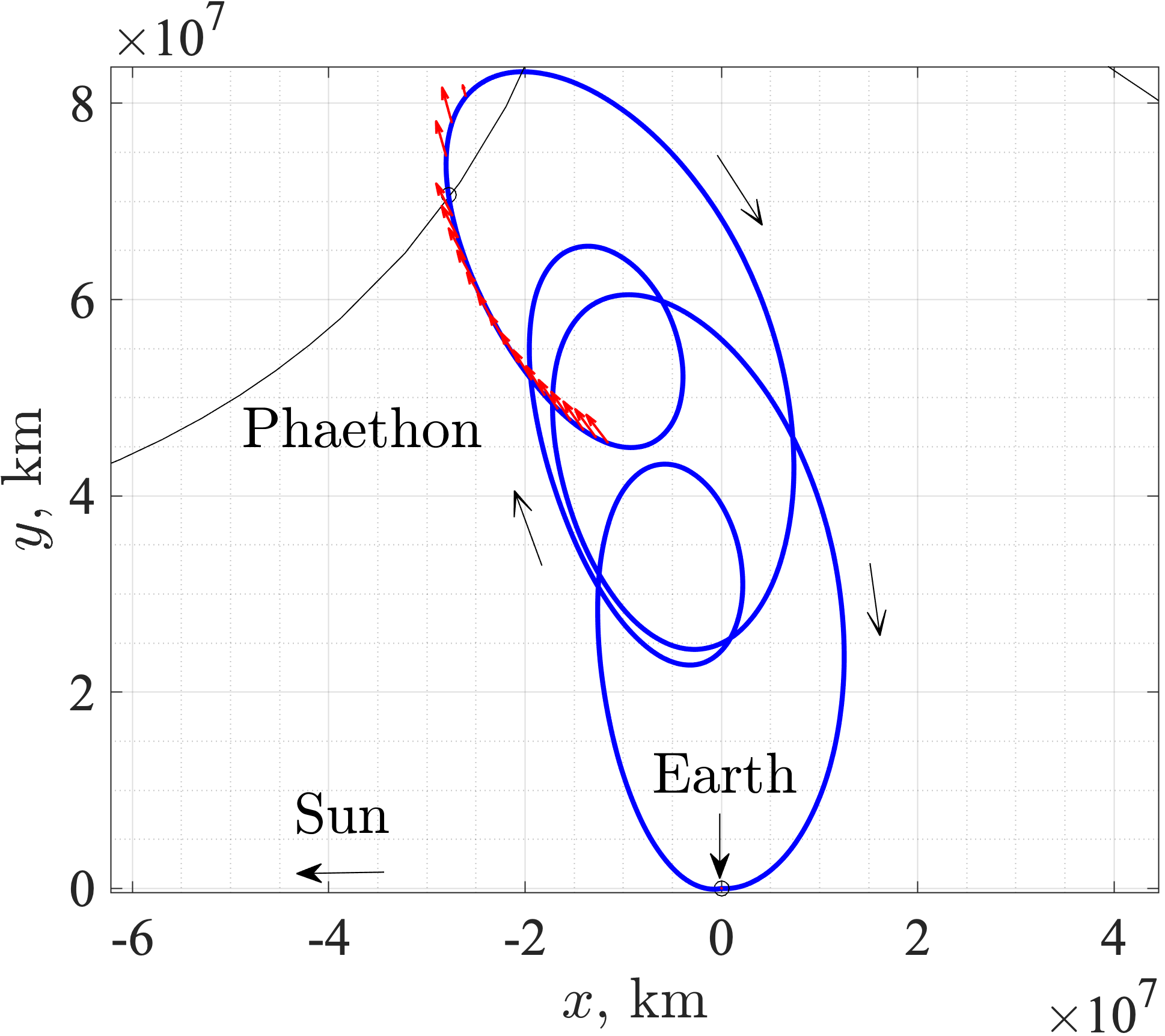}
         \caption{Case \#8, 4:4 full}
         \label{fig:interp_2030_case8}
     \end{subfigure}
     \begin{subfigure}[b]{0.23\textwidth}
         \centering
         \includegraphics[width=\textwidth]{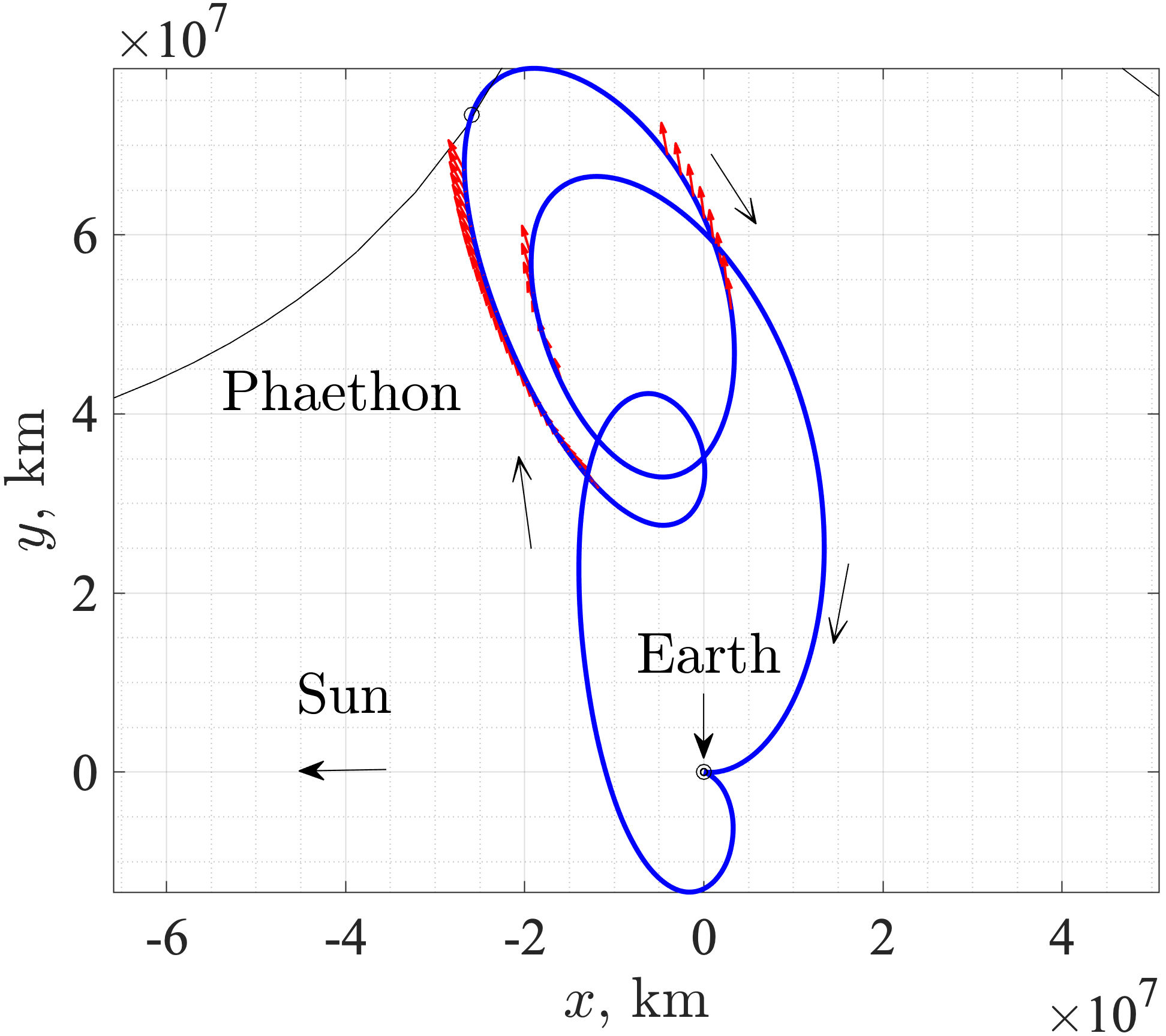}
         \caption{Case \#9, 3:3 generic}
         \label{fig:interp_2030_case9}
     \end{subfigure}
     \begin{subfigure}[b]{0.23\textwidth}
         \centering
         \includegraphics[width=\textwidth]{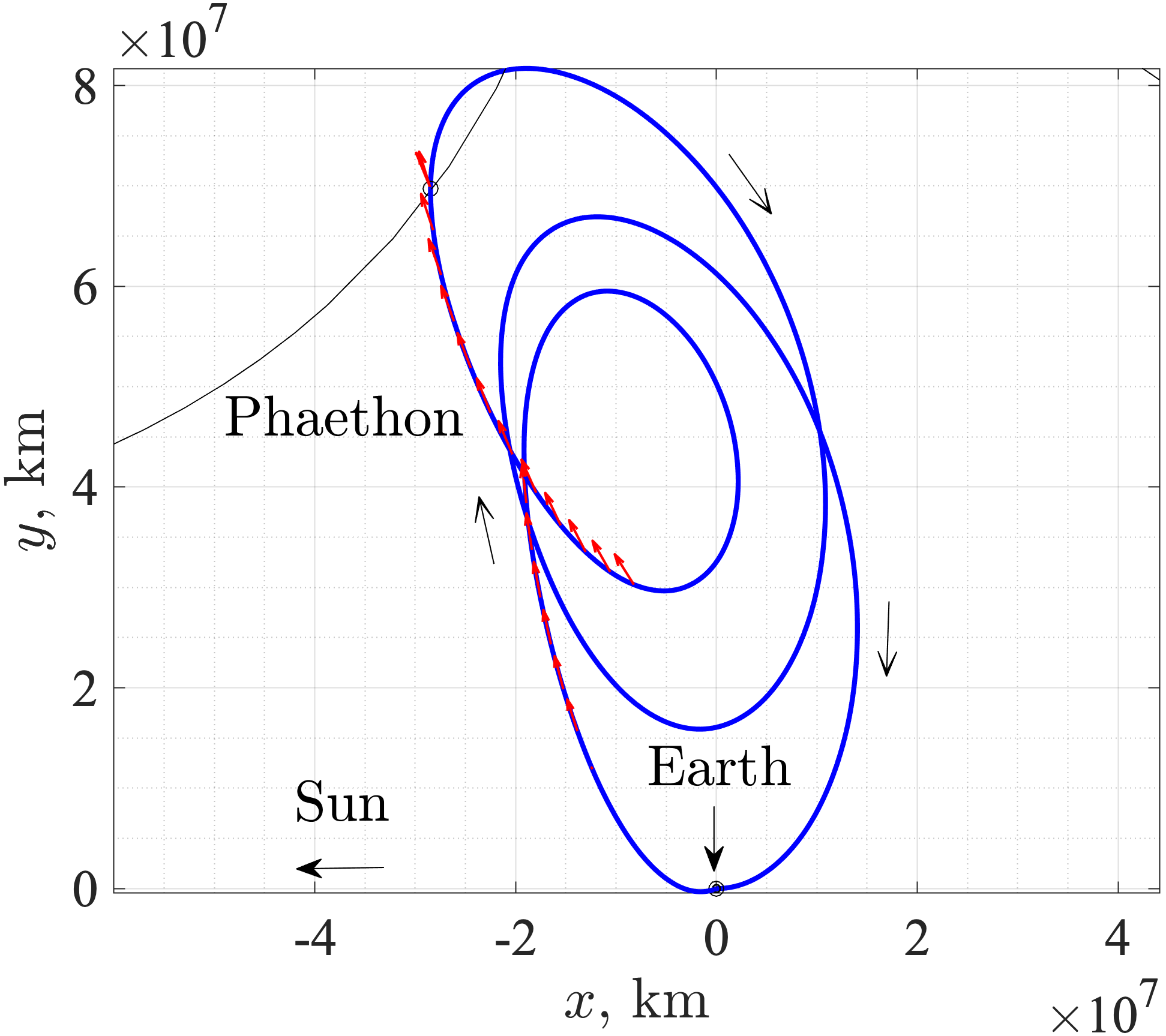}
         \caption{Case \#10, 3:3 full}
         \label{fig:interp_2030_case10}
     \end{subfigure}
     \caption{Representative IPT trajectories (Earth-centered, Sun-Earth line fixed rotational frame. Black solid line denotes Phaethon trajectory. Blue solid line describes \destiny trajectory. Red arrow indicates thrust vectors.)}
     \label{fig:interp_all}
\end{figure}
\fi

\begin{figure}[h!]
\centering
\ifreview
\includegraphics[width=\textwidth]{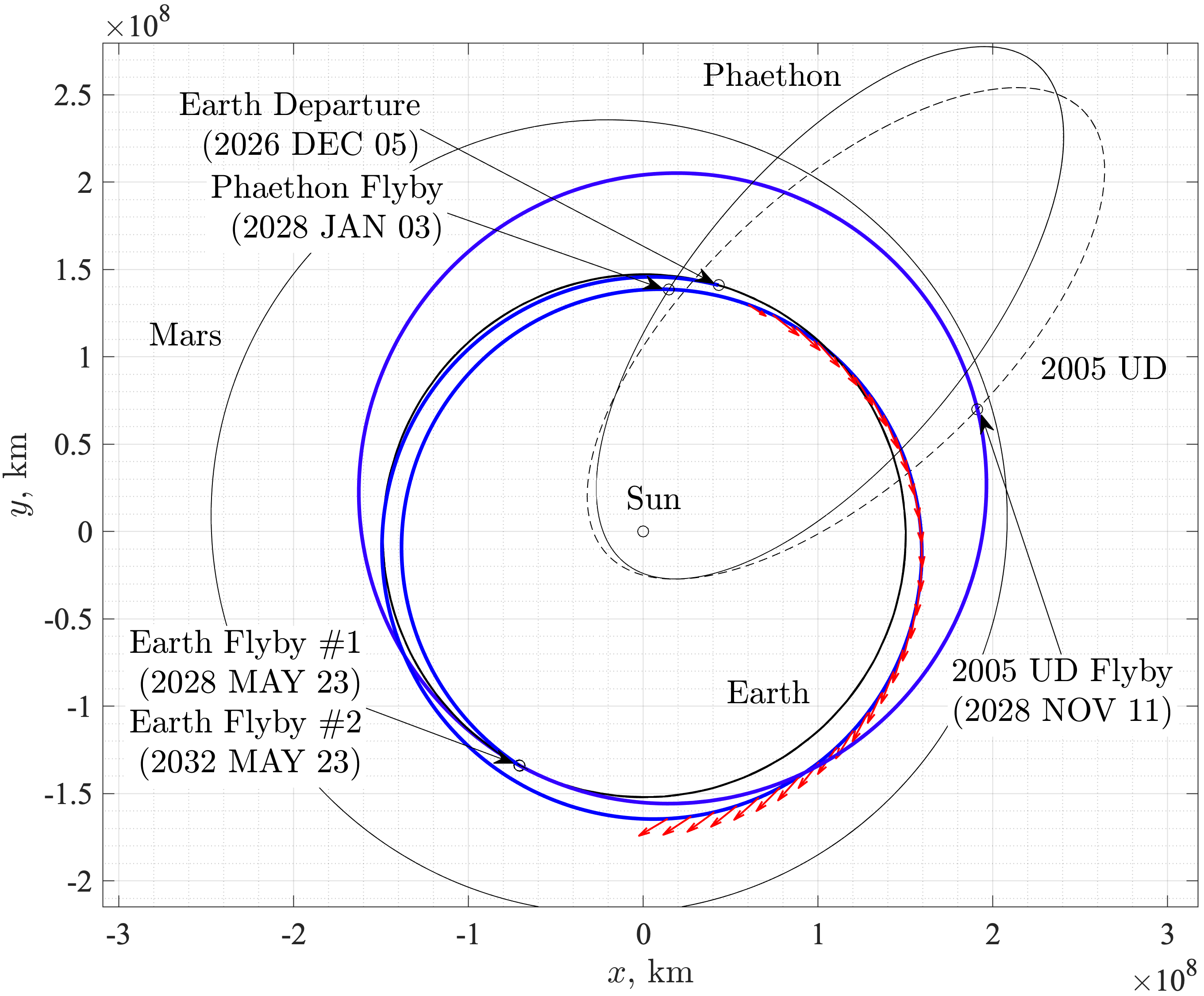}
\else
\includegraphics[width=0.46\textwidth]{img/destiny_baseline_trj_iner.png}
\fi
\caption{Example interplanetary trajectory (Sun-centered, ECLIPJ2000 inertial frame)}
\label{fig:interplanetary_baseline_iner}
\end{figure}

\begin{figure}[h!]
\centering
\ifreview
\includegraphics[width=\textwidth]{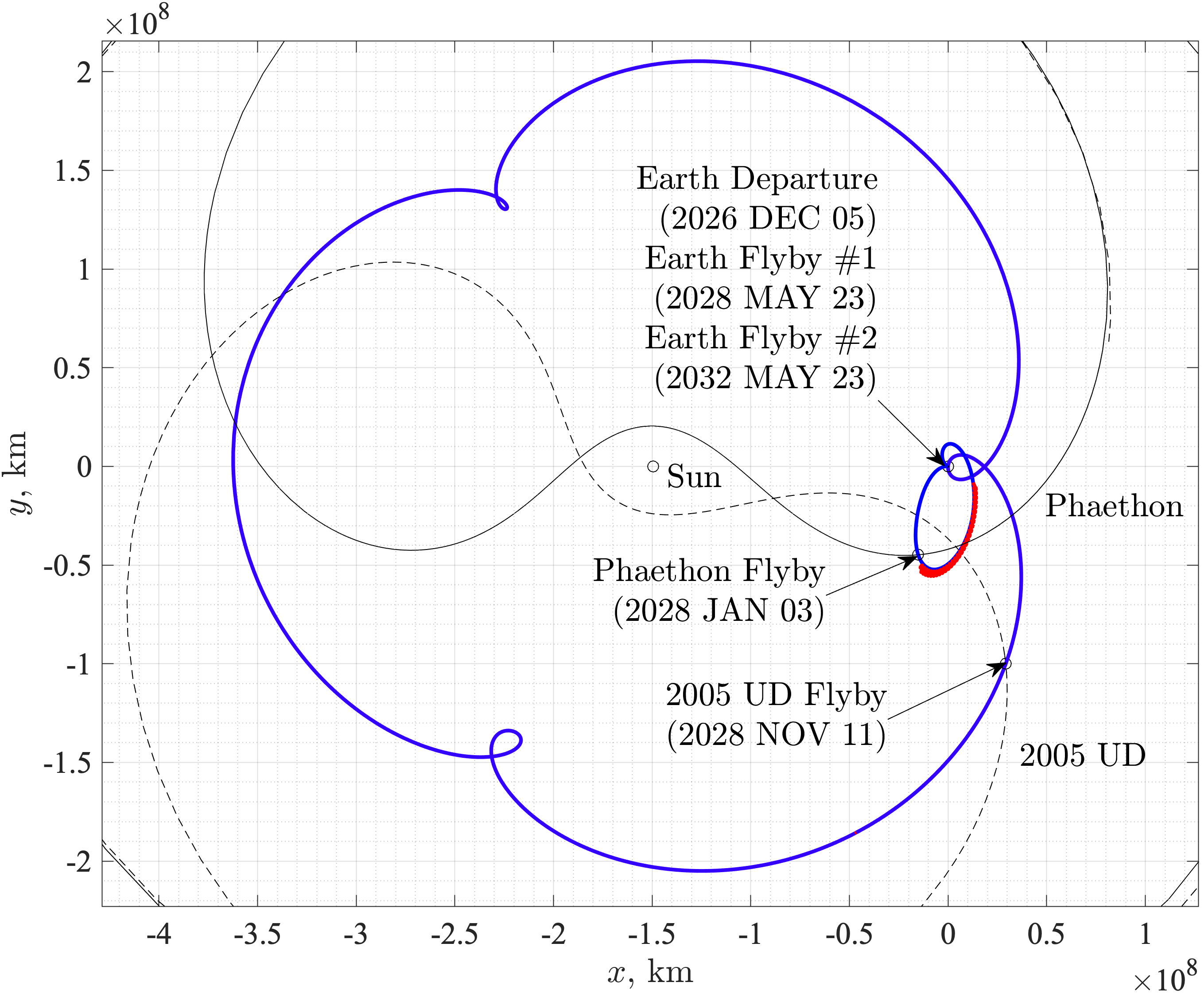}
\else
\includegraphics[width=0.46\textwidth]{img/destiny_baseline_trj_rot.png}
\fi
\caption{Example interplanetary trajectory (Earth-centered, Sun-Earth line fixed rotational frame)}
\label{fig:interplanetary_baseline_rot}
\end{figure}

\begin{figure}[h!]
\centering
\ifreview
\includegraphics[width=\textwidth]{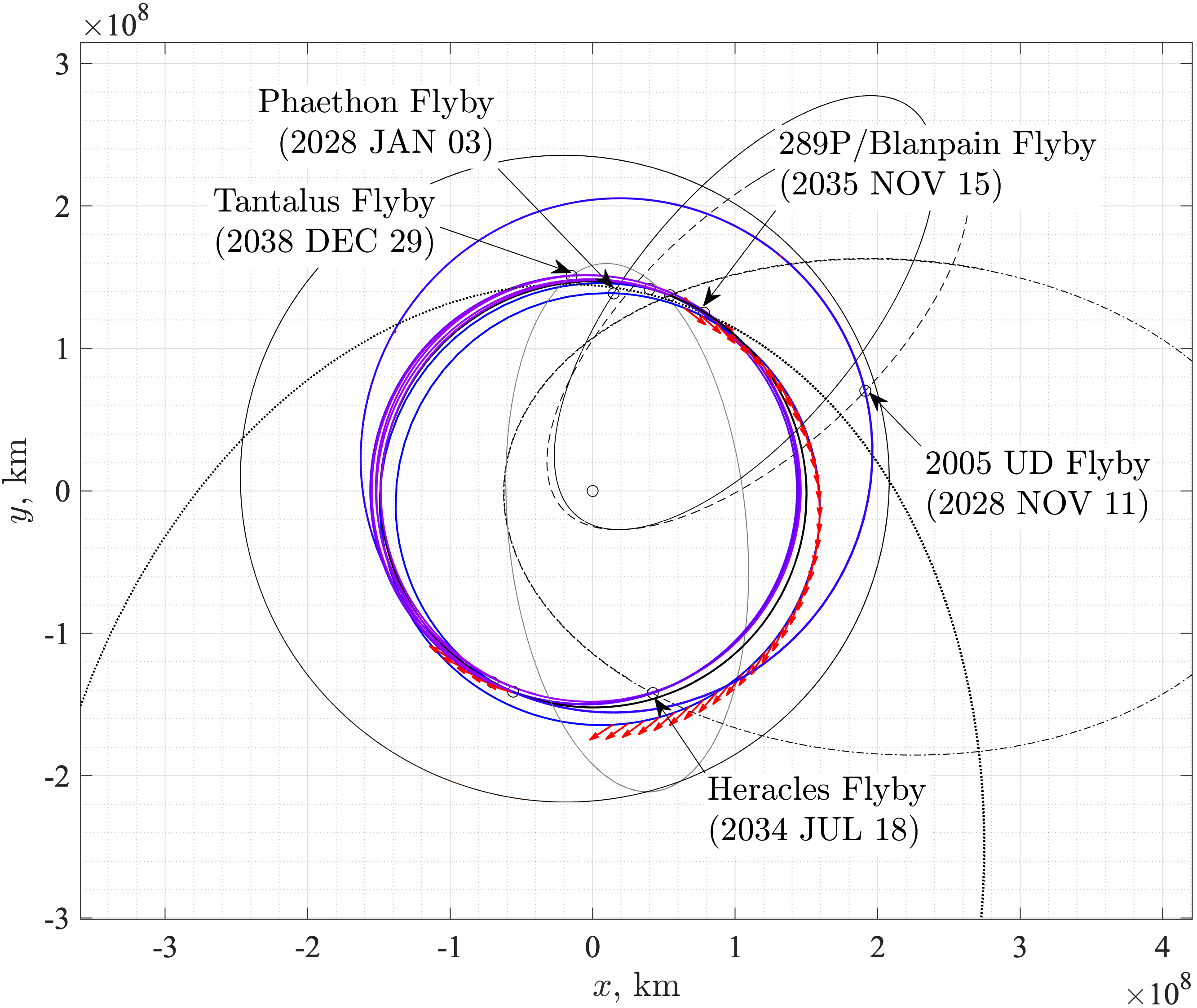}
\else
\includegraphics[width=0.46\textwidth]{img/destiny2028_multiflyby.png}
\fi
\caption{Example of multiple asteroid flyby trajectory (Sun-centered ECLIPJ2000 inertial frame)}
\label{fig:multiple_flyby_example}
\end{figure}

\begin{figure}[h!]
\centering
\ifreview
\includegraphics[width=\textwidth]{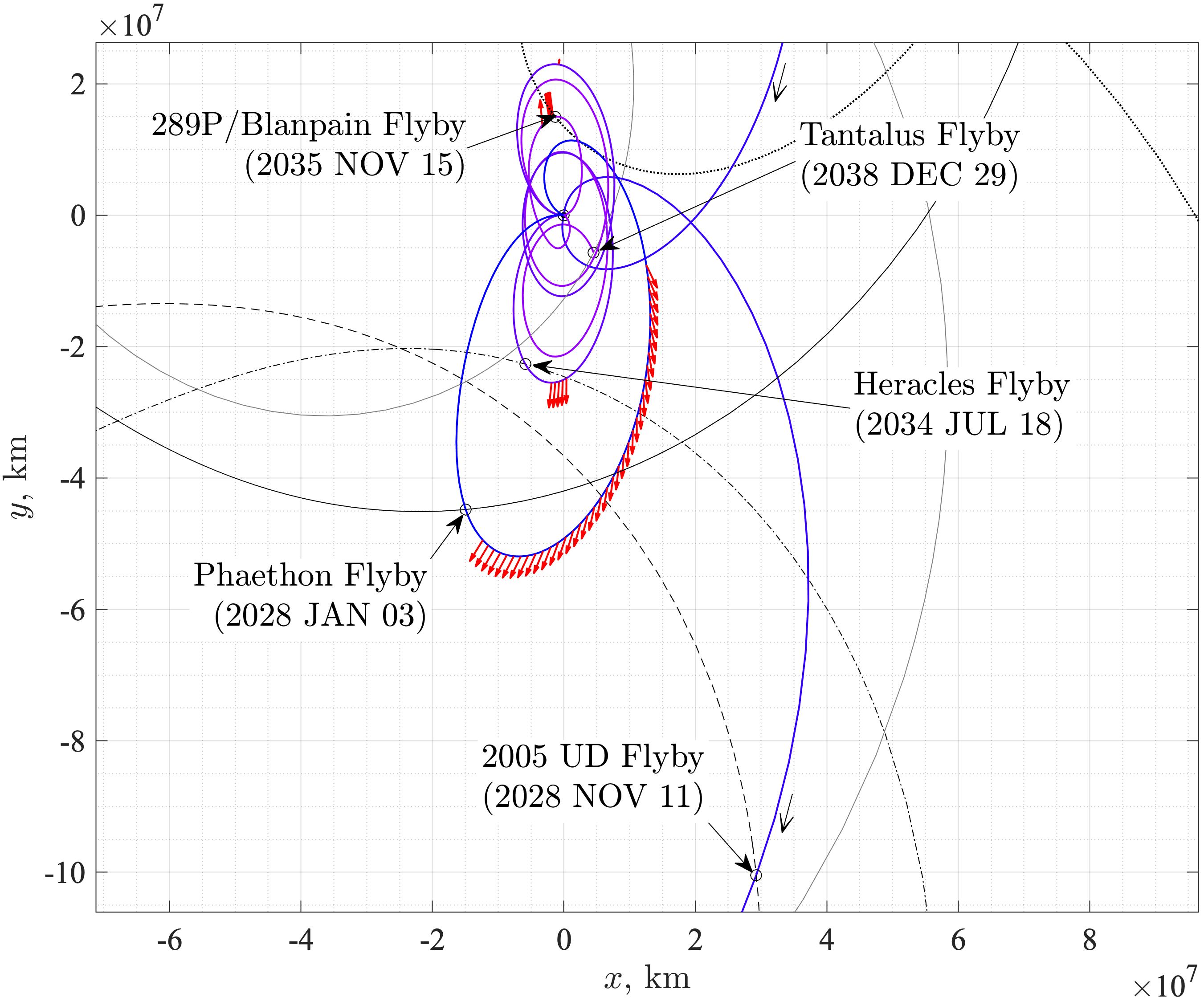}
\else
\includegraphics[width=0.46\textwidth]{img/destiny2028_multiflyby_rot.png}
\fi
\caption{Example of multiple asteroid flyby trajectory (Earth-centered, Sun-Earth line fixed rotational frame)}
\label{fig:multiple_flyby_example_rot}
\end{figure}

Applying the same technique to design the asteroid flyby cyclers for the 2005 UD flyby, we can also find the Earth-2005 UD-Earth transfer solutions. As the result, we found the Earth-Phaethon-Earth-2005 UD-Earth solutions for Cases \#1, \#3, \#5, where a 4:3 free-return transfer is used for the Earth-2005 UD-Earth arc. Figures \ref{fig:interplanetary_baseline_iner} and \ref{fig:interplanetary_baseline_rot} show the example interplanetary trajectory of Case \#3. Utilizing the asteroid flyby cyclers design method that relies on deep neural networks\cite{Ozaki2021}, we can also design some trajectories that allow for visiting multiple targets with a relatively small amount of $\Delta V$($\Delta V < 1$ km/s in the example), as shown in Figs.\ref{fig:multiple_flyby_example} and \ref{fig:multiple_flyby_example_rot}. \destiny can visit these targets if it leaves enough propellant after a successful nominal mission.

%
%
\subsection{Moon Flyby Phase}

After the SOR and before the IPT, \destiny\ will rely on a number of lunar flybys to escape from the gravitational influence of the Earth and proceed to the interplanetary phase.

The design of this phase is performed backwards in time. The interface with the IPT consists of the Earth departure $V_{\infty}$ vector and epoch. We analytically computed the set of hyperbolic escape trajectories from different Moon positions that patch with the Earth departure $V_{\infty}$ vector at the right epoch; these steps are explained in detail in Appendix~\minorblue{A}. The computed trajectories provide the velocity vector right after the Moon flyby. There are two groups of trajectories: those that reach the escape state directly (shown in Figure~\ref{fig:escape_trajectory_short}) and those that reach the escape state after a final Earth flyby (shown in Figure~\ref{fig:escape_trajectory_long}); these two scenarios are called short and long arc transfer, respectively. The color map in Fig.~\ref{fig:escape_trajectory_all} illustrates the outgoing $V_{\infty}$ with respect to the Moon required to connect to the IPT trajectory. Since the required outgoing $V_{\infty}$ with respect to the Moon is too high to connect it with the incoming $V_{\infty}$ that can be achieved at the end of the SOR, we designed a Moon-to-Moon sequence to bridge the gap. The multi-lunar-flyby approach has been successfully implemented in past missions\cite{DICARA2005250,UESUGI1991347}, and has been shown to provide the necessary energy boost to patch the SOR and IPT\cite{Yarnoz2016,Suda2017AAS,Yamamoto2019IAC}. In the present work, we designed the Moon-to-Moon sequence through a grid search \minorblue{in the full ephemeris model}: for each state at the final Moon flyby, we grid the direction of $\bm{v}_{\infty, \textrm{Moon}}^{in}$ and we propagate each case backward in time until they cross the Moon's sphere of influence again. The final states of this propagation become the set of states that may be patched with the spiral phase. In summary, the spiral phase is followed by a number of lunar flybys which then lead to a escape from the Earth directly or through a final Earth flyby.

\begin{figure}[h!]
\centering
\ifreview
\includegraphics[width=\textwidth]{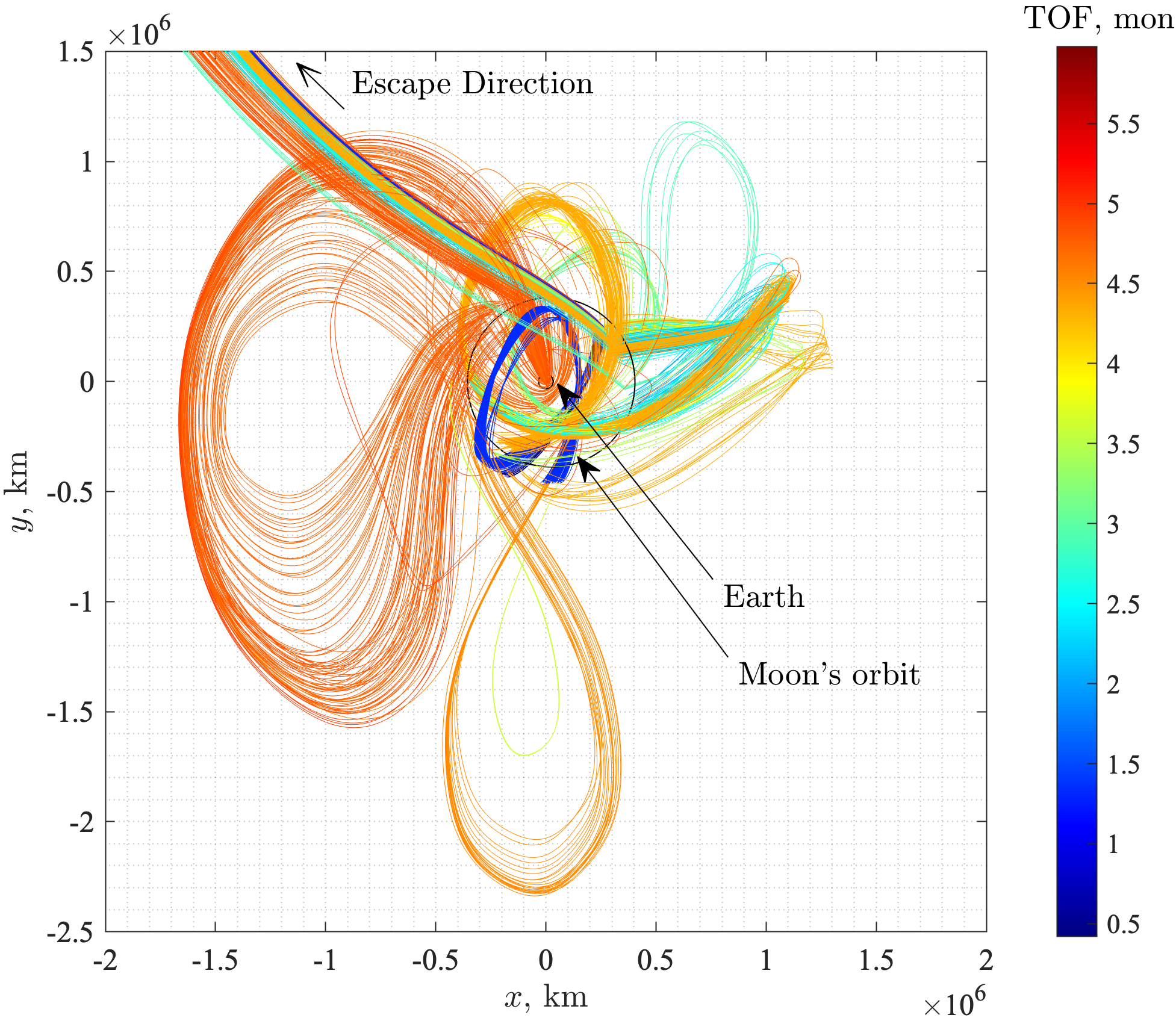}
\else
\includegraphics[width=0.46\textwidth]{img/m2m_24547_314_baseline_rev1.png}
\fi
\caption{Moon flyby trajectories for $V_{\infty,0}<1.0$~km/s and TOF$<5$~mon (Earth-centered, Sun-Earth-line-fixed rotational frame).}
\label{fig:m2m_trj}
\end{figure}

\begin{figure}[h!]
\centering
\ifreview
\includegraphics[width=\textwidth]{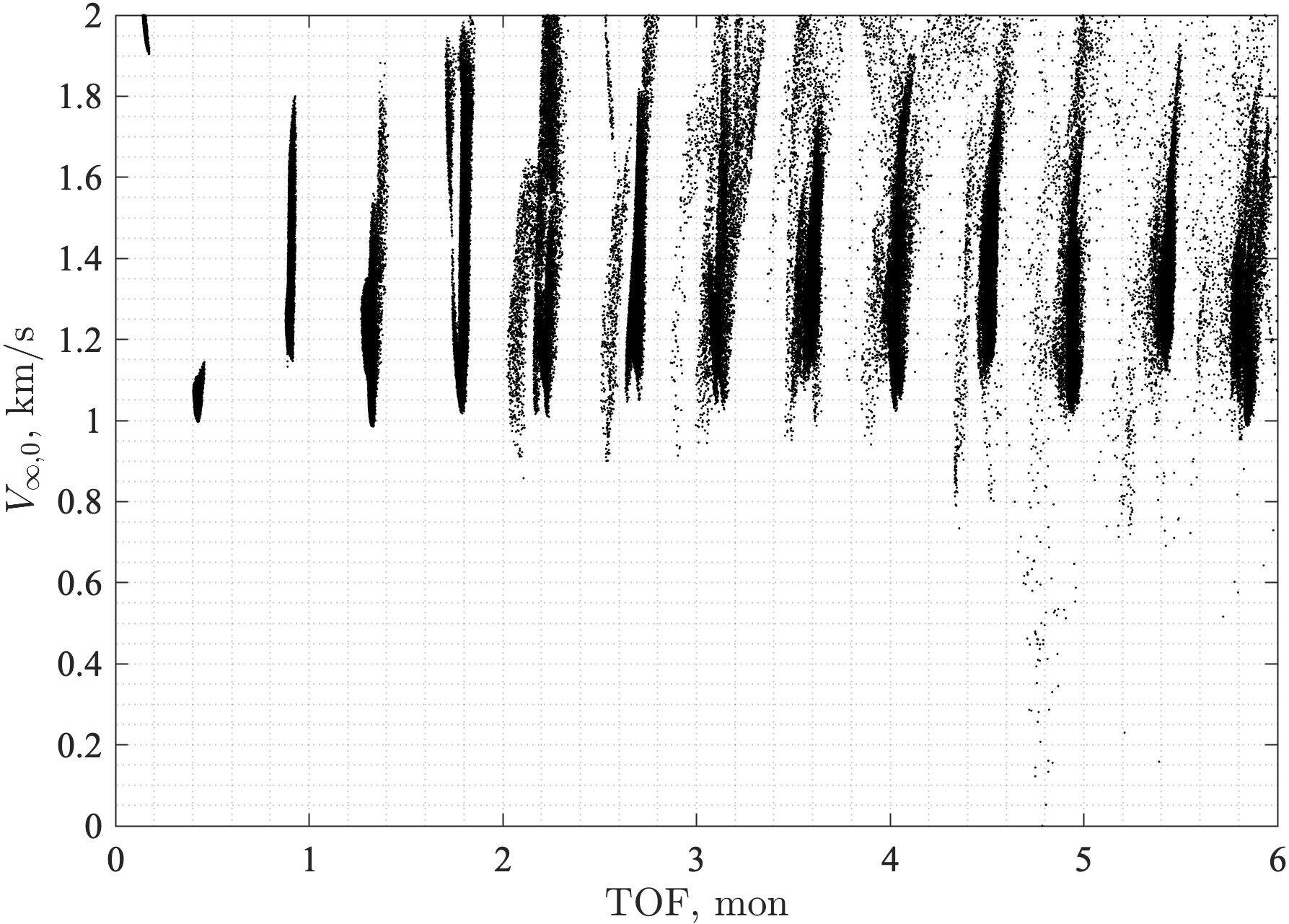}
\else
\includegraphics[width=0.46\textwidth]{img/m2m_24547_314_baseline_tof-vs-vinf0_rev1.png}
\fi
\caption{Initial $V_{\infty,0}$ required for Moon flyby trajectories for different time of flight.}
\label{fig:m2m_tof_vs_vinf0}
\end{figure}

The backward propagation is performed using an $n$-body integrator considering \minorblue{the full ephemeris model described previously}. Figure~\ref{fig:m2m_trj} illustrates a number of successful Moon-to-Moon transfers for up to 5~months of duration. Figure~\ref{fig:m2m_tof_vs_vinf0} analyses the range of both short and long arc solutions obtained via this method. There exist multiple instances of $v_{\infty}$ lower than 1~km/s for different times of flight. This is the $V_{\infty}$ magnitude that the spacecraft must achieve at the end of the spiral phase. Among the feasible solutions, we prefer shorter times of flight.


\ifreview
\begin{figure}
     \centering
     \begin{subfigure}[htb]{\textwidth}
         \centering
         \includegraphics[width=\textwidth]{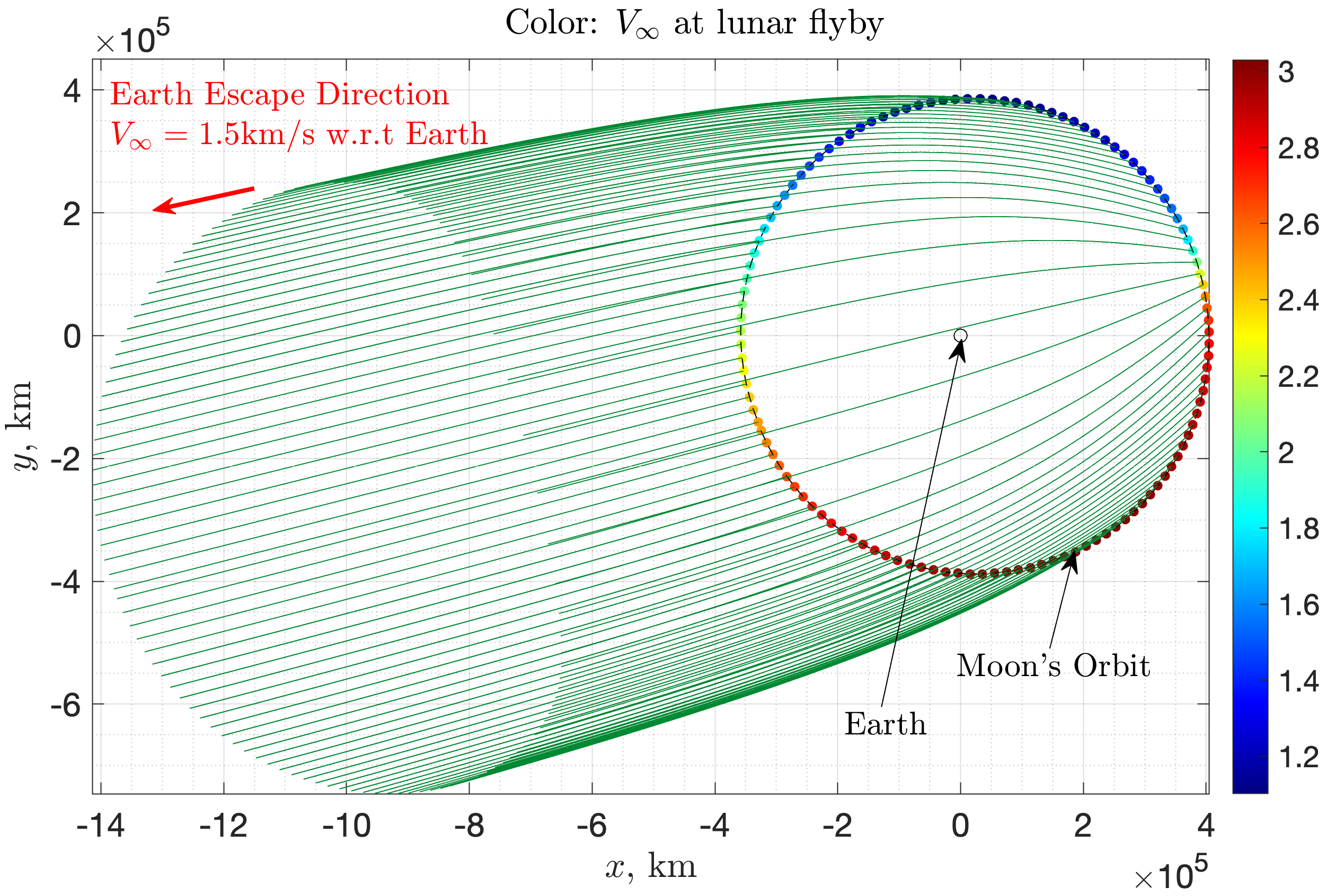}
         \caption{Short arc transfers.}
         \label{fig:escape_trajectory_short}
     \end{subfigure}
     \begin{subfigure}[htb]{\textwidth}
         \centering
         \includegraphics[width=\textwidth]{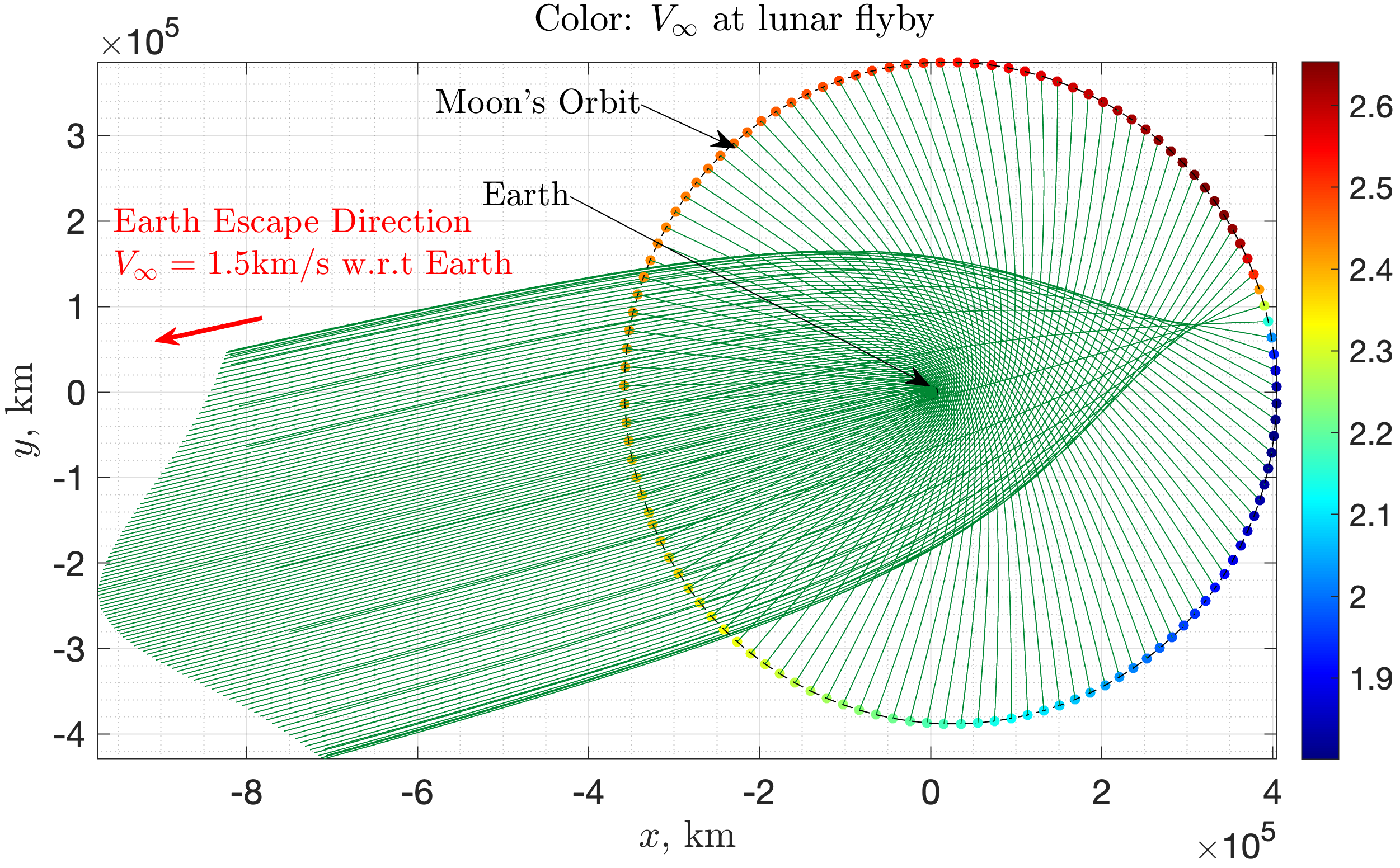}
         \caption{Long arc transfers.}
         \label{fig:escape_trajectory_long}
     \end{subfigure}
     \caption{Earth escape trajectory with Moon flyby (Earth-centered, ECLIPJ2000).}
     \label{fig:escape_trajectory_all}
\end{figure}
\else
\begin{figure}
     \centering
     \begin{subfigure}[htb]{0.46\textwidth}
         \centering
         \includegraphics[width=\textwidth]{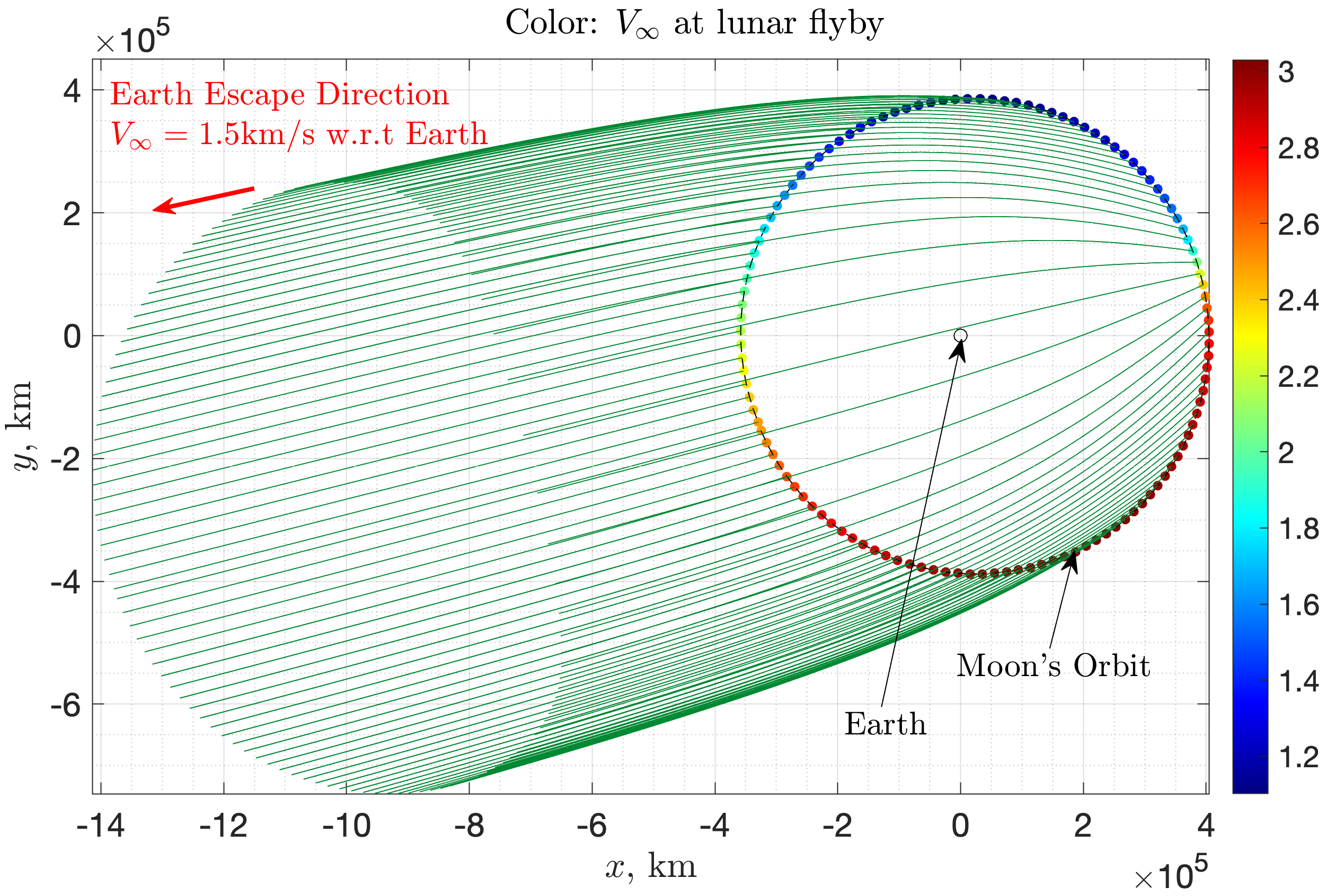}
         \caption{Short arc transfers.}
         \label{fig:escape_trajectory_short}
     \end{subfigure}
     \begin{subfigure}[htb]{0.46\textwidth}
         \centering
         \includegraphics[width=\textwidth]{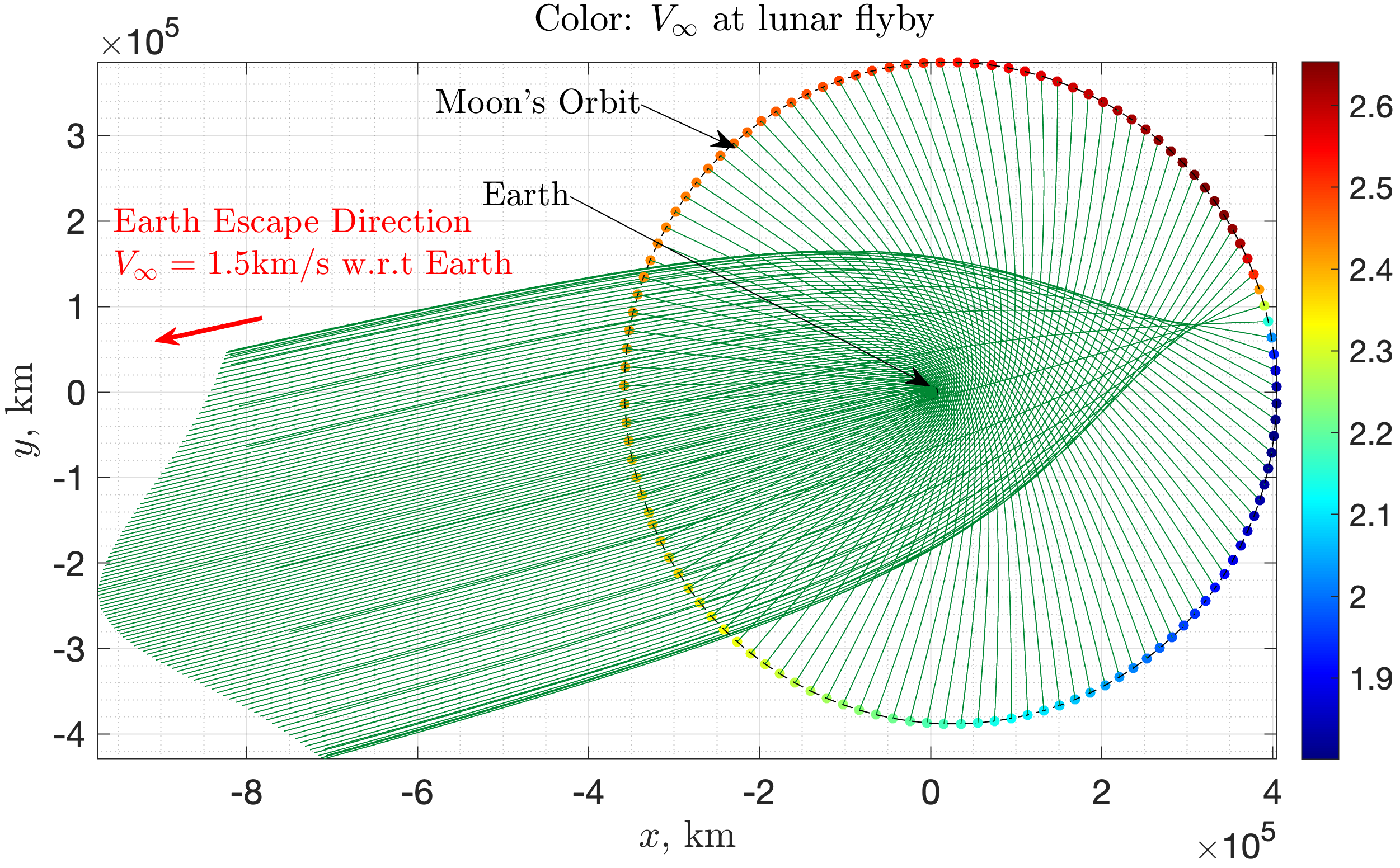}
         \caption{Long arc transfers.}
         \label{fig:escape_trajectory_long}
     \end{subfigure}
     \caption{Earth escape trajectory with Moon flyby (Earth-centered, ECLIPJ2000).}
     \label{fig:escape_trajectory_all}
\end{figure}
\fi

%
%
\subsection{Patched Trajectory Optimization}

\begin{figure}[h!]
\centering
\ifreview
\includegraphics[width=\textwidth]{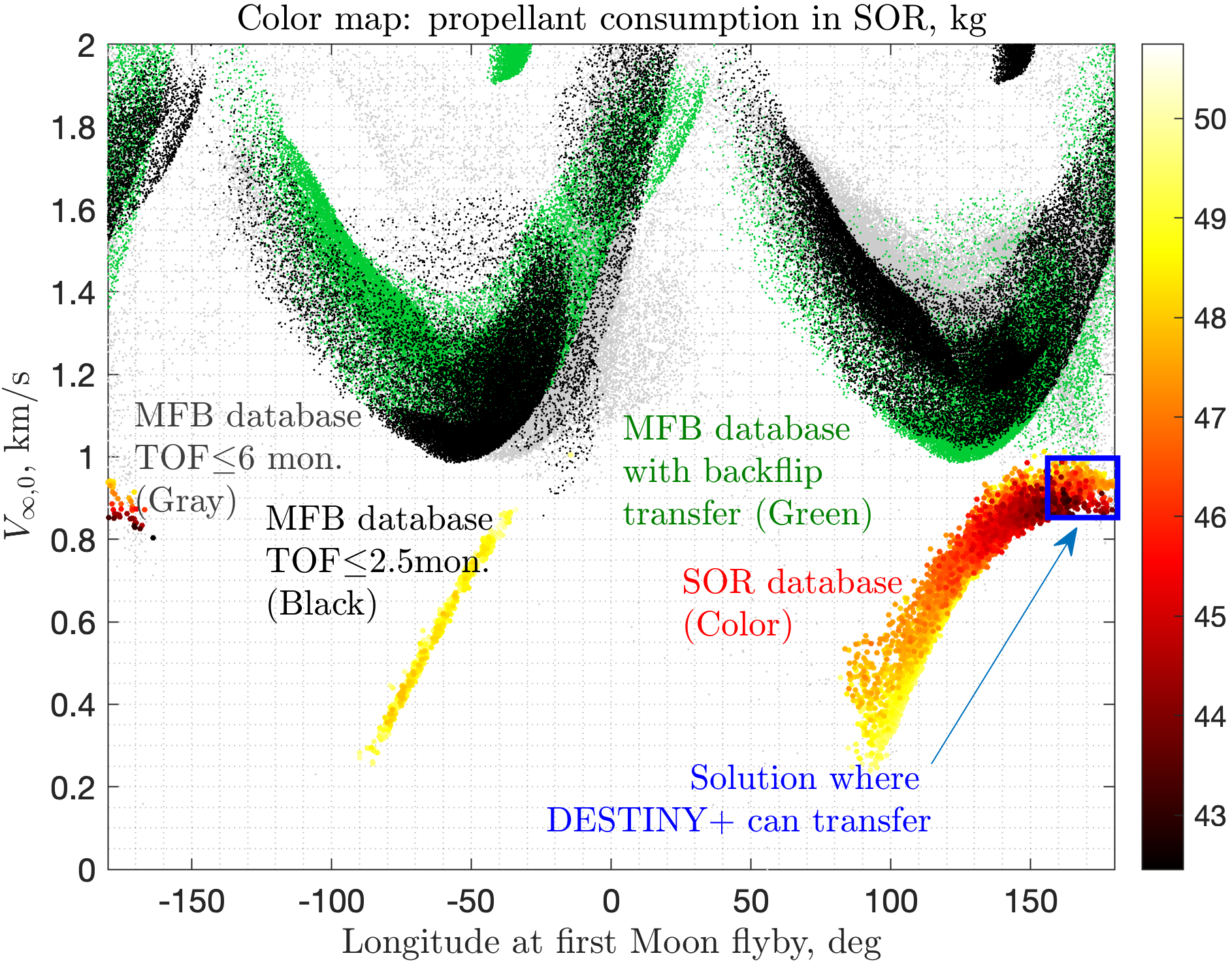}
\else
\includegraphics[width=0.46\textwidth]{img/patch_long_vs_vinf_rev1.png}
\fi
\caption{Longitude-$V_{\infty}$ plot for the first Moon flyby.}
\label{fig:patch_long_vs_vinf}
\end{figure}

Finally, we patch the full trajectory \minorblue{, including the last tens of revolutions of SOR, all phases of MFB, and all phases of IPT,} through a low-thrust trajectory optimization problem\cite{Campagnola2015} \minorblue{in the full ephemeris model}. We search for a suitable pair of the SOR and MFB trajectories with a small difference in longitude and $V_{\infty}$ at the first Moon flyby. Figure~\ref{fig:patch_long_vs_vinf} shows the longitude and $V_{\infty}$ at the first Moon flyby for both the SOR and the MFB trajectories. Note that this plot ignores the constraint of the deflection angle of the Moon flyby. We can also insert some additional Moon free-return transfers if time allows. The green dots in Fig.~\ref{fig:patch_long_vs_vinf} are the solutions for the cases of adding a backflip free-return transfer\cite{Strange2008}, which can change the longitude at the first Moon flyby by 180~degrees in half of a lunar orbital period. Figure~\ref{fig:patch_long_vs_vinf} shows an overlap in the region of 160-180~degrees of longitude\minorblue{, and MFB trajectories with backflip transfers provide the shortest transfer to patch the entire trajectory.} 

Using a solution pair as an initial guess, we optimize the patched low-thrust trajectory \minorblue{in the full ephemeris model}. The optimization includes the last 10-20~revolutions of the SOR trajectory because of the phasing and targeting of the Moon. In the final stage of the SOR, the perturbation caused by the Moon's gravity can no longer be ignored. In particular, if a distant Moon flyby occurs throughout the phase, we cannot perform the first Moon flyby under the expected $V_{\infty}$ and longitude conditions. Therefore, we design a patched trajectory that avoids distant Moon flybys. Future work will positively exploit the effects of the distant moon flybys.

\minorred{We first patch the MFB and IPT trajectories with the constraints of the Earth escape $V_{\infty}$ magnitude and the longitude of the first lunar flyby. Meanwhile, we optimize the final phase of the SOR trajectory so that the spacecraft can perform the lunar flyby at the proper longitude. Finally, we solve the trajectory optimization problem of the fully patched trajectory. In a preliminary optimization step of the fully patched trajectory, we still impose the constraint that the Earth departure $V_{\infty}$ is less than 1.5~km/s, which is the interface condition used in the IPT trajectory design.} The patched trajectory optimization, particularly the part of the MFB trajectory, tends to diverge without this constraint. Once we find a feasible patched trajectory, we perform the final optimization without this constraint on the Earth departure $V_{\infty}$ and reduce the fuel consumption in the IPT.

\begin{figure}[h!]
\centering
\ifreview
\includegraphics[width=\textwidth]{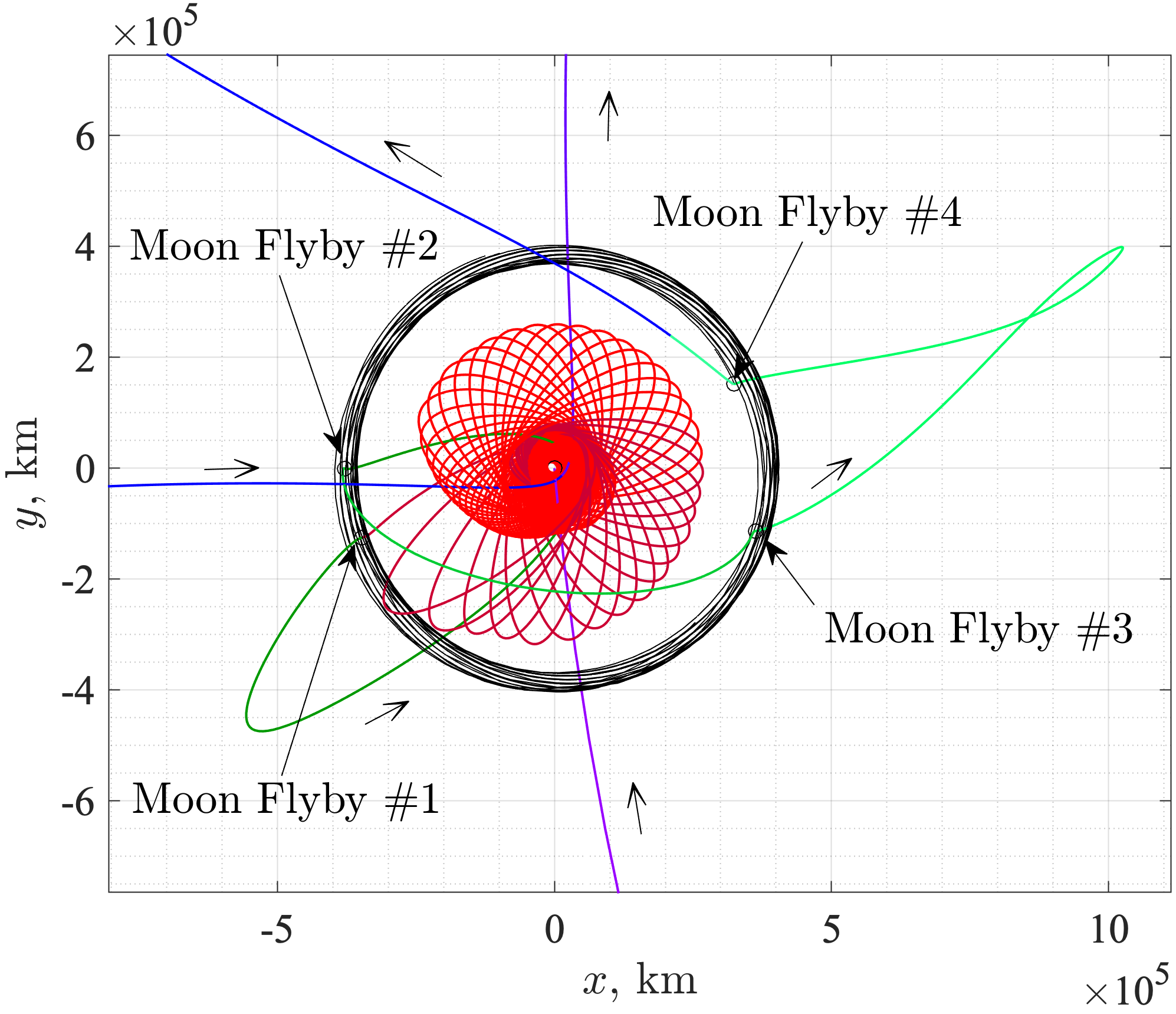}
\else
\includegraphics[width=0.45\textwidth]{img/destiny_trj_se_rot_ec.png}
\fi
\caption{Baseline near-Earth trajectory of \destiny\ in the Earth-centered Sun-Earth line fixed rotational frame.}
\label{fig:baseline_earthescape_trajectory_rot}
\end{figure}

\begin{figure}[h!]
\centering
\ifreview
\includegraphics[width=\textwidth]{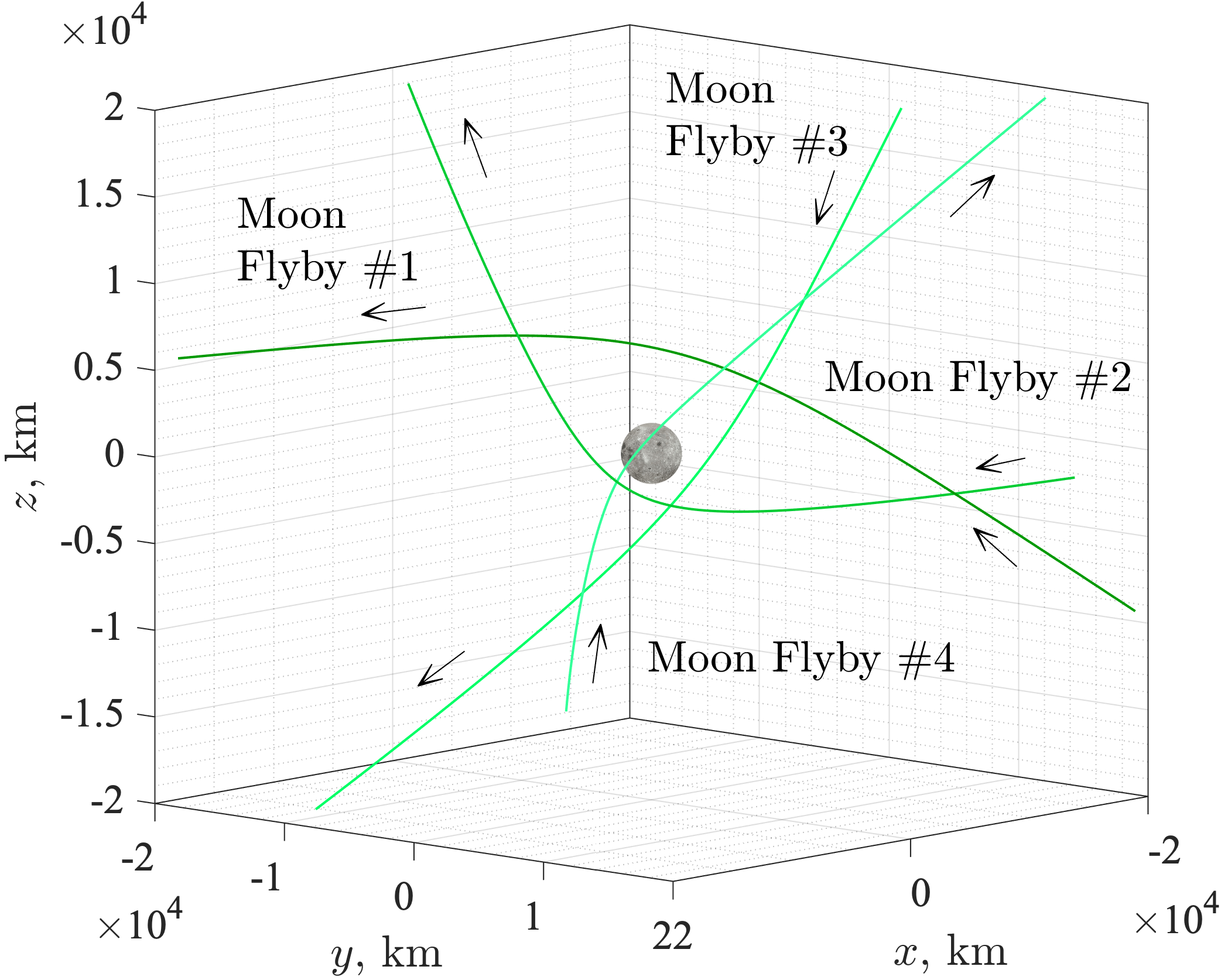}
\else
\includegraphics[width=0.45\textwidth]{img/moon_flyby_baseline.png}
\fi
\caption{Moon flyby in the baseline trajectory in the Moon-centered ECLIPJ2000 inertial frame.}
\label{fig:baseline_moon_flyby}
\end{figure}

The complete trajectory is shown in Figs. \ref{fig:baseline_earthescape_trajectory}, \ref{fig:baseline_interplanetary_trajectory}, and \ref{fig:baseline_earthescape_trajectory_rot}. The date and time, the spacecraft mass, and $V_{\infty}$ at the major event are summarized in Table~\ref{tab:baseline_trajectory}. In this trajectory, we inserted three Moon-to-Moon transfers with four Moon flybys. The first one is a 1:1 resonance transfer, which can be removed if the arrival time at the Moon is delayed. We insert this first transfer to ensure robustness against delays in the SOR operation. The second one is a backflip transfer that changes the longitude of Moon flyby by 180~degrees. The final transfer is the one obtained from the MFB database, shown in Fig.~\ref{fig:m2m_trj}, and has a flight time of about 2 months. The geometry of these lunar flybys is shown together in Fig.~\ref{fig:baseline_moon_flyby}, where the lowest flyby altitude is 200~km in the flyby \#4.


%
%
\section{Flight Operation Plan}
\label{sec:flight_operation}

This section describes the flight operation plan in two critical operations including the spiral orbit-raising phase and Phaethon flyby phase.

%
%
\subsection{Spiral Orbit-raising Operation}

The SOR requires \minorblue{a total of 1.5 years of ion engine operation, which} consumes 80\% of the onboard propellant. During the SOR, we plan to determine the orbit of \destiny\ by radio navigation. There will be up to three days of non-visible time from Uchinoura Space Center, the main ground station used for \destiny, and therefore we will make arrangements to use a backup station if necessary.

\destiny\ can control its orbit in two different ion engine operation modes: onboard automatic orbit control mode and manual orbit control mode. In the onboard automatic orbit control mode, the spacecraft can accelerate along the tangential direction of the orbit using the autonomous onboard Attitude and Orbit Control System (AOCS)\minorblue{, where the tangential direction is predicted from the instantaneous orbital elements via the propagator, without using an on-board navigation system}. This mode is mainly used in SOR-2, the sub-phase after the initial checkout operation and before the spacecraft exits the radiation belt. In the latter part of the spiral orbit-raising phase, SOR-3, we specify the time series of thrust directions using the manual orbit control mode, because more efficient orbit control and phase adjustment for the lunar flyby is required. 

\minorred{Regardless of the ion engine operation mode, the spacecraft can always generate maximum power by rotating the single-axis gimbal of the solar array panel, and the spacecraft can accelerate in any direction except during an Earth or Moon eclipse. During the eclipse of up to 1.5 hours, we plan to suspend the ion engine acceleration due to power constraints.}

By determining thrust direction and Sun direction, we can uniquely define the attitude of the spacecraft. However, the presence of the Earth and Moon in the star tracker's field of view may degrade the attitude determination accuracy. Regarding the communication capabilities, the spacecraft is equipped with a low-gain antenna able to transmit data with Earth \minorblue{and to perform radio navigation} even in periods when unfavorable attitudes for communication have to be maintained. 


We need to consider the counterplan against uncertainties from navigation errors and contingency operations in the SOR operation. In particular, we plan a contingency operation so that the ground station antenna can acquire the spacecraft's signal again even if the ion engine unexpectedly stops during a time window of no communication. These contingency operations will lead to a delay for the first Moon flyby. Therefore, we plan to leave enough coasting arcs until the first Moon flyby \minorblue{(more than several months)} and switch them to thrusting arcs when the contingency operation occurs.



As a science mission, the DDA will observe the dust environment around the Earth during the entire SOR phase. We will also perform optical calibration operations for TCAP and MCAP.

%
%
\subsection{Phaethon Flyby Operation}

\begin{figure}[h!]
\centering
\ifreview
\includegraphics[width=\textwidth]{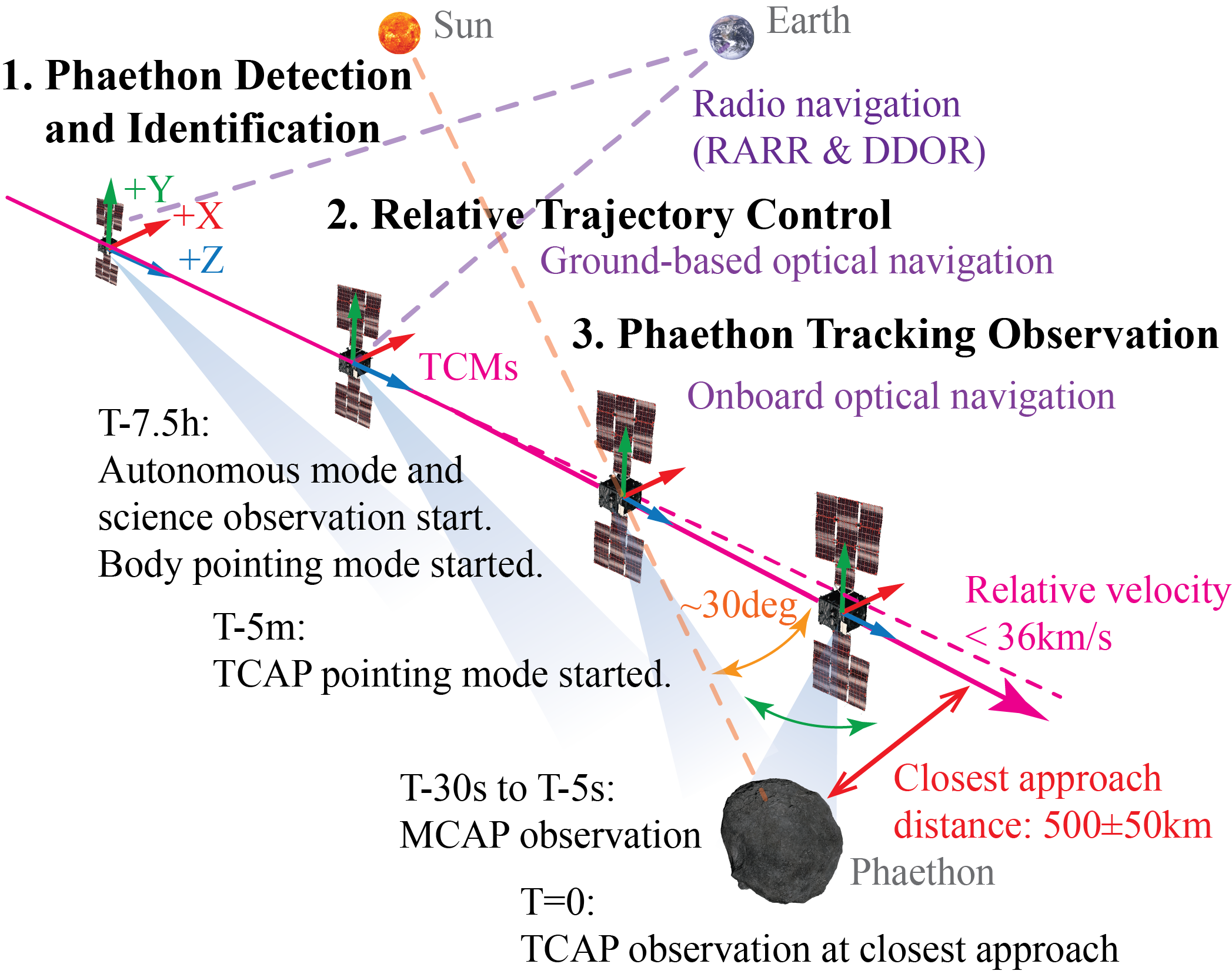}
\else
\includegraphics[width=0.45\textwidth]{img/phaethon_flyby_operation.png}
\fi
\caption{Overview of Phaethon flyby operation.}
\label{fig:phaethon_flyby_operation}
\end{figure}

For the Phaethon flyby phase (PFB), we define the three sub-phase: Phaethon detection and identification phase, relative orbit control phase, and Phaethon tracking observation phase. Figure~\ref{fig:phaethon_flyby_operation} is an overview of the Phaethon flyby operation. Table~\ref{tab:flyby_soe} shows the details of the sequence of events. \destiny\ will perform flyby observations on the most illuminated side at the closest approach distance of 500$\pm$50~km, where the maximum allowed relative velocity is 36~km/s. This closest approach position accuracy is achieved by hybrid radio-optical navigation.

In the Phaethon detection and identification sub-phase, the TCAP detects and identifies the Phaethon against the background stars. In this sub-phase, we estimate the misalignment between the TCAP and the star tracker using optical images of the stars. We use a medium-gain antenna with a 2-axis gimbal to downlink images at a high data rate during the operation. Precise misalignment estimation is essential for onboard optical navigation and tracking control. After the Phaethon identification is completed, we perform optical navigation and trajectory correction maneuvers (TCMs). As a result of the TCMs, we achieve 50~km (3$\sigma$) closest approach position accuracy on the B-plane. In the Phaethon tracking observation operations, \destiny\ controls the TCAP and its attitude using autonomous onboard optical navigation so that the TCAP can track Phaethon with a pointing accuracy of 0.15-0.20~deg (3$\sigma$) and a pointing stability $1.2\times 10^{-3}$~deg during 0.3~ms (3$\sigma$) at the closest approach. 

There are two tracking control modes: the body pointing mode and the TCAP pointing mode. In the body pointing mode, the rotation angle of the TCAP is fixed at the origin, and the spacecraft's attitude is controlled so that the TCAP tracks Phaethon. Although the tracking speed is slow, the TCAP disturbances can be ignored in this mode, so we can achieve higher pointing accuracy. This mode is used until the spacecraft can accurately estimate the Phaethon direction in order to fix the former's attitude. In the TCAP pointing mode, the spacecraft attitude is fixed to that of the closest approach, and the TCAP is controlled to follow Phaethon. Because the tracking speed of this mode is faster than the body pointing mode, this mode is used for tracking control near the closest approach.

\begin{table*}[t]
\caption{\label{tab:flyby_soe} Sequence of events in Phaethon flyby phase}
\centering
\ifreview
\scalebox{0.7}{
\fi
\begin{tabular}{lll}
\toprule
\textbf{Time} & \textbf{Sub-phase} & \textbf{Events}\\
\midrule
\multirow{4}{*}{T-30d to T-5d} & \multirow{4}{*}{Phaethon detection and identification} & Precise orbit determination by radio navigation\\
& & Misalignment estimation between TCAP and star tracker\\
& & Phaethon detection completed (T-10d at the latest)\\
& & Phaethon identification completed (T-5d at the latest)\\
\midrule
T-5d to T-2.5d & \multirow{7}{*}{Relative trajectory control} & Downlink Phaethon images for optical navigation \\
\multirow{2}{*}{T-2.5d to T-2d} &  & Relative orbit determination using optical images\\
& & TCM1 planning and execution\\
\multirow{2}{*}{T-2d to T-1d} & & Dust trail observation by DDA started\\
& & Downlink Phaethon images for optical navigation\\
\multirow{2}{*}{T-1d to T-7.5h} &  & Relative orbit determination using optical images\\
& & TCM2 planning and execution if necessary\\
\midrule
\multirow{3}{*}{T-7.5h to T-5m} & \multirow{5}{*}{Phaethon tracking observation} & Autonomous mode and science observation started\\
& & Body pointing mode started\\
T-5m to T+15m & & 3D attitude fixed and TCAP pointing mode started\\
\multirow{2}{*}{T+15m to T+1d} & & Attitude maneuver for high speed communication\\
& & Phaethon flyby operation completed\\
\bottomrule
\end{tabular}
\ifreview
}
\fi
\end{table*}

%
%
\section{Conclusions}
\label{sec:conclusions}

\destiny\ is an upcoming JAXA Epsilon medium-class mission to be launched in 2024 using Epsilon~S launch vehicle. The mission is the world's first mission to escape from a near-geostationary transfer orbit to deep space using low-thrust propulsion. The primary engineering mission objective is the demonstration of advanced technologies that include efficient solar electric propulsion and low-energy astrodynamics; these demonstrations will play an important role in enabling future small-scale spacecraft to carry out deep space exploration. As the primary science mission objective, \destiny\ will perform a high-speed flyby of (3200) Phaethon and potentially more small-body targets.

This paper presented an overview of the mission profile, the spacecraft systems, the trajectory design approach for each phase of the mission, and the flight operation plan for the relevant phases.

Different methods of trajectory design are used for different parts of the mission. In the first of its three phases, the spacecraft will depart from a near-geostationary transfer orbit and gradually raise its apogee in a spiral trajectory, a low-thrust many-revolution trajectory. In this phase, we employ a multi-objective evolutionary algorithm that minimizes flight time, propellant consumption, and the duration of the longest eclipse. In the second phase, the final conditions from the spiral phase and the initial conditions of the interplanetary phase are bridged through several Moon flybys that provide the required energy boost and address phasing problems. We generate a database of high-fidelity Moon flyby transfers that are computed via backward propagation from the Earth escape condition. The third phase, the interplanetary portion of the mission, is designed using the trajectory design method of asteroid flyby cyclers under two-body dynamics. Joining all three phases, we perform low-thrust trajectory optimization in \minorblue{full} ephemeris of the baseline mission scenario. In this scenario, \destiny\ will be launched in July 2024 and will perform a flyby of Phaethon in January 2028.

Operational considerations for the critical parts of the mission profile are discussed. The flight operation plan for the spiral phase accounts for space environment effects, eclipses, communication windows, and navigation errors. A timeline of planned events for the Phaethon flyby sequence is provided; it groups the operations into asteroid detection, relative orbit control, and asteroid tracking.

\section*{Acknowledgment}
This work was supported by the Japan Society for the Promotion of Science KAKENHI grant number 19K15214. Ferran Gonzalez-Franquesa, Roger Gutierrez-Ramon, and Nishanth Pushparaj are supported by the Monbukagakusho Scholarship from 
\ifreview
\\
\fi
Japan's Ministry of Education, Culture, Sports, Science and Technology (MEXT). The authors would like to thank \destiny's project team members for their valuable comments.

\appendix
%
%
\section*{Appendix} \label{sec:appendix}

\subsection*{A. Final Moon Flyby Condition}\label{sec:final_lungar_swingby_condition}
\renewcommand{\theequation}{A.\arabic{equation} }
\setcounter{equation}{0}

We assume that the final Moon flyby epoch equals the Earth departure epoch in the IPT trajectory design. For each final Moon flyby epoch, we can obtain the Earth departure V-Infinity $\bm{v}_{\infty\oplus}^{\textrm{out}}$ from the IPT results and the final Moon flyby position $\bm{r}_{M}$ from the ephemeris. Given $\bm{v}_{\infty\oplus}^{\textrm{out}}$ and $\bm{r}_{M}$, let us calculate the velocity of the spacecraft after the final lunar flyby.

The magnitude of the velocity after the final flyby is
\begin{equation}
    v_M = \sqrt{\|\bm{v}_{\infty\oplus}^{\textrm{out}}\|^2 + \frac{2\textrm{GM}_{\oplus}}{r_M}},
\end{equation}
and the semi-major axis is 
\begin{equation}
    a = -\frac{\textrm{GM}_{\oplus}}{v_{\infty\oplus}^{\textrm{out}2}}.
\end{equation}
The velocity direction can be calculated by the following procedure.

\begin{figure}[htb]
\centering
\ifreview
\includegraphics[width=0.5\textwidth]{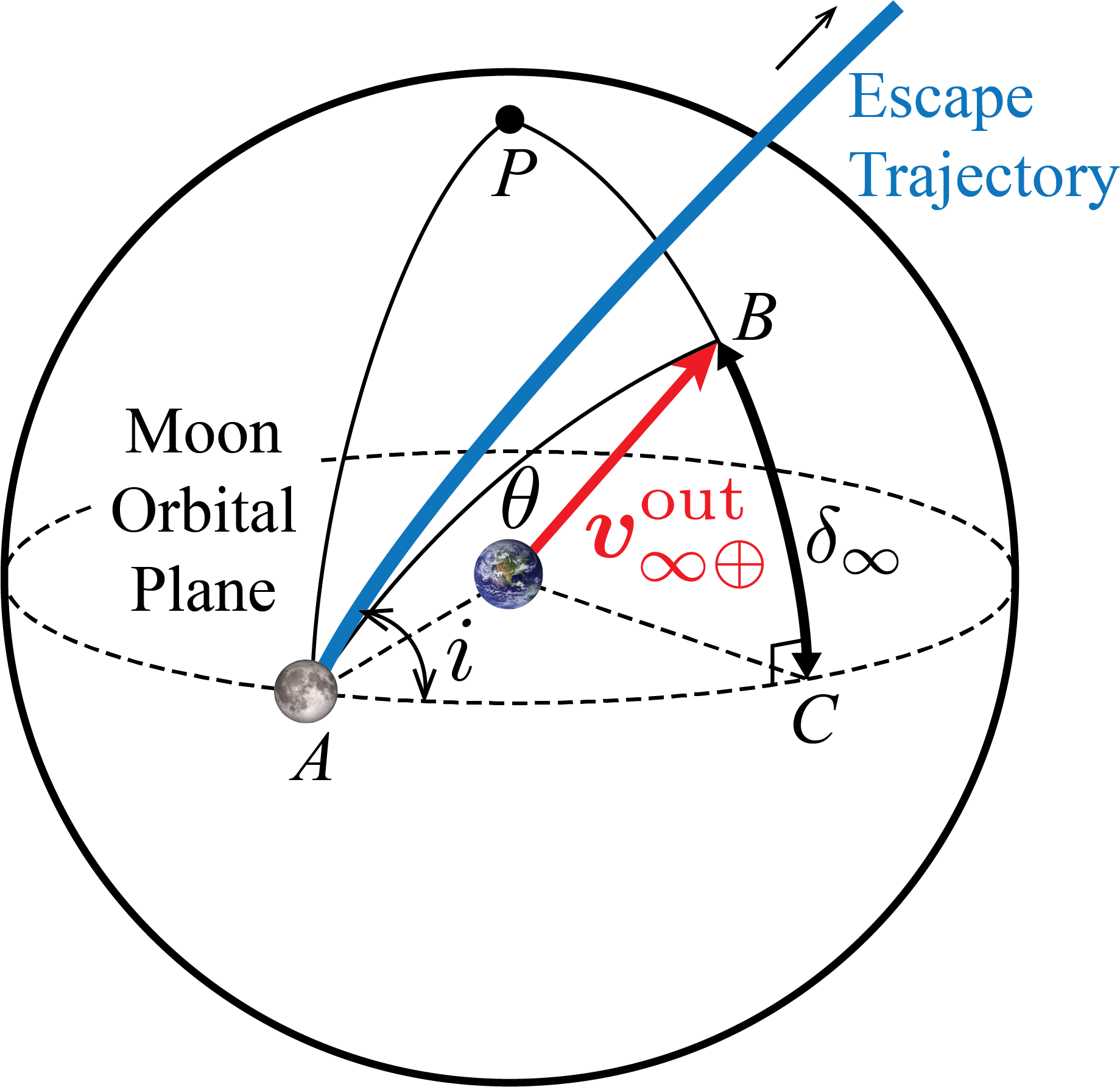}
\else
\includegraphics[width=0.25\textwidth]{img/final_lunar_flyby_outofplane.png}
\fi
\caption{Final lunar flyby geometry (out-of-plane view); the spacecraft leaves from the Moon and escapes the Earth through a hyperbolic trajectory.}
\label{fig:final_lunar_flyby_outofplane}
\end{figure}

The inclination of the hyperbolic orbit is given by applying the spherical trigonometry to $\triangle{ABC}$ as shown in Fig.~\ref{fig:final_lunar_flyby_outofplane}.
\begin{equation}
    i = \sin^{-1}\left(\frac{\sin\delta_{\infty}}{\sin\theta}\right)
\end{equation}
where $\delta_{\infty}$ is the longitude of $\bm{v}_{\infty\oplus}^{\textrm{out}}$ in the moon reference plane, and $\theta$ is defined by
\begin{equation}
    \theta = 
    \begin{cases}
    \cos^{-1}\left(\frac{\bm{r}_M\cdot\bm{v}_{\infty\oplus}^{\textrm{out}}}{\|\bm{r}_M\| \|\bm{v}_{\infty\oplus}^{\textrm{out}}\|}\right) & \textrm{ \ if short arc transfer}\\
    2\pi-\cos^{-1}\left(\frac{\bm{r}_M\cdot\bm{v}_{\infty\oplus}^{\textrm{out}}}{\|\bm{r}_M\| \|\bm{v}_{\infty\oplus}^{\textrm{out}}\|}\right) & \textrm{ \ if long arc transfer}
    \end{cases}
\end{equation}

\begin{figure}[htb]
\centering
\ifreview
\includegraphics[width=0.6\textwidth]{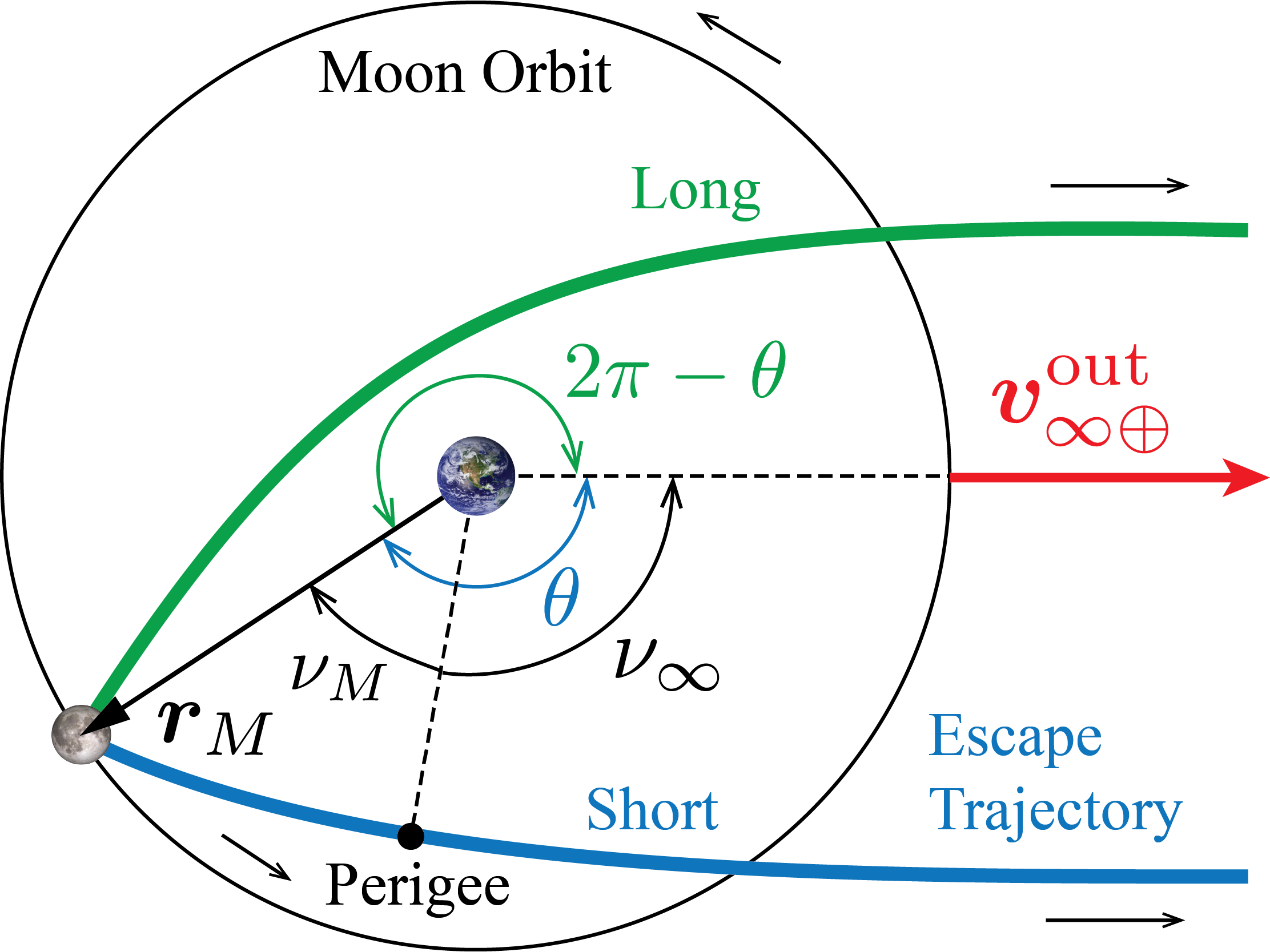}
\else
\includegraphics[width=0.3\textwidth]{img/final_lunar_flyby_inplane.png}
\fi
\caption{Final lunar flyby geometry for long and short escape trajectories (in-plane view).}
\label{fig:final_lunar_flyby_inplane}
\end{figure}

The eccentricity of the hyperbolic orbit is calculated by the following equations:
\begin{equation}
    e = \sqrt{\chi^2+1}
\end{equation}
where $\chi(\geq 0)$ is given by solving the following quadratic equation
\begin{equation}
    a\chi^2 + r_M\sin\theta \chi + (1-\cos\theta)r_M = 0
\end{equation}
This quadratic equation is obtained from 
\begin{align}
\begin{cases}
1+e\cos \nu_{\infty} &= 0\\
1+e\cos(\nu_{\infty}-\theta) &= \frac{a(1-e^2)}{r_M}
\end{cases}
\end{align}
where $\theta=\nu_{\infty}-\nu_{M}$ as shown in Fig. \ref{fig:final_lunar_flyby_inplane}.

Finally, we can calculate the radial and tangential components of the velocity.
\begin{align}
    v_r &= \sqrt{\frac{\textrm{GM}_{\oplus}}{a(1-e^2)}}e\sin(\nu_{\infty}-\theta)\\
    v_{\theta} &= \sqrt{\frac{\textrm{GM}_{\oplus}}{a(1-e^2)}}\left\{1+e\cos(\nu_{\infty}-\theta)\right\}
\end{align}

\bibliographystyle{elsarticle-num}
\bibliography{destiny_reference}

\end{document}